\RequirePackage{etoolbox}
\providetoggle{inBook}
\providetoggle{inJournal}
\toggletrue{inBook}
\togglefalse{inJournal}

\RequirePackage{etoolbox}
\providetoggle{inBook}
\providetoggle{inJournal}
\iftoggle{inBook}{}{\toggletrue{inJournal}}
\providetoggle{inLLNCS}

\iftoggle{inBook}{
}{}
\iftoggle{inLLNCS}{}{
    \documentclass[acmsmall,\iftoggle{inBook}{authorversion,nonacm}{}]{acmart}

    \author{Li Huang}
    \orcid{https://orcid.org/0000-0003-3531-4045}
    \email{li.huang@constructor.org}
    \affiliation{%
    	\institution{Constructor Institute of Technology}
    	\country{Switzerland}
    	\city{Schaffhausen}
    }
    
    \author{Sophie Ebersold}
    \orcid{https://orcid.org/0000-0002-0957-2844}
    \email{Sophie.Ebersold@irit.fr}
    \affiliation{%
    	\institution{IRIT CNRS UMR5505, University of Toulouse}
    	\country{France}
    	\city{Toulouse}
    }
    
    \author{Alexander Kogtenkov}
    \orcid{https://orcid.org/0000-0003-4873-8306}
    \email{kwaxer@mail.ru}
    \affiliation{%
    	\institution{Previously at Constructor Institute of Technology}
    	\country{Switzerland}
    	\city{Schaffhausen}
    }
    
    \author{Bertrand Meyer}
    \orcid{https://orcid.org/0000-0002-5985-7434}
    \email{Bertrand.Meyer@inf.ethz.ch}
    \affiliation{%
    	\institution{Constructor Institute of Technology}
    	\country{Switzerland}
    	\city{Schaffhausen}
    }
    
    \author{Yinling Liu}
    \orcid{https://orcid.org/0000-0002-9711-2118}
    \email{Yinling.Liu@univ-lorraine.fr}
    \affiliation{%
    	\institution{CRAN CNRS UMR 7039, University of Lorraine}
    	\country{France}
    	\city{Epinal}
    }
    
    \setcopyright{acmcopyright}
    \acmJournal{CSUR}
    \acmYear{2025} \acmVolume{1} \acmNumber{1} \acmArticle{1} \acmMonth{1}
    \acmDOI{10.1145/3448975}
    
}

\usepackage{hhline}
\usepackage{xcolor,colortbl}
\usepackage[utf8]{inputenc}
\usepackage{microtype}
\usepackage{glossaries}
\usepackage{enumitem}
\usepackage{graphicx}
\usepackage{url}
\usepackage{hyperref}
\usepackage{subcaption}
\usepackage{listings}
\usepackage{comment}
\usepackage{eiffel}
\usepackage{boogie}
\usepackage{dafny}
\usepackage{array,graphicx}
\usepackage{booktabs}
\usepackage{multirow}
\iftoggle{inLLNCS}{\usepackage{amsmath, amssymb}}{}
\usepackage{stackrel}

\providecommand{\Description}[2][]{}

\usepackage{calc}
\usepackage{tikz}
\usetikzlibrary{arrows.meta}
\usetikzlibrary{calc}
\usetikzlibrary{positioning}
\usepackage{tikzscale}

\usepackage{pgfplots, pgfplotstable}
\usepgfplotslibrary{dateplot}

\pgfplotsset{compat=1.17}

\pgfplotstablesort[sort key={Project},sort cmp={string <},col sep=tab]{\SystemsByProject}{years.txt}
\pgfplotstablesave[col sep=tab]{\SystemsByProject}{p_years.txt}
\pgfplotstablesort[sort key={Year},col sep=tab]{\SystemsByYear}{p_years.txt}
\pgfplotstablesave[skip rows between index={0}{4},columns={Project,Year,Approach}]{\SystemsByYear}{fs_years.txt}

\definecolor{ekwcolor}{HTML}{222277}

\definecolor{lightgray}{rgb}{0.92, 0.92, 0.92}

\usepackage{color}      
\newcommand{\gt}[1]{{\small\texttt{#1}}}

\newcommand{\e}[1]{\mbox{\lstinline[basicstyle=\normalsize]|#1|}}

\newcommand{\ekw}[1]{{\color[HTML]{222277}\bfseries\ttfamily{#1}}}

\begin{document}

\title{Lessons from Formally Verified Deployed Software Systems
\iftoggle{inJournal}{}%
    {\iftoggle{inBook}{(Extended Version)}{(Unknown version)}}%
}

\iftoggle{inLLNCS}{
    \author{Li Huang \and Sophie Ebersold \and Alexander Kogtenkov \and Yinling Liu \and Bertrand Meyer}

    \institute{Constructor Institute of Technology, Schaffhausen, Switzerland \and University of Toulouse, Toulouse, France, \and IRIT, Toulouse, France
}
}{

}

\begin{abstract}

The technology of formal software verification has made spectacular advances, but how much does it actually benefit the development of practical software? Considerable disagreement remains about the practicality of building systems with mechanically-checked proofs of correctness. Is this prospect confined to a few expensive, life-critical projects, or can the idea be applied to a wide segment of the software industry? To help answer this question, the present survey examines a range of projects, in various application areas, that have produced formally verified systems and deployed them for actual use. It considers the technologies used, the form of verification applied, the results obtained, and the lessons that the software industry should draw regarding its ability to benefit from formal verification techniques and tools.

\iftoggle{inBook}{
    Note: this version is the extended article, covering all the systems identified as relevant. A shorter version, covering only a selection, is also available \cite{shortversion}.
}{}

\end{abstract}

\maketitle


\newacronym{ACRE}{ACRE}{Approach for Context-based Requirement Engineering}
\newacronym{BON}{BON}{Business Object Notation}
\newacronym{CAS}{CAS}{Complex Adaptive System}
\newacronym{CSP}{CSP}{Constraint Satisfaction Problem}
\newacronym{DBC}{DbC}{Design by Contract}
\newacronym{DSL}{DSL}{Domain Specific Language}
\newacronym{DSML}{DSML}{Domain Specific Modeling Language}
\newacronym{EAST-ADL2}{EAST-ADL2}{Electronic Architecture \& Software Tools - Architecture Description Language}
\newacronym{GEMOC}{GEMOC}{Generic Model of Computation}
\newacronym{GORE}{GORE}{Goal-Oriented Requirements Engineering}
\newacronym{HDL}{HDL}{Hardware Description Language}
\newacronym{INCOSE}{INCOSE}{International Council on Systems Engineering}
\newacronym{LGS}{LGS}{Landing Gear System}
\newacronym{MARTE}{MARTE}{Modeling and Analysis of Real Time and Embedded systems}
\newacronym{MBSE}{MBSE}{Model-Based System Engineering}
\newacronym{MDE}{MDE}{Model Driven Engineering}
\newacronym{NL}{NL}{Natural Language}
\newacronym{OCL}{OCL}{Object Constraint Language}
\newacronym{RE}{RE}{Requirements Engineering}
\newacronym{SysML}{SysML}{Systems Modeling Language}
\newacronym{SoS}{SoS}{System of Systems}
\newacronym{STD}{STD}{State Transition Diagrams}
\newacronym{UML}{UML}{Unified Modeling Language}
\newacronym{URML}{URML}{User Requirements Modeling Language}
\newacronym{VV}{V\&V}{Verification and Validation}


\section{Introduction}\label{sec:introduction}

The ever more central role that software plays in all processes of the modern world brings to the forefront the question of program correctness. How do we know that software systems perform as expected? \textit{Formal verification} is the task of proving that a program fulfills its specification. Although is a long-established research area, disagreement persists on its relevance to mainstream system development. Many practitioners have not even heard of formal methods, and those who have often dismiss them as too hard to apply to mainstream projects. Against this view, proponents of formal verification argue that the technology is now mature and deployable. Which of these two views is correct? In other words, how realistic is the prospect of applying formal methods to production projects? Answering this question requires objective evidence: an assessment of existing attempts to apply formal methods in practice. The present survey provides such an assessment, by reviewing software systems fulfilling two properties: they have actually been deployed; and they were subjected, during their development, to formal verification.

Two or three decades ago, one could legitimately phrase the underlying question simply as: ``\textit{Can formal verification be applied in industry?}''. In that form, it is no longer open: a number of well-publicized industrial projects have used formal verification, even if initially for relatively small programs in mission-critical and particularly life-critical areas such as transportation and defense, where the consequences of incorrectness in programs are so great as to justify any difficulties and extra costs that formal verification might imply. The contemporary version of the question does not ask any more about feasibility (which has been established) but about practical aspects, such as: \textit{How large a system} can formal verification handle? \textit{What} \textit{special qualifications} or training does it require for the development team? \textit{What extra cost}s, if any, does it imply?

While it is beyond the scope of this survey to provide definitive answers to these and other questions on the practicality of formal verification, it introduces a factual basis for discussing them. It analyzes a number of deployed, formally verified systems according to a set of criteria listed in section \ref{guidelines}. The lessons learned from this analysis appear in section \ref{discussion_lessons_learned}.

\color{black}
\noindent The present article, which does go into technical details, resulted from studying a significant set of formally verified and deployed software systems of widely different kinds, extending across a variety of implementation languages and verification techniques. 
The projects were selected using a combination of literature review and responses to a questionnaire widely circulated by the authors.
This article helps answer the following questions: 
\begin{itemize}
\item 
In what areas of the industry have formally verified IT systems been deployed?
\item
What are the properties of the formally verified systems' projects in terms of required initial developer expertise, learning effort and effect on the software process? 
\item
What approaches (programming languages, mathematical basis, verification techniques, verification tools, verification schemes) have been applied to verify deployed systems?       
\item
What are the potential of and obstacles to generalizing the results to the software industry as a whole?  Are there specific kinds of systems that do not lend themselves to formal verification? 
\end{itemize}
The rest of the article is structured as follows:
\begin{itemize}
 \item 
 Section \ref{related-work} relates the findings to those of previous surveys on neighboring topics.
\color{black}     
\item 
Section \ref{selected_systems} presents the selection criteria for the analyzed systems, and the analysis criteria.
    \item 
Sections \ref{verification_concepts} and \ref{verification_approaches} provide a short tutorial on formal verification, focusing on the methods actually used in the projects under study.
\iftoggle{inBook}{
    \item 
Section \ref{sec:architecture} describes the structuring and application of a typical verification process applying these techniques.
}{}
    \item 
Section \ref{Descriptions} is the core of the article; it describes and assesses the individual projects and their use of formal verification, according to the criteria of section \ref{guidelines}. \item
Section \ref{discussion_lessons_learned} draws the general lessons of the analysis and examines its limitations as well as its significance for the generalization of formal verification in industry.
\end{itemize}

\section{Previous surveys} \label{related-work}
Table \ref{table:surveys} provides a point-by-point comparison with earlier surveys covering part of the scope of the present one. These articles and some of their main results are the following:

\color{black}
\begin{itemize}
\item Surveys of systems verified with a specific approach (Event-B in \cite{Abrial07formalmethods}, 
SPARK in \cite{chapman2014we}) conclude that regulators seem skeptical on the use of strong static analysis and theorem proving in critical software, and that the primary determinant of the practical success of B-style formal methods is the quality and power of the supporting tools.
\color{black}
\item Surveys of verified systems in a specific application area or of a specific kind (separation kernels \cite{zhao2017high}, 
distributed systems \cite{fonseca2017study}) 
 show that full formal verification has attracted large efforts in recent years. They conclude that verification effectively prevents protocol bugs. Verification relies on assumptions that must be tested, typically through testing toolchains. These studies do not cover multi-core verification or automatic code generation.
\color{black}

\item
A survey of approaches based on the concept of proof assistant \cite{ringer2019qed} has as its main finding that proof engineers succeed in continually increasing  productivity by adapting the results of research on programming practices and development tools.

\color{black}


\item
A 2009 survey \cite{Woodcock2009formal} presented a questionnaire-based summary of projects that had applied formal methods to some degree, not necessarily at the level of formally verified code.

\color{black}
\item
Recent surveys \cite{zhangsurvey, zhang2019survey} present (with less detail than here) a number of formally verified systems. 
They conclude that formal methods can be deployed in many industrial applications.
\item 
\color{black}The most recent survey \cite{ter2024formal} provides a broad scope of successful formal methods applications, including lithography manufacturing and cloud security in e-commerce. 
\end{itemize}
\color{black}



\vspace{-0.4cm}
\begin{table}[htb!]
\newcolumntype{L}[1]{>{\raggedright\let\newline\\\arraybackslash\hspace{0pt}}p{#1}}
\newcolumntype{C}[1]{>{\centering\let\newline\\\arraybackslash\hspace{0pt}}p{#1}}
\scriptsize
  \centering
  \caption{Comparison with major prior surveys}
  \begin{minipage}{\columnwidth}
    \vspace{-2ex}
      \textbf{Sel}: selection method (L: Literature review, Q: Questionnaire) 
      || \textbf{Num}: Number of systems surveyed
      || \textbf{Dep}: Deployability
      || \textbf{FML}: Formally verified (Yes|No)
      || \textbf{App}: approaches 
        (B: B-method, I: abstract Interpretation, M: Model checking, P: theorem Proving, R: Refinement, \mbox{S: axiomatic Semantics}, T: Testing)
      || \textbf{Discussion}: Project discussion aspects 
          (C: Components, V: Verified properties, X: project conteXt, D: Decisions, \mbox{T: Tool stack}, Y: stYle, W: softWare characteristics, P: Project characteristics) 
  \end{minipage}
  
 \begin{tabular}{C{15pt} L{14pt} C{12pt} C{60pt} C{1em} L{85pt} C{1em} L{50pt} L{58pt}}
    \toprule
    Ref & Sel & \makebox[1em][c]{Num} & Languages & \makebox[1em][c]{Dep} & Areas covered & \makebox[1em][c]{Fml} & App &  Discussion\\
    \midrule
    \cite{Abrial07formalmethods} & \makebox[1em][c]{L} & 1 & B & Y & Transport & Y & B 
    &
    \makebox[1em][c]{C}%
    \makebox[1em][c]{V}%
    \makebox[1em][c]{X}%
    \makebox[1em][c]{}%
    \makebox[1em][c]{T}%
    \makebox[1em][c]{}%
    \makebox[1em][c]{W}%
    \makebox[1em][c]{P}%
    \\
    \cite{chapman2014we} & \makebox[1em][c]{L} & 1 & SPARK & Y & Navy, Aeronautics & Y & %
    \makebox[1em][c]{}
    \makebox[1em][c]{}
    \makebox[1em][c]{}
    \makebox[1em][c]{P}
    &
    \makebox[1em][c]{}%
    \makebox[1em][c]{V}%
    \makebox[1em][c]{X}%
    \makebox[1em][c]{}%
    \makebox[1em][c]{T}%
    \makebox[1em][c]{}%
    \makebox[1em][c]{W}%
    \makebox[1em][c]{}%
    \\
    \cite{fonseca2017study} & \makebox[1em][c]{L} & 12 & Coq, Ivy, Dafny, TLA+ & N & Distributed systems  & Y &%
    \makebox[1em][c]{}
    \makebox[1em][c]{}
    \makebox[1em][c]{M}
    \makebox[1em][c]{P} 
    &
    \makebox[1em][c]{}%
    \makebox[1em][c]{V}%
    \makebox[1em][c]{}%
    \makebox[1em][c]{}%
    \makebox[1em][c]{T}%
    \makebox[1em][c]{}%
    \makebox[1em][c]{}%
    \makebox[1em][c]{}%
    \\
    \cite{ringer2019qed}& \makebox[1em][c]{L} & 41 & Coq, HOL4, Isabelle... & Y & Industry & Y &
    \makebox[1em][c]{}
    \makebox[1em][c]{}
    \makebox[1em][c]{}
    \makebox[1em][c]{P}
    &
    \makebox[1em][c]{}%
    \makebox[1em][c]{}%
    \makebox[1em][c]{X}%
    \makebox[1em][c]{}%
    \makebox[1em][c]{}%
    \makebox[1em][c]{}%
    \makebox[1em][c]{W}%
    \makebox[1em][c]{}%
    \\
    \cite{Woodcock2009formal} & \makebox[1em][c]{}\makebox[1em][c]{Q} & 10+ & Z, B, VDM, CSP & Y & National infrastructure, Microcode, Finance, Security & N  &
    \makebox[1em][c]{}
    \makebox[1em][c]{}
    \makebox[1em][c]{M}
    \makebox[1em][c]{P}
    \makebox[1em][c]{R}
    \makebox[1em][c]{}
    \makebox[1em][c]{T}
    &
    \makebox[1em][c]{C}%
    \makebox[1em][c]{}%
    \makebox[1em][c]{X}%
    \makebox[1em][c]{D}%
    \makebox[1em][c]{T}%
    \makebox[1em][c]{}%
    \makebox[1em][c]{}%
    \makebox[1em][c]{P}%
    \\
    \cite{zhangsurvey} & \makebox[1em][c]{L} & 13 & B, VDM, Z & Y & Transport, Military, Defense& Y & %
    \makebox[1em][c]{B}
    \makebox[1em][c]{}
    \makebox[1em][c]{}
    \makebox[1em][c]{P}
    &
    \makebox[1em][c]{}%
    \makebox[1em][c]{}%
    \makebox[1em][c]{X}%
    \makebox[1em][c]{}%
    \makebox[1em][c]{T}%
    \makebox[1em][c]{}%
    \makebox[1em][c]{}%
    \makebox[1em][c]{P}%
    \\
    \cite{zhang2019survey} & \makebox[1em][c]{L} & 15 & Coq, HOL, Isabelle & N & Mobile, Web servers & Y & 
    \makebox[1em][c]{}
    \makebox[1em][c]{}
    \makebox[1em][c]{}
    \makebox[1em][c]{P}
    &
    \makebox[1em][c]{}%
    \makebox[1em][c]{}%
    \makebox[1em][c]{}%
    \makebox[1em][c]{}%
    \makebox[1em][c]{T}%
    \makebox[1em][c]{}%
    \makebox[1em][c]{W}%
    \makebox[1em][c]{P}%
    \\
    \cite{zhao2017high} & \makebox[1em][c]{L} & 8+ & Isabelle, Coq & Y & Aeronautics, Military, Medicine & N &%
    \makebox[1em][c]{B}
    \makebox[1em][c]{}
    \makebox[1em][c]{M}
    \makebox[1em][c]{P}
    &
    \makebox[1em][c]{}%
    \makebox[1em][c]{V}%
    \makebox[1em][c]{X}%
    \makebox[1em][c]{}%
    \makebox[1em][c]{T}%
    \makebox[1em][c]{}%
    \makebox[1em][c]{W}%
    \makebox[1em][c]{}%
    \\

    \cite{ter2024formal}& \makebox[1em][c]{L}\makebox[1em][c]{Q} & 18 & B, Event-B, Frama-C, Coq & Y & Transport, Automotive, Aeronautics, OS, Network, Hardware, Manufacturing& Y &%
    \makebox[1em][c]{B}
    \makebox[1em][c]{}
    \makebox[1em][c]{M}
    \makebox[1em][c]{P}
    \makebox[1em][c]{}
    \makebox[1em][c]{}
    \makebox[1em][c]{T}
    &
    \makebox[1em][c]{}%
    \makebox[1em][c]{}%
    \makebox[1em][c]{}%
    \makebox[1em][c]{D}%
    \makebox[1em][c]{T}%
    \makebox[1em][c]{}%
    \makebox[1em][c]{}%
    \makebox[1em][c]{}%
    \\
    \textbf{This}& \makebox[1em][c]{L}\makebox[1em][c]{Q} & 30+ & Coq, SPARK, Frama-C, CBMC... & Y & Compiler, Library, OS, Aeronautics, Transport, Nuclear, Network, Database&Y&%
    \makebox[1em][c]{B}
    \makebox[1em][c]{I}
    \makebox[1em][c]{M}
    \makebox[1em][c]{P}
    \makebox[1em][c]{}
    \makebox[1em][c]{S}
    &
    \makebox[1em][c]{C}%
    \makebox[1em][c]{V}%
    \makebox[1em][c]{X}%
    \makebox[1em][c]{D}%
    \makebox[1em][c]{T}%
    \makebox[1em][c]{Y}%
    \makebox[1em][c]{W}%
    \makebox[1em][c]{P}%
    \\
    \bottomrule
    \end{tabular}%

  \label{table:surveys}%
\end{table}

\vspace{-0.35cm}
\noindent

 \noindent While earlier studies provide critical insights into tool capabilities, proof structures, and empirical correctness, they tend to focus on early-stage verification and engineering properties, under-emphasizing social, economic, and maintenance aspects. 
 The current survey stands out by also addressing what happens after deployment, drawing pragmatic, team-level and lifecycle lessons with direct advice for practitioners.

Other differences include the following:
\begin{itemize}
    \item This article is, to our knowledge, the most comprehensive as to the number approaches reviewed and goes into more detail about the surveyed projects' features and results.

    \item It highlights two major ways of dealing with unverified parts.

    \item It analyzes the reasons behind the performance impact of verification.

    \item It highlights obstacles to widespread adoption beyond the difficulty of verification itself.
\end{itemize}

\section{Selected systems} \label{selected_systems}
The software systems under review must be both ``formally verified'' and ``deployed''. A system is \emph{formally verified} if it has been mathematically proven to possess properties specified as part of its requirements.
A system is \emph{deployed} if it is either:
\begin{itemize}
    \item Publicly available on the Web (in which case the authors of this article were able to use it).
\item
Not publicly available, but with strong evidence that it is used in production by several users.
\end{itemize}

\noindent Omitting the second category would have excluded commercial, proprietary systems, which account for some of the most significant applications of formal verification. The downside is that analysis of these systems has to rely on available documents often  written by the authors of the respective systems, or interviews with these authors, rather than direct examination of the software. 






\subsection{Selection process}

While this article is not a Systematic Literature Review, it follows the recommended review protocol for SLRs \cite{keele2007guidelines}. The selection of relevant systems used a combination of literature search and a questionnaire  (which remains available \cite{questionnaire}). 
A key reason for including a questionnaire is that many formally verified systems from industry, relevant to the questions listed in Section \ref{sec:introduction}, do not appear in the scholarly literature. 
The questionnaire was distributed to numerous communities and Internet channels starting in December 2020 to identify and document systems that have received little or no attention in the literature. It also asked for contacts to enable further interviews with authors. 
\color{black}
It yielded responses on 20 systems, of which 10 were deemed relevant. Examination of the documentation on these systems, suggestions from various sources, a list of companies that use formal verification methods in software engineering \cite{practical-fm}, and the general literature on formal verification led, by transitive closure on the references, to the identification of 65 potentially relevant systems from December 2020 to October 2021.

\iftoggle{inJournal}{
After application of the selection criteria described below, 32 systems remained (Table \ref{table:systems}). 21 out of the 32 systems (highlighted in gray) are confirmed to be active --- they are either still being actively used or their new features are under development.
Systems that come from the replies of the questionnaire are \underline{underlined} in the second column of Table \ref{table:systems}. Systems were classified according to their application domain, yielding 9 categories.

For reasons of space, this article focuses on 11 of them, chosen for their representativeness (scale of project, levels of activity, verification and deployment) and to ensure good coverage of application fields and verification approaches.
The other systems listed in the table are discussed in the extended version of this article \cite{longversion}.
} 

At {\textcolor{blue}{\gt{https://github.com/huangl223/verified-software-systems}}}, readers will find a list of the relevant sites of the surveyed systems, including software release pages and software repositories. In addition, the authors maintain a living review of deployed and formally verified systems, continuing to collect new information as it comes out \cite{livingreview}.

\iftoggle{inBook}{
In the shorter version of this article \cite{shortversion}, only 11 of them appear, selected for their representativeness.
}

\begin{table*}[htbp]
    \iftoggle{inLLNCS}{%
        \tiny%
    }{%
        \scriptsize%
    }%
  \centering
  \caption{Deployed and Verified Systems Surveyed}
    \begin{tabular}{|p{30pt}|p{33pt}|p{70pt}|p{23pt}|p{28pt}|p{31pt}|p{18pt}|c|p{12pt}|l|}
    \hline
    \multicolumn{1}{|c|}{Cat\-e\-go\-ry} & \multicolumn{1}{c|}{Name} & \multicolumn{1}{c|}{Verified properties} & Input PL & Output PL & Proof engine & Spec/\discretionary{}{}{}Imp & KLoC & Ef\-fort (py) & \multicolumn{1}{c|}{Ref\-er\-ences} \\
    \hhline{|=|=|=|=|=|=|=|=|=|=|}
    \multirow[t]{6}{=}[-1ex]{Compiler} & \cellcolor{lightgray} \textbf{CompCert} & semantics preservation & Coq & OCaml, C  &  Coq & 1  & 135 & 6 & \cite{leroy2009compcert,leroy2016compcert,kastner2018compcert}  
    \\ \cline{2-10}
           & \cellcolor{lightgray} \underline{CakeML} & semantics preservation & CakeML & CakeML & HOL4 & -- & 100 & -- & 
           \cite{kumar2014cakeml,tan2015verified,gueneau2017verified} 
            \\\cline{2-10}
           & \cellcolor{lightgray} Vellvm &  semantics preservation & Coq & OCaml, C  & Coq & 1 & 32 & -- & \cite{zhao2012mechanized,zhao2012formalizing,zhao2013formal}
           \\ \cline{2-10}
           & \cellcolor{lightgray} Vericert & semantics preservation & Coq & OCaml & Coq & 3.63 & 2.5 & 1.5 & \cite{herklotz2021formal}
          \\ \cline{2-10}
           & \cellcolor{lightgray} Vermillion & functional correctness & Coq & OCaml & Coq & -- & 8 & 0.75 & \cite{Sam:19}
          \\ \cline{2-10}
           & \cellcolor{lightgray} \underline{Q*Cert}
           & functional correctness & Coq & OCaml & Coq & -- & -- & -- &
           \cite{Joshua:17}
    \\  \hhline{|=|=|=|=|=|=|=|=|=|=|}

    \multirow[t]{3}{=}[-1ex]{Library} & \cellcolor{lightgray}\textbf{Eiffel\-Base2}
    & functional correctness & Eiffel & Eiffel & AutoProof, Boogie, Z3 & -- & 9.37 & 6 & \cite{polikarpova2015fully}
           \\ \cline{2-10}
           & \cellcolor{lightgray} \textbf{\textbf{HACL*}}
           & security, functional correctness, memory safety & F* & C
           & Z3 & 3.3 & 31 & $<$ 1 &
           \cite{zinzindohoue2017hacl,haclrepo}
            \\\cline{2-10}
              & DICE*
           & security, functional correctness, memory safety & F* & C
           & Z3, Meta-F* & 4.7 & 29 & - &
           \cite{tao2021dice}
            \\\cline{2-10}
              & Signal*
           & security, functional correctness, memory safety & F* & Web\-Assembly
           & Z3, ProVerif & - & 4 & - &
           \cite{protzenko2019signal}
            \\\cline{2-10}
           & \cellcolor{lightgray} Amazon s2n & functional correctness, security &
           C & C & Coq, SAW & 0.86 & 0.7 & $>$ 3
           & \cite{chudnov2018continuous,s2nrepo}
    \\ \hhline{|=|=|=|=|=|=|=|=|=|=|}
    \multirow[t]{14}{=}[-1ex]{OS} & \cellcolor{lightgray} \textbf{\underline{seL4}} & functional correctness, security & C & C & Isabelle, Z3, Sonolar & 100 & $>$ 10 & 31.2 &
    \cite{seL4-whitepaper,klein2014comprehensive}
    \\ \cline{2-10}
           & \cellcolor{lightgray} Proven\-Core & security & C & C & Proven\-Tools & -- & -- & 3 & \cite{lescuyer2015provencore}
            \\\cline{2-10}
           & Verve & safety & Beat & C\#, TAL & Boogie, Z3 & 3
           & 7.55 & 0.75 &
           \cite{yang2010safe} 
           \\ \cline{2-10}
           & \cellcolor{lightgray} \textbf{mCerti\-KOS} & security, functional correctness & Coq & ClightX, LAsm & 
           Coq & 6 & 3 & 1 &
           \cite{gu2015deep,gu2016certikos,costanzo2016end} 
          \\ \cline{2-10}
           & \cellcolor{lightgray} mC2 & security, functional correctness & Coq
           &  ClightX, LAsm 
          & Coq & 17 & 6.5 & 2 & \cite{gu2016certikos}
          \\ \cline{2-10}
           & \cellcolor{lightgray} Hyper-V & functional correctness, safety & C & Assembly, C & Boogie, Z3 & 5 & 105 & 1.5 & \cite{leinenbach2009verifying}
          \\ \cline{2-10}
           & \cellcolor{lightgray} PikeOS & functional correctness, security & C & Assembly, C 
           & Boogie, Z3  & 5 & -- & 1.5 & \cite{baumann2012lessons}
          \\ \cline{2-10}
           & \textbf{Express\-OS} & security & C\#, Dafny & C\# & Z3 & 0.028 & -- & -- & \cite{Mai2013Verifying}
         
    \\ \hhline{|=|=|=|=|=|=|=|=|=|=|}
    \multirow[t]{4}{=}[-1ex]{Aero\-nau\-tics}
           & \underline{SHOLIS} & functional correctness, security, tim\-ing\discretionary{}{}{}/\discretionary{}{}{}memory constraints & SPARK & SPARK & 
Simplifier, Proof Checker & 3.07 & 27 & 19 &
\cite{king1999value}  
           \\ \cline{2-10}
           & \cellcolor{lightgray} Lockheed C130J & functional correctness & 
           SPARK & SPARK & Simplifier, Proof Checker& -- & 350 & -- &
           \cite{croxford1995breaking}
            \\\cline{2-10}
           & \cellcolor{lightgray} \textbf{\underline{NATS} \underline{iFACTS}} & absence of run-time exceptions, functional correctness, memory safety & SPARK & SPARK & Simplifier    & 
           0.3 & 250 & $>$~50 & 
           \cite{chapman2014we,chapman2016industrial}
           \\ \cline{2-10}
           & \underline{Euro\-Fighter} \underline{Typhoon} & functional correctness & Ada & Ada & Supertac,
ProofPower & -- & 35 & 1.5 &
\cite{stich2012clearance,o2013automated}
    \\ \hhline{|=|=|=|=|=|=|=|=|=|=|}
     \multirow[t]{2}{=}[-1ex]{Trans\-port} & \textbf{Roissy Shuttle} &  security & B, Ada & Ada & EDiTh B, Bertille & 1.16 & 158 & -- & \cite{badeau2005using}
    \\ \cline{2-10}
           & \cellcolor{lightgray} Dutch Tunnel CS &  safety and liveness & mCRL2 & Java & 
           mCRL2, VerCors & 0.15 & 37 & 8 
           &\cite{DutchTunnelControlSystem}
    \\ \hhline{|=|=|=|=|=|=|=|=|=|=|}
    Nuclear Power & \textbf{Sizewell B} & functional correctness & B & PL/M-86, Assembly & MALPAS & -- & 150 & 250 & \cite{10.1007/978-1-4471-2061-2_19}
    \\ \hhline{|=|=|=|=|=|=|=|=|=|=|}
        \multirow[t]{4}{=}[-1ex]{Network}
           & \cellcolor{lightgray} Ironclad Apps & functional correctness, security & Dafny & Dafny & Boogie, Z3 & 0.5 & 85 & 3 &
           \cite{floyd1993assigning}
            \\\cline{2-10}
            & \underline{Quark} & security & Coq & OCaml & 
            Coq, Ynot & 0.2 & 976 & 0.83 &
            \cite{jang2012establishing,jang2014language,quarkweb} 
           \\ \cline{2-10}
           & \cellcolor{lightgray} \textbf{CoCon}
           & security & Isabelle/\discretionary{}{}{}HOL & Scala & 
           Isabelle/\discretionary{}{}{}HOL, BD-Security & 0.67 & 15.2 & 0.25 
           & \cite{popescu2021cocon,kanav2014conference}  
           \\ \cline{2-10}
           & \cellcolor{lightgray} CoSMed & security & Isabelle/\discretionary{}{}{}HOL & Scala & 
           Isabelle/\discretionary{}{}{}HOL, BD-Security & 1 (ker)
0.33 (app) & 7.4 & 0.33
           & \cite{Cosmed1}
    \\  \hhline{|=|=|=|=|=|=|=|=|=|=|}
        Database 
           & \textbf{Ynot} & functional correctness & Coq & OCaml & Coq & 1 & 3.03 & 8 &
         \cite{malecha2010toward}
    \\ \hhline{|=|=|=|=|=|=|=|=|=|=|}
       {Verifica\-tion Tool} & Flover
       & soundness with respect to standard semantics & Coq & HOL4 &
Coq, HOL4 & 10 & 3 & 2 & \cite{becker2018verified,floverrepo}
    \\ \hline
    \end{tabular}%
  \label{table:systems}%
\end{table*} 




\subsection{Criteria for the analysis}
\label{guidelines}
The description of the selected systems (in section \ref{Descriptions}) uses the following criteria: 
\begin{itemize}
\item \textbf{Scope}: What system was verified, in what application domain and for what purpose?  
\item \textbf{Components}: What are the verified components and their roles?
\item \textbf{Verified properties}: What properties of the system were verified?
\item \textbf{Project context}: What is the motivation behind the project? 
\item \textbf{Decisions}: What key decisions were made to help the verification process?
\item \textbf{Tool stack}: What tools were used for developing and verifying the software?
\item \textbf{Style}: What is the underlying approach to specification?
\item \textbf{Software characteristics}: What results did the verification effort produce? 


\item \textbf{Project characteristics}: What amount of resources was spent to verify the software?
\item \textbf{Lessons}: What are the conclusions about the verification experience?
\end{itemize}

\footnotesize
\noindent Blank entries in Table \ref{table:systems} express that the corresponding information was not available.



\normalsize
\section{Verification: basic concepts} \label{verification_concepts}

\subsection{Definitions}

The term ``verification'' covers techniques that ascertain the correctness of programs. It has both broad and narrow meanings:

\begin{itemize}
\item In practice, the software industry mostly uses ``verification'' to mean testing. In the programming research literature, the term is used instead to denote techniques for \textit{proving} correctness mathematically; the meaning also retained in the present article.
\item Software engineering textbooks often distinguish verification from \textit{validation} (using ``V \& V'' for their combination): verification is internal (checking consistency, ``the systems does things right'') and validation is external (checking the program against its specification, ``the system does the right things''). In line with the usual sense of ``formal verification'' in the programming research community, this article uses ``verification'' to cover both parts.
\item Verification is ``dynamic'' if it needs to execute the program, as in the case of testing. It is ``static'' if it works only from the program text. This article focuses on static techniques (which, since they do not perform any execution, require neither a compiler nor input data).
\item Program proving is the most ambitious form of static verification, meant to ascertain that the program satisfies its specification. If the specification covers all relevant aspects of the program behavior, the proof will demonstrate ``full functional correctness''.
\item Other forms of static analysis only analyze the program for the presence or absence of specific faults, such as deadlock or memory overflow. 
\end{itemize}

The following characteristics apply to both static and dynamic forms of verification:
\begin{itemize}
\item Verification is not debugging. It may find faults (``bugs''), but correcting them is not its job.
\item The words ``succeed'' and ``fail'' mean something else for verification tools and for programs. A verification tool \textit{succeeds} if it either finds faults or ascertains
that the program contains no faults of the specified kinds. The tool \textit{fails} if it is unable to decide either way. (In contrast, a program succeeds if it completes its job according to its specification, and fails otherwise. Verification can succeed on a failing program --- by identifying the fault --- and conversely.)
\end{itemize}

\subsection{Inherent theoretical and practical limitations} \label{limitations}
It is well known that a passing test, or any number of them, do not demonstrate that the program is correct. (A non-passing test does prove that the program is incorrect.) In principle, then, mathematical proofs are superior to tests. Limitations remain, however.
First, a failed proof (as defined above) does not indicate that the program is incorrect, only indicates that the proof tool is unable to decide either way (correct or incorrect). This case is frequent for both theoretical and practical reasons:

\begin{itemize}
\item Theoretically: program correctness for any useful programming language is an undecidable problem, in the sense that it is not possible to produce a tool that will always prove or disprove the correctness of any program in finite time. (This property does not preclude the production of tools that will yield proofs for \textit{some} programs.)
\item Practically: a failed proof attempt is not the end of the story, just a step. (``Losing a battle, not the war''.) It  is in fact the common experience in the day-to-day practice of verification: try to verify; find that the tool either reveals a fault or produces no result; in either case, tweak the program, the specification or the proof to try to correct the problem; repeat.
\end{itemize}

\noindent Such an iterative scheme is reminiscent of the ``run a test, fix a bug, repeat'' of traditional non-formal development using tests. The difference is that, in static approaches, the basic iterative step involves analyzing the text of the program (rather than executing it). But in both cases, the process is iterative  (as opposed to an ideal two-step process of writing the program then running a tool to prove its correctness). As a result, even though modern verification tools are described as permitting ``automatic''  or ``mechanical'' verification, their actual use still involves human effort. How much effort is an important practical factor to consider when assessing verification tools and techniques. This survey attempts, whenever information is available, to provide assessments of the effort that was involved in the verification of the reviewed systems.

\subsection{The annotation issue}
It is only meaningful to verify a program against specified properties. As noted, these properties can specify the entire relevant behavior of the program (``full correctness'') or only some of its characteristics, for example that the computation will not produce an arithmetic overflow.

The first approach obviously yields more benefits, but it requires an extra \textit{annotation} effort in addition to the program, since it needs a specification of the intended properties. The programmer or whoever is performing the verification must equip the program with such properties and often, in practice, intermediate ``proof obligations''. The choice between verification approaches is in part a tradeoff between the amount of annotation effort and the extent of properties proved. 

\subsection{Maintining realistic expectations} \label{expectations}

Any verification success is relative: we verify a certain program element against a certain specification under certain hypotheses. The \textit{specification} may be wrong (in the sense of not correctly expressing the desired behavior) or incomplete. The \textit{hypotheses} may be violated; for example a program written in a high-level language, even if certified correct by a verifier for that language, can still fail if the compiler does not respect the semantics of the language. Only a few compilers have themselves been certified correct (one example is CompCert, is the topic of section \ref{system:compcert}), subject to the same observations. These limitations generically apply to all descriptions of  verification successes and to any statement that a program has been proved ``bug-free'' (a qualification which does not appear in this article).  To guarantee that a system is ``bug-free'' one would have to prove that the  specifications are correct and complete, the models are consistent with the code, and the compilation preserves the exact semantics of the source code without introducing any bugs.





\section{Verification approaches}\label{verification_approaches}

The verification of the systems reviewed here relies on a variety of approaches and frameworks.

\subsection{\textbf{Axiomatic semantics}}

Axiomatic semantic (also called 
 Hoare logic or Floyd-Hoare-Dijkstra semantics) is one of the most widespread formal frameworks. It uses the most annotations and addresses full correctness.
\vspace{-0.4cm}
\begin{lstlisting}[float=h,xleftmargin=2em]
sqrt (x: REAL; epsilon: REAL): REAL
		-- Non-negative square root of 'x' with precision 'epsilon'
	require
		x >= 0
	do
		... Algorithm to compute into Result the square root approximation ...
	ensure 
		Result >= 0
		abs (Result ^ 2 - x) <= epsilon
	end
\end{lstlisting}

\vspace{-0,3cm}
Axiomatic semantics assumes that the program is equipped with ``verification conditions'', also called ``assertions'', which may include routine preconditions and postconditions, loop invariants, class invariants (in object-oriented programming), as well as conditions included at arbitrary program places. 

A verification condition is a boolean-valued function on program states (or, in the case of postconditions, on two program states, initial and final). Proofs of termination of loops and recursive routines additionally require integer ``variants''. An example of a routine annotated with a precondition and a postcondition is, in the notation of the Eiffel programming language

\noindent The precondition (\e{require}) expresses a property of \e{x} in the original state, at the time of a call; the postcondition (\e{ensure}) expresses a property of the \e{Result}, in relation to the value of \e{x}. The combination of the precondition and postcondition of a routine is its specification, or ``contract''.

Verification consists of proving that the implementation matches this specification. It relies on a set of rules (axioms and inference rules) associated with the programming language. An example inference rule (with \{P\} A \{Q\} stating that if P, a verification condition, is satisfied prior to the execution of A, then Q will hold afterwards) characterizes the sequencing of instructions as usually represented by the ``;'' symbol in programming languages:

\begin{equation}
    \frac{\{P \} A \{Q\}  \{Q \} B \{R\}}{\{P \} A ; B \{R\}}
\end{equation}

\noindent The part above the line is a hypothesis; if that hypothesis is satisfied, the inference rule makes it possible to infer the conclusion below the line. The rule states that the sequence A ; B, assuming P on start, will produce R on end if there is an intermediate condition Q such that A ensures Q from P and B ensures R from Q. This sequencing rule is typical of how the rules of axiomatic semantics enable reasoning formally about programs. They can be automated and fed into a proof tool. They make it possible to specify the effect of a program, typically through preconditions and postconditions, and to use the proof tool to prove that the program actually produces that effect. To achieve this result, it is necessary to provide enough verification conditions to guide the proof tool.

\subsection{\textbf{Abstract interpretation}}

Abstract interpretation is a mathematical framework for performing static analyses of programs, based on mapping concrete domains of execution values onto more abstract domains, so that analyses of important properties (such as arithmetic overflow, pointer nullness or safety constraints), which would be prohibitive on a concrete domain, can be performed on the abstract domain with its results still valid at the concrete level. Abstractions satisfying such properties are so-called \textit{Galois connections}, enjoying conservation properties both ways (concrete to abstract and back). 

Determining specific properties of relevant program properties involves writing a set of equations applying on the variables of the program, based on its control flow graph, data flow graph or both. These equations are mutually recursive, implying that the way to obtain a solution is to compute a \textit{fixpoint} of the equations through an iterative method. Such fixpoint computation, and prior to it the construction of the graph and its conversion into a set of equations, are usually impractical or even impossible for the program being verified, if only because the ``concrete domains'' in which program variables take their values are very large or infinite. Abstract interpretation will instead perform the computation on an abstracted version of the program, in an abstract domain.  To ensure that the fixpoint on the abstract domain is reached after a finite set of iterations, it may be necessary to simplify the abstracted computation further through \textit{narrowing} and \textit{widening} operations.

A simple example of abstraction could serve to determine whether a variable \textit{x} can ever be zero at a certain program point, where the instruction involves a division by \textit{x}. Assuming for simplicity that the original values (concrete domain) are integers, the abstract domain will only include five values, representing zero (Z), Positive (P), Negative (N), Bottom (representing impossible values) and Top (representing all possible values), with a partial order relation ``less defined than'' defining a lattice structure. The operations on numbers transpose in the abstract domain: for example Z + Z = Z - Z = Z, P + P = P - N = P, N + N = N - P = N, but P + N = N + P = N - N = P - P = Top (since the last operations may yield a result of any of the categories). The static analysis in the abstract domain can determine whether the abstract value of \textit{x} can ever be Z, much more easily than if we attempt to perform a similar analysis on the concrete program and domain. Under the appropriate conditions, results obtained in the abstract domain can be mapped back to the original.



\subsection{\textbf{Model-checking}}
Model-checking is the result of a ``why not?'' reaction to the accepted wisdom about the impossibility of certain tasks. Exhaustive testing is well known to be impossible in practice, since the number of cases would be infinite. On further analysis, however:
\begin{itemize}
\item Computers are finite automata; integers as represented by computers, for example, do not form an infinite set but one limited to (typically) 2$^{32}$ or 2$^{64}$ values. Similarly, the number of times the body of an ordinary ``while'' loop can be executed is unbounded in principle; but in practice the number of iterations is, in any program execution,  not only finite but bounded (since a loop taking 100 years to execute would be of no interest).
\item These sizes are extraordinarily large at the human scale, making the number of possible states for any realistic program appear, at first sight, intractable. But modern computers are very powerful, executing billions of operations a second, which may make it possible to achieve the seemingly absurd goal of exhaustive state-space search, perhaps not for the program itself (except in elementary cases) but for a simplified version known as a model.

\end{itemize}

\vspace{-0.1cm}\noindent An elementary example of a program model is a ``boolean model'' which replaces every integer variable by a boolean one, with False standing for 0 and True for any other integer value, dramatically reducing the number of possible states (an integer variable now yields two states instead of e.g. 2$^{64}$). Another example of a technique for fighting ``state explosion'' is loop unfolding, which replaces every loop by a scheme executing the body 1 to N times, often for a small N (typically 2 or 3). 

The model-checking algorithm will then verify a specified property, such as liveness (non-deadlock) or non-starvation for a concurrent program, by constructing the set of all possible states of the model program and trying to find a ``counter-example'': an execution path that leads to a state violating the desired property. This step may or may not be the end of the story:
\begin{itemize}
    \item If there is a violation of the property (a fault) in the original program, it will (for the appropriate kinds of property) persist in the model. Then, if the state space exploration does not produce a counter-example, it definitely proves that the original program is free from the fault.
   \item Finding the fault in a state of the model is, however, not conclusive: the fault might be in the original, or it might be an artifact of the reduction to a model. For example, m $\neq$ n between integer variables holds for m = 1 and n = 2, but in a boolean model both values map to True, allowing the model-checker to find a counter-example. Such cases require refining the model. 
\end{itemize}



\subsection{B} \label{Bmethod}

Most approaches to verification analyze a program to determine whether it is correct, regardless of how it was produced in the first place. ``Correctness by construction'' denotes a different process: build software so that it is correct, intertwining the construction and verification efforts. The B method \cite{EventB} applies this idea, combined with the notion of \textit{refinement}. Most systems using B rely on the ``Event B'' variant, which adds to the basic framework the notion of event. 

A refinement process starts with a very high-level view of the program, involving variables defining a state, abstract events affecting that state, and invariants. As an example, in a system controlling access to a road segment undergoing repair work, an invariant could state that all cars in the system are either stopped or traveling one-way (all East-West or all West-East). The events might be ``Let a car enter East'', ``Let a car exit East'', the same for West, ``Car arrives East''  and ``Car arrives West''. Variables include: numbers $we$ and $ww$ of cars waiting on the East and West sides, current direction $d$ of travel (boolean), number $n$ of cars currently traveling on the road segment, maximum number $N$ on it. Each event is defined by its effect on the variables (and hence the state); for example each ``Enter''  event increases $n$ by one, decreases the respective waiting variable ($we$ or $ww$) by one, and leaves the other variables unchanged.
Invariants in this example include $n \geq 0$ and $n \leq N$. Events can have guards, meaning conditions that must be satisfied for an event to occur; for example, the guard for ``enter east''  is that the direction of travel is East to West, $we > 0$ and $n < N$. The fundamental correctness rule is that every event, when executed with its guard satisfied as well as the invariants, must ensure that the invariant is satisfied again.

The B method works by stepwise refinement. Each step, refining an existing (``abstract'') model into a new (``concrete'') one,  may introduce new events, new invariant properties, and more specific versions of the abstract events. The refinement is correct if the new events preserve both the concrete and abstract invariants. For example, we might refine the road construction model by introducing traffic lights on both ends, with events such as going from green to orange, orange to red etc. on either of them. New invariant properties appear (for example, if the East light is green the West light must be red). We may refine ``Let a car enter East'' by including the change of light to green. All new versions must satisfy the preceding properties and the new ones.

Refinement proceeds until it has reached a level of detail where the result is explicit enough to be directly implemented in a programming language. With the possible exception of this last translation step, the result is correct by construction, since every step has been proved correct in the sense of invariant preservation. 

B-specific proof tools support the method, to prove at each step that new events preserve invariants and that the new invariants imply those of the preceding refinement level. 

\vspace{-0.3 cm}

\subsection{SMT solvers} \label{SMT}

One way to prove a ``universal'' property, stating that all elements of a set $E$ satisfy a property $P$, is to prove the absence of a counter-example; in other words, to prove that it is impossible to find an element of $E$ that satisfies $\lnot P$, the negation of $P$.

The properties $P$ of interest, and their negations, are boolean formulae. Disproving $P$ means finding a variable assignment that satisfies  $\lnot P$. The boolean satisfiability problem is NP-complete, potentially requiring unrealistic computation time, but for theories meeting specific criteria, known as satisfiability modulo theories (SMT), effective algorithms are possible. SMT solvers apply this discovery and lie at the basis of many modern proof tools.

\vspace{-0.3 cm}

\subsection{Z} \label{Z}

The Z specification language, a predecessor of B, is a formal specification language based on set theory. Z makes it possible to specify systems in terms of sets and operations on them, represented by mathematical functions and relations. Associated tools support both consistency proofs of the specifications themselves and proofs that programs in specific languages satisfy the specification.

\iftoggle{inBook}{
\subsection{Simulink}
Simulink \cite{simulink} is a graphical modeling tool of MathWorks\textsuperscript{\textregistered} for simulating and analyzing dynamic systems. It provides various types of \emph{blocks}, such as primitive-, control flow-, and temporal-blocks, as well as predefined libraries for modeling dynamic systems and code generation. Simulink also supports the definition of custom blocks via Stateflow \cite{stateflow} diagrams or \emph{user-defined function} blocks, namely \emph{S-Functions}, written in C, C++, or Fortran. Simulink models are executable and their execution results are displayed as simulation traces.
Fig. \ref{fig:simulink_example} shows a Simulink modeling a dynamic system: it uses Newton’s second law to calculates acceleration based on force, and integrates to get the velocity, and then further integrates to obtain the position.

 \begin{figure}[htbp]
  \centering
  \includegraphics[width=9cm]{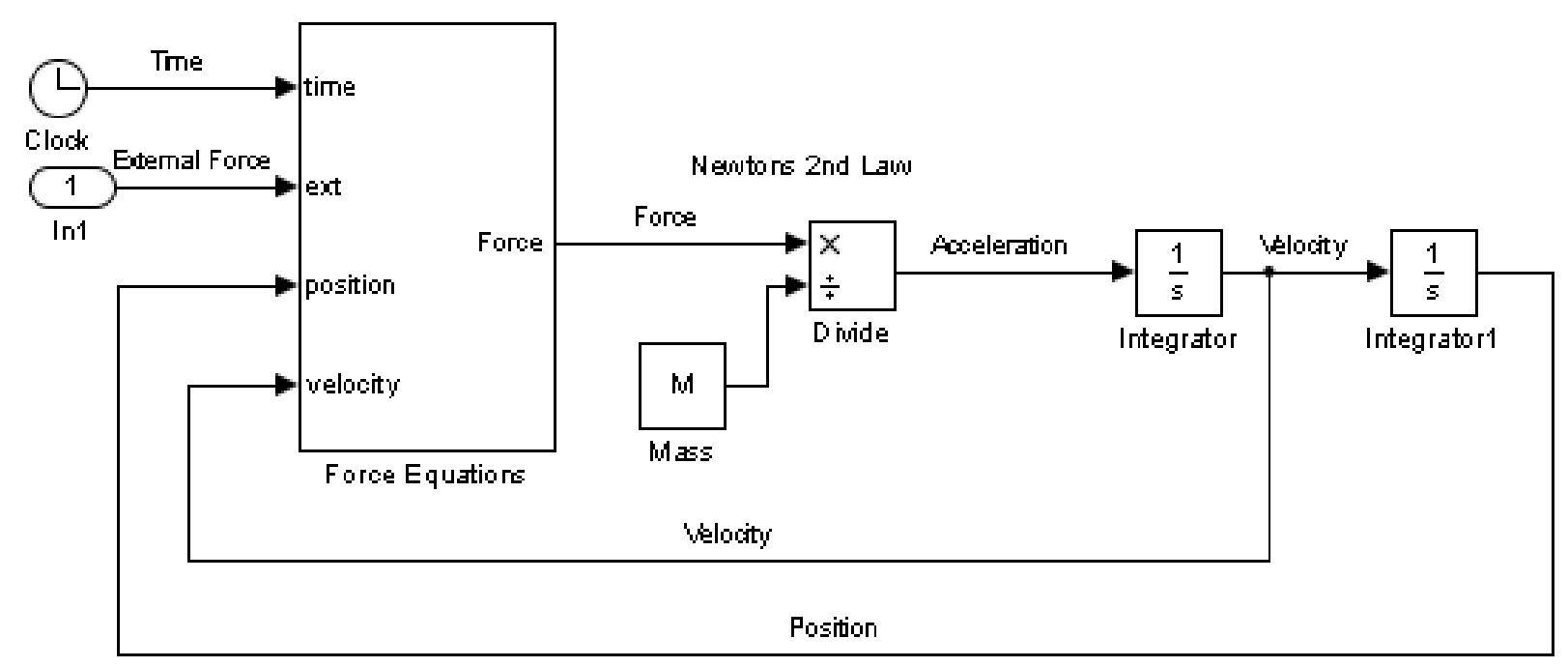}
  \caption{An example Simulink model}
  \label{fig:simulink_example}
\end{figure}
}{}


\vspace{-0.3 cm}

\subsection{HOL}
HOL (Higher-Order Logic) is a mathematical-logic framework underlying by two of the frameworks used by systems in this survey, HOL4 \cite{hol-itp} and Isabelle/HOL \cite{nipkow2002isabelle,isabelle}. As reflected by the name, it can include logic of several increasing orders: propositional (zero-order), predicate calculus, second-order (with quantification over relations), third-order (with quantification over sets of sets). It also supports typed $\lambda$-calculus with object-level polymorphism \cite{history2014itp}.

\vspace{-0.3 cm}

\subsection{Coq (now Rocq)} \label{Coq}
The Rocq framework \cite{Rocq:25}, referred to in this article's system descriptions as Coq, its name until March 2025,  relies on a different logical basis: constructive intuitionistic type theory. The Rocq language serves as both a specification language and a programming language. A formal proof in Rocq, according to the Curry–Howard correspondence, has an associated program with a suitable type.
It is possible to extract a program, guaranteed correct, directly from the proof.

\subsection{Labeled Transition Systems} \label{lts}
A Labeled Transition System (LTS) \cite{leroy2016compcert} is a mathematical relation $current\ state \stackrel{trace}{\longrightarrow} next\ state$ that describes one step of execution of a program and its effect on the program state. The sequence of transitions from an initial state defines the observable behavior. A finite sequence describes a terminating program execution, where the program terminates either normally or with a run-time error. An infinite sequence of transitions describes a  program execution that runs forever.

\subsection{Other verification approaches} \label{others}

The approaches covered above do not represent the full extent of verification techniques. In particular, an interesting approach has garnered attention in recent years: runtime verification (RV) \cite{bartocci_rv} \cite{falcone_rv} \cite{sanchez_survey_2019}, a light-weight formal verification approach that analyzes execution traces of systems. RV usually involves running the system under analysis, monitoring the system's executions, and checking the system's conformance to formal specifications based on the runtime behaviors. A typical RV process consists of three stages: 
\begin{itemize}
\item Monitor synthesis: generate, from a formally specified property, a \emph{monitor}, a decision procedure that receives \emph{events} (describing system actions or states) during execution and emits associated \emph{verdicts} (indicating whether the property is satisfied). 
\item System instrumentation: modifying the system to enable production of events during execution; those events are the inputs to monitors. 
\item Execution analysis: analyze execution, either step-wise or a posteriori. 
\end{itemize}
Compared to static formal verification approaches (such as model checking and theorem proving), RV is more applicable, as most of the systems are inherently executable; RV may, however, produce false negatives if not all execution traces are monitored.

\iftoggle{inBook}{\section{The context of verification} \label{sec:architecture}

The framework in which a project achieves the formal verification of its code consists of more than just a single verifier tool. It includes a whole compendium of tools and relies on a specific scheme for carrying out the verification. This section presents:

\begin{itemize}
    \item The roster of tools on which the verifier depends, which we may call the ``verification profile'' (\ref{sec:verification profile}).
    \item The overall ``verification scheme'' presiding over the verification process (\ref{sec:verification scheme}).
\end{itemize}


\subsection{Verification profiles}
\label{sec:verification profile}


A verification profile provides a higher-level picture of the verification status of all the components, related to a given software. 
It is so a (recursive) collection of all the dependencies, at compile-time and run-time, of a verified system. 
For convenience, a verification profile is documented in a tabular format as depicted below. 
Every column reflects dependencies of a particular program component; each component appears on a dedicated line.

\newcommand{\verified}{\textbf}
\newcommand{\NA}{\multicolumn{1}{c}{$-$}}
\begin{center}
	\footnotesize
	\begin{tabular}{l c c c c c} 
		\toprule
		\multicolumn{1}{c}{Program}  & \multicolumn{1}{c}{Verifier} & \multicolumn{1}{c}{Compiler} & \multicolumn{1}{c}{Execution} & \multicolumn{1}{c}{OS} & \multicolumn{1}{c}{Other} \\
		\midrule
		\verified{Application} & V$_1$ &  C$_1$ & E$_1$ & O$_1$ & D$_1$ \\
		V$_1$ & V$_2$ &  C$_2$ & E$_2$ & O$_2$ & D$_2$ \\
		C$_1$ & V$_3$ &  C$_3$ & E$_3$ & O$_3$ & D$_3$ \\
		\multicolumn{6}{c}{...} \\
		D$_n$ & V$_m$ &  C$_m$ & E$_m$ & O$_m$ & D$_m$ \\
		\bottomrule
	\end{tabular}
	\label{table:verification_dependencies}
\end{center}

The program Application is verified by V$_1$, compiled by C$_1$, executed on E$_1$ with operating system O$_1$ using a software D$_1$. The verifier V$_1$ itself is verified by V$_2$, compiled by C$_2$ etc.
The software involved in developing and running a verified system can be classified as follows:

\begin{enumerate}
 \item Compilers;  
\item Resource managers; 

\item Verifiers, 
which fall into two major categories:
\begin{enumerate}
    \item Auto-active verification tools \cite{HATCLIFF2012} (like VCC \cite{dahlweid2009vcc} or SPARK \cite{barnes:03}) which either report that the program is correct or that they cannot prove it. 
    \item Interactive theorem provers (like Coq or Isabelle) which give the details of the proof at every step and so provide means to update the verification context. They use a proof script or a special proof language (as Isar in Isabelle \cite{Isabelle/Isar}). In addition, such environments provide code extraction utilities to generate code from the proof-oriented language to a general-purpose programming language like OCaml \footnote{\url{https://ocaml.org/docs/}} or Scala \cite{odersky2004overview}. 
\end{enumerate}

\item Execution engines. 
The program can run in a virtual machine, in interpreted mode, compiled at the load time, or mixed. 
Such a virtual machine becomes the dependency that affects the whole system correctness. 

\item Other software: software outside the verified system that is still essential to perform the functions of this last one. They include libraries, components and services such as databases, network tunnels, logging facilities, device drivers and so on.
\end{enumerate}

As an example, the following table \verified{highlights} the verified components in the verification profile of HACL* (Section \ref{system:hacl*}): 
%
HACL* is a cryptographic library written and verified in F*. The code extractor KreMLin generates C code which can be compiled, among others, by the verified CompCert compiler (cf. section \ref{system:compcert}). Apart from HACL* and CompCert, everything else is unverified.

    \begin{center}
    	\footnotesize
    	\begin{tabular}{l c c c c c}
    		\toprule
    		\multicolumn{1}{c}{Program}  & \multicolumn{1}{c}{Verifier} & \multicolumn{1}{c}{Compiler} & \multicolumn{1}{c}{Execution} & \multicolumn{1}{c}{OS} & \multicolumn{1}{c}{Other} \\
    		\midrule
    		\verified{HACL*} & F* &  KreMLin, \verified{CompCert} & ARM-64 & Linux & libc \\
    		F* & \multicolumn{1}{c}{$-$} &  F*, OCaml & ARM-64 & Linux & libc \\
    		KreMLin & \NA &  OCaml & ARM-64 & Linux & libc \\
    		\verified{CompCert} & Coq &  Coq, OCaml & ARM-64 & Linux & libc \\
    		Coq & \NA &  OCaml & ARM-64 & Linux & libc \\
    		OCaml & \NA & OCaml& ARM-64 & Linux & libc \\
    		libc & \NA & gcc &ARM-64 & Linux & \NA \\
    		gcc & \NA & gcc & ARM-64 & Linux & libc \\
    		Linux & \NA & gcc & ARM-64 & \NA & \NA\\
    		\bottomrule
        \end{tabular}
    \end{center}

\iftoggle{inBook}{
    Hyper-V (section \ref{system:HyperV}): the correctness of this hypervisor is checked by VCC that translates the verification checks into the Boogie verification language, Boogie is written in C\# and uses Z3 as the backend to do the checks. The correctness of the verification relies on the correctness of MSVC and associated run-time libraries, .NET and Windows operating system, none of which are completely formally verified.

    \begin{center}
    	\footnotesize
    	\begin{tabular}{l c c c c c}
    		\toprule
    		\multicolumn{1}{c}{Program}  & \multicolumn{1}{c}{Verifier} & \multicolumn{1}{c}{Compiler} & \multicolumn{1}{c}{Execution} & \multicolumn{1}{c}{OS} & \multicolumn{1}{c}{Other} \\
    		\midrule
    		\verified{Hyper-V} & VCC &  MSVC* & x86-64 & Windows & libc\\
    		VCC & \NA &  csc & x86-64 & Windows & Boogie\\
    		MSVC & \NA &  MSVC & x86-64 & Windows & \NA\\
    		csc & \NA &  csc & x86-64 & Windows & \NA\\
    		csc & \NA & csc & x86-64 & Windows & .NET\\
    		Boogie & \NA & csc & x86-64 & Windows & Z3\\
    		.NET & \NA & csc, MSVC & x86-64 & Windows & \NA\\
    		Z3 & \NA & MSVC & x86-64 & Windows & \NA\\
    		Windows & \NA & csc, MSVC & x86-64 & \NA & \NA\\
    		\bottomrule
        \end{tabular}
    \end{center}
}{}

    grep in CakeML \iftoggle{inBook}{Section \ref{system:CakeML}}{\cite{tan2019verified,cakemlrepo,kumar2014cakeml,tan2015verified}}: this Unix-like standard utility is written in CakeML, the verified compiler that --- unlike the other verified compiler, CompCert (Section \ref{system:compcert}) --- is bootstrapped.

   \begin{center}
    	\footnotesize
    	\begin{tabular}{l c c c c c}
    		\toprule
    		\multicolumn{1}{c}{Program}  & \multicolumn{1}{c}{Verifier} & \multicolumn{1}{c}{Compiler} & \multicolumn{1}{c}{Execution} & \multicolumn{1}{c}{OS} & \multicolumn{1}{c}{Other} \\
    		\midrule
    		\verified{grep (CakeML)} & HOL4 &  \verified{CakeML} & x86-64 & Linux & libc\\
    		\verified{CakeML} & HOL4 &  \verified{CakeML} & x86-64 & Linux & libc\\
    		HOL4 & \NA &  Poly/ML & x86-64 & Linux & libc\\
    		Poly/ML & \NA & Poly/ML & x86-64 & Linux & libc\\
    		libc & \NA &  gcc & x86-64 & Linux & \NA\\
    		gcc & \NA & gcc & x86-64 & Linux & libc\\
    		Linux & \NA & gcc & x86-64 & \NA & \NA\\
    		\bottomrule
        \end{tabular}
    \end{center}

    Milawa \cite{myreen2014reflective} (not included in the survey because it does not satisfy all selection criteria (Section \ref{selected_systems}), but mentioned here as a good example of the current achievements in building an unusually deep verification profile). This theorem prover is formally verified by several levels of Milawa (from 2 to 11) which support gradually more complex versions of the proof language, with at Level 2 the Milawa Kernel, the simplest one, and at Level 11 the full Milawa prover, the most complicated one. The soundness of Milawa Kernel is formally proved in HOL4. All levels of Milawa are compiled with Jitawa \cite{myreen2011verified}, formally verified in HOL4 as well.

    \begin{center}
    	\footnotesize
    	\begin{tabular}{l c c c c c}
    		\toprule
    		\multicolumn{1}{c}{Program}  & \multicolumn{1}{c}{Verifier} & \multicolumn{1}{c}{Compiler} & \multicolumn{1}{c}{Execution} & \multicolumn{1}{c}{OS} & \multicolumn{1}{c}{Other} \\
    		\midrule
    		\verified{Milawa} & \verified{Milawa L11} &  \verified{Jitawa} & x86-64 & Linux & libc\\
    		\verified{Milawa L11} & \verified{Milawa L10} &  Jitawa & x86-64 & Linux & libc\\
    		\multicolumn{6}{c}{\textellipsis}\\
    		\verified{Milawa L2} & \verified{Milawa Kernel} &  Jitawa & x86-64 & Linux & libc\\
    		\verified{Milawa Kernel} & HOL4 &  \verified{Jitawa} & x86-64 & Linux & libc\\
    		\verified{Jitawa} & HOL4 & x86 text & x86-64 & Linux & libc\\
    		HOL4 & \NA & Poly/ML & x86-64 & Linux & libc\\
    		Poly/ML & \NA & Poly/ML & x86-64 & Linux & libc\\
    		libc & \NA & gcc & x86-64 & Linux & \NA\\
    		gcc & \NA & gcc & x86-64 & Linux & libc\\
    		Linux & \NA & gcc & x86-64 & \NA & \NA\\
    		\bottomrule
        \end{tabular}%
    \end{center}

Is it feasible to verify all components in a given verification profile? There are several examples of verified compilers (CompCert, CakeML \iftoggle{inBook}{section \ref{system:CakeML}}{\cite{tan2019verified,kumar2014cakeml,tan2015verified}}), libraries (HACL*), OS kernels (Sel4 section \ref{system:seL4}, mCertiKOS section \ref{system:mCertikOS}, ExpressOS section \ref{system:ExpressOS}, PikeOS \iftoggle{inBook}{section \ref{system:PikeOS}}{\cite{baumann2011proving,baumann2012lessons}}). Moreover, compilers, libraries and OSes can be involved in the process of bootstrapping where they are compiled using themselves and reaching a so-called program fixed point. A practical example of this possibility is CakeML \cite{cakemlrepo}.




\iftoggle{inBook}{
How far can we go? What about the verifier? Quis custodiet ipsos custodes? (Latin: Who will guard the guards?) 
An experiment with Milawa demonstrates that verification of the verifier is possible to some extent, but in the end, Milawa Kernel is verified by another theorem prover, unverified. 
Why can it not verify itself? The answer depends on what we mean by verification. The second Gödel incompleteness theorem states that a formal system cannot demonstrate its own consistency (it is impossible to prove that the verifier does not lead to a contradiction using the verifier itself). It should be possible, howevere, to demonstrate that the verifier follows the given formal system to check the theorems. Such a verifier could be called "bootstrapped" similar to compiler bootstrapping process. At the time of writing, we are unaware of any verifier that is fully bootstrapped (in the sense that it takes into account the Gödel's restriction). The work in progress, Metamath Zero \cite{mm0repo,carneiro2020metamath} lists this as one of its goals. When completed, it would be potentially possible to mark all cells in the verification profile as verified.
}{}

\subsection{Verification schemes}
\label{sec:verification scheme}

\begin{figure}
	\centering
    \begin{subfigure}{0.33\textwidth}
        \resizebox{\textwidth}{!}{\begin{tikzpicture}[-{Stealth}, rectangle, very thick]
	\tikzstyle{P} = [draw, rounded corners=1ex]
	\tikzstyle{T} = [draw, fill=green!20]
	\node[P, fill=yellow!20] (C) {Proof + Code\vphantom{p}};
	\node[T, below left=3ex and -2em of C,anchor=north east] (I) {
		\begin{tabular}{c}
			FM-based\\
			\midrule
			B\\
			Event-B
		\end{tabular}
		\begin{tabular}{c}
			ITP\\
			\midrule
			Coq\\
			Isabelle
		\end{tabular}
	};
	\node[T, below right=3ex and -2em of C,anchor=north west] (E) {Code generator};
	\node[P, below=3ex of E] (G) {Generated code\vphantom{p}};
	\node[T, below=3ex of G] (R) {Compiler};
	\node[P, below=3ex of R] (B) {Binary};
	\draw (C) -> (I);
	\draw (C) -> (E);
	\draw (E) -> (G);
	\draw (G) -> (R);
	\draw (R) -> (B);
\end{tikzpicture}\unskip}
    	\caption{Proof-based: the code is extracted from the proof.}
    	\label{fig:scheme-proof}
    \end{subfigure}
    \hfill
    \begin{subfigure}{0.25\textwidth}
    	\centering
        \resizebox{\textwidth}{!}{\begin{tikzpicture}[-{Stealth}, rectangle, very thick]
	\tikzstyle{P} = [draw, rounded corners=1ex]
	\tikzstyle{T} = [draw, fill=green!20]
	\node[P, fill=yellow!20] (C) {Code + Specification};
	\node[T, below left=4ex and -4ex of C,anchor=north east] (I) {
		\begin{tabular}{c}
			Verifier\\
			\midrule
			Boogie\\
			VCC\\
			Dafny
		\end{tabular}
	};
	\node[T, below right=4ex and -4ex of C] (R) {Compiler};
	\node[P, below=3ex of R] (B) {Binary};
	\draw (C) -> (I);
	\draw (C) -> (R);
	\draw (R) -> (B);
\end{tikzpicture}\unskip}
    	\caption{Specification-based: the code is annotated with specifications.}
    	\Description[The code annotated with specifications serves as in input for both the verifier and the compiler.]{The verifier uses the code annotated with specifications in the single source for checking. The compiler generates an executable (binary) file directly from the same source.}
    	\label{fig:scheme-spec}
    \end{subfigure}
    \hfill
    \begin{subfigure}{0.33\textwidth}
    	\centering
        \resizebox{\textwidth}{!}{\begin{tikzpicture}[-{Stealth}, rectangle, very thick]
	\tikzstyle{P} = [draw, rounded corners=1ex]
	\tikzstyle{T} = [draw, fill=green!20]
	\node[P, fill=yellow!20] (C) {Code\vphantom{p}};
	\node[T, right=2em of C] (R) {Compiler};
	\node[P, right=2em of R] (B) {Binary};
	\node[T, below=3ex of R] (D) {Decompiler};
	\node[P, below=3ex of D] (A) {Abstract code\vphantom{p}};
	\node[P, left=of D, fill=yellow!20, yshift=-3ex] (S) {Specification};
	\node[T, below left=2ex of A] (I) {
		\begin{tabular}{c}
			ITP/Verifier\\
			\midrule
			HOL\\
			VerCors
		\end{tabular}
	};
	\draw (C) -> (R);
	\draw (R) -> (B);
	\draw (C) -> (D);
	\draw (B) -> (D);
	\draw (D) -> (A);
	\draw (S) -> (I);
	\draw (A) -> (I);
\end{tikzpicture}\unskip}
    	\caption{Separate: the code and the specification are separate.}
    	\Description[The code and the specification come from different sources.]{A decompiler produces the format suitable for the verifier from the source or generated code. The verifier takes this output and a separate specification to perform the checks.}
    	\label{fig:scheme-split}
    \end{subfigure}
    \caption{Verification schemes (programs are rectangles with rounded corners and tools are rectangles with sharp corners).}
\end{figure}

\begin{table}
\centering
\begin{tabular}{l c c c}
\toprule
\multicolumn{1}{c}{\multirow{2}[2]{*}{Property}} & \multicolumn{3}{c}{Verification scheme} \\ \cmidrule(){2-4}
& Proof-based & Specification-based & Separate \\ \midrule
Are code and specification coupled? & Yes & Yes & No \\
Is verification mandatory? & Yes & No & No \\
Is code/model extraction required? & Yes & No & Yes \\
Interface architecture & Interactive & Auto-active & Both \\ \bottomrule
\end{tabular}
\caption{Classification of verification schemes.}
\label{tab:verification_schemes_classification}
\end{table}

Software verification is the process of proving that a certain program satisfies certain properties. The three elements involved are the program, the specification of its properties, and the proof. Depending on the relative role of these three parts, we can distinguish three overall strategies: proof-based, specification-based, and separate, defined as follows (see also the summary in Table \ref{tab:verification_schemes_classification}):

\begin{itemize}
    \item In the proof-based approach (Fig. \ref{fig:scheme-proof}), the starting point is a possibly abstract program equipped with a proof; the verification process consists of making the program more concrete, while preserving the proof at every step. The original program can be expressed at a very high level of abstraction, not suitable for direct use, but amenable to proof through a mechanized theorem prover. It can then be refined progressively into more and more concrete versions, up to a final, realistic version. The key property of this approach is that at every step the current version, regardless of its level of abstraction, is entirely proven correct, under the control of the proof tool. Examples of the proof-based approach include the Event-B method \cite{EventB} and supporting tools, which support a refinement-based software development process starting from a high level of abstraction and introducing more details in each step, and techniques based on the Coq and Isabelle/HOL proof systems.
    
    \item In the specification-based approach (Fig. \ref{fig:scheme-spec}), the starting point is a program equipped with a specification of its intended properties; the verification process consists of developing the proof. The specification can be expressed in the same programming language as the program (using its mechanisms for expressing boolean properties), or in a different language intended only for specification. The verification process consists of proving that the program satisfies the specification. In practice, this process is typically interactive, requiring the programmer to add extra annotations (``verification conditions'') to help the prover. Examples of specification-based approach are ``Design by Contract'' techniques associated with languages such as Eiffel and JML, where program can include specification elements such as pre-conditions, post-conditions and class invariants. Proof tools such as Boogie then check the implementations of routines and classes against these contracts. 
    
    \item In the separate approach (Fig. \ref{fig:scheme-split}), the starting point is just a program; the specification of expected properties is provided separately, using any adequate formalism. The verification process consists of proving that the program, as given, satisfies the properties. Examples of this approach include abstract interpretation and model-checking tools, which produce abstracted versions of a program on which certain properties can be automatically checked.
\end{itemize}

Each approach has advantages and limitations:

\begin{itemize}
    \item In the proof-based approach, correctness is guaranteed by construction: it is simply not possible to run the code without having run the proof as well. The disadvantage is that this scheme promotes a development scheme that is radically different from the way most of the software industry works: instead of writing a program in a standard programming language, developers have to write a proof-equipped program in a highly abstract, mathematical-style language, then subject it to a process of refinement until it becomes realistic enough for direct use.
    
    \item In the specification-based approach, the language can be closer to the mainstream, although it still has to support special specification constructs. The process of getting the prover to validate the code against the specification can be arduous.
    
    \item In the separate approach, there is no disruption to standard processes of software development, and the scheme can support programming languages that have not been designed with verification in mind. The downside is that the range of provable properties can be smaller, since the specification is disconnected from the code; model-checking, for example, typically verifies specific properties of the code, such as liveness, rather than full functional correctness.
\end{itemize}
}{}

\section{The systems: Descriptions} \label{Descriptions}
The present section describes the selected systems and reviews them through the ten criteria of the columns in table \ref{table:systems} (\ref{guidelines}). The criteria are mostly self-explanatory, but note the following:

\begin{itemize}
\item Verified Properties: properties being verified, e.g. some properties only (such as termination or liveness), or full functional correctness.
\item Programming language(s): languages used for the implementation (not the specification). 
\item Specification/implementation (Spec/Imp) ratio: estimate of  ratio between lines of specification/annotation and lines of implementation code. 

\item Effort (py): estimate of development and verification effort, in person-years. 

\end{itemize}


\subsection{CompCert}
\label{system:compcert}
\iftoggle{inBook}{
\begin{table*}[htbp]
 \scriptsize
  \centering
  {\renewcommand{\arraystretch}{1.3}
        \begin{tabular}{|p{40pt}|p{45pt}|p{55pt}|p{55pt}|p{45pt}|p{43pt}|p{40pt}|}
    \hline
     Nature & {Properties} & {Programming language(s)} & {Programming style} & {Verification framework} &{ Specification language(s)} & {Number of LOC} \\
    \hline
     Software development tool
 & Semantics preservation &  Coq, OCaml, C & Dependently-typed functional & Higher-order type theory & Coq & 135k 
    \\    \hline \hline
      {Spec/Imp}  & {Proof Engine} & {Team Background} & {Institution(s)} & {Development period} & {Effort}  & {Size of user community} \\
    \hline
    1 & Coq  & Researchers  & Inria Paris, University of Rennes 1 - IRISA, AbsInt Angewandte Informatik GmbH, Saarland University & 2005 - present (commercially available in 2015) & 6 & 25 contributors, 67 watchers, 972 stars, 150 forks
    \\ \hline
    \end{tabular}%
    }
  \label{compcert_id}%
\end{table*}%

}{}

\subsubsection{Scope}
CompCert is a compiler for the C programming language, intended for compiling life-  and mission-critical software that must meet high levels of assurance.

\subsubsection{Components}
The CompCert project focuses on compilation, excluding preprocessor, assembler, and linker. The latter components are unverified and come from a legacy compilation tool chain. The compiler supports almost all of the C language (ISO C99) \cite{leroy2016compcert}, generating code for the PowerPC, ARM, RISC-V and x86 processors.

\subsubsection{Verified properties}
The compiler includes multiple passes. It formally verifies semantic preservation between the input and output of every pass. To that end it provides formal semantics for every source, intermediate and target language, from C to assembly; such semantics defines the set of all possible program behaviors \cite{leroy2016compcert}, including termination (normal or abnormal). 
These formally verified properties only apply to the correctness of the compiler itself, without guaranteeing the correctness of the compiled software or the absence of harmful events such as null-pointer de-referencing. CompCert's formal semantics for C and assembly were hand-crafted; while there is a high degree of confidence that it accurately describes these languages, this is not formally guaranteed since the languages themselves are not formally defined. 

\subsubsection{Project context}
The  goal of CompCert is to avoid incorrect compilation, which would generate incorrect machine code from a correct source program. Many production compilers have bugs due to the complexity of code generation and optimization. Then, even if the source code of a program has been proved correct, the version actually executed may produce the condition of violations that the program's formal proof was supposed to have rooted out.

\subsubsection{Decisions}

The compiler is split into 20 passes using 11 intermediate languages \cite{kastner2018compcert}. The unverified parts deal with initial preprocessing, type refinement and language-specific simplifications. The formally verified passes follow the conventional multi-pass compiler structure and include parsing, front-end compiler, back-end compiler and assembling. The final assembling and linking steps involve standard non-verified tools. 

The staged implementation makes it possible to reason in a modular way about each pass and postpone the verification of certain passes (such as parsing, assembly and linking) to a later step. 
As long as the intermediate language between two successive passes is the same, the proof of semantics preservation for both passes guarantees semantics preservation for their combination.

The verified compiler code is extracted automatically from the proofs in Coq. The code of the extractor has been proved only on paper, not mechanically \cite{letouzey2002new}. The Coq documentation explicitly states that in some cases the translation from Coq to OCaml may be unsound, for example out of difference between the type systems, and that CompCert developers are responsible for making sure no bad events (such as integer overflow) occur during compiler runs.

A  validation tool called Valex compensates for current absence of formal proofs for assembling and linking. It reads and disassembles the generated executable, then compares it with the output produced by CompCert to ensure that there are no injections or other changes.

\subsubsection{Tool stack}
The verified components of the compiler are written in Coq to guarantee their correctness. Given the formal specification and the associated constructive proof, 
Coq extracts a certified program from the proof in OCaml. The extracted code becomes part of CompCert.

\subsubsection{Style}
The semantics of every language is specified in small-step operational style as a labeled transition system (section \ref{lts}). Because the behavior of a program can be nondeterministic, 
the specification relies on a refinement of the allowed behaviors, replacing every non-deterministic behavior with a more deterministic one  (by proving 15 simulation diagrams for each intermediate translator independently and then composing them to establish semantic preservation for the whole compiler).

The specifications and proofs are written in Coq. The soundness theorem \cite{leroy2009compcert} states that if the source code $S$ satisfies the specification $Spec$ (written $S \models Spec$): all observable behaviors $B$ of $S$ satisfy $Spec$ ($\forall B.\ S \Downarrow B \implies Spec(B)$), so does the compiled program $C$:
\begin{displaymath}
    S \models Spec \implies C \models Spec
\end{displaymath}

This result comes from two other theorems. One states that every intermediate translator guarantees that the target program $P_{tgt}$ simulates the source program $P_{src}$: $P_{src} \succsim P_{tgt}$. Another one \cite{song2019compcertm} shows that in this case, the behavior of the target program (the set of its execution traces) reproduces the one of the source program: $\mathsf{Beh}(P_{src}) \supseteq \mathsf{Beh}(P_{tgt})$.

\subsubsection{Software characteristics}

The source code had 100,000 lines of Coq in 2016 \cite{leroy2016compcert}. Various checks indicate that CompCert is more reliable than GCC and LLVM. While the 
test-generation tool Csmith \cite{yang2011finding} found 79 bugs in GCC and 202 in LLVM with randomized differential testing,  after 6 CPU-years of testing CompCert, focusing on bugs in its code generator, it found none in the verified parts. The potentially more effective tool Orion \cite{le2014emi} found 147 unique bugs in GCC and LLVM with the ``Equivalence Modulo Input'' compiler validation methodology, but none in CompCert. Although these results do not guarantee the absence of bugs, they give very high confidence in the compiler.

\subsubsection{Project characteristics}
The project started around 2005 \cite{compcertweb} and resulted in more than 30 publications (conferences, journals and books), as well as 3 PhDs. The first public version of CompCert (1.2) was released in 2008. The first commercial version has been available since 2015 \cite{leroy2016compcert}. As of 2016, the project had required 6 person-years (code + verification). 

\subsubsection{Lessons}
The project demonstrates the feasibility of writing a formally verified compiler for a popular programming language. Correctness comes at a cost: generated programs run 10–20\% slower compared to GCC4 with optimizations turned on.
Verified compilation alone, however, turns out to be insufficient for industrial applications. Established tool chains require further extension of the compiler (e.g., to produce debug information \cite{leroy2016compcert}), as well as refactoring of existing source code (to move compiler-dependent pragmas and hand-coded inline assembly code to the run-time environment \cite{kastner2018compcert}). The pipelined nature of compilation supports almost independent development and verification of more than 10 compilation passes. Unverified parts of the compiler rely on additional validation tools to guarantee correctness of the output.

\iftoggle{inBook}{
\subsection{CakeML}
\label{system:CakeML}

\begin{table*}[htbp]
 \scriptsize
  \centering
  {\renewcommand{\arraystretch}{1.3}
        \begin{tabular}{|p{40pt}|p{45pt}|p{55pt}|p{55pt}|p{45pt}|p{43pt}|p{40pt}|}
    \hline
     Nature & {Properties} & {Programming language(s)} & {Programming style} & {Verification framework} &{ Specification language(s)} & {Number of LOC} \\
    \hline
    Software development tool
 & Semantics preservation &  CakeML & Functional & HOL in a functional big-step style & HOL4 & 100k  
    \\    \hline \hline
      {Spec/Imp}  & {Proof Engine} & {Team Background} & {Institution(s)} & {Development period} & {Effort}  & {Size of user community} \\
    \hline
    Inf (specification and proofs are also the implementation) & HOL4  & Researchers  & University of Cambridge, University of Kent, ANU, NICTA, UNSW, A*STAR, ENS de Lyon, Inria, Chalmers University of Technology, Carnegie Mellon University & 2012 - present & N/A (53 contributors during 9 years) & Used mostly in research projects, 170 subscribers in channel \#general of CakeML slack group, 50 contributors, 53 watchers, 671 stars, 59 forks
    \\ \hline
    \end{tabular}%
    }
  \label{cakeml_id}%
\end{table*}%

\subsubsection{Scope}
The CakeML compiler is a verified realistic compiler for a functional programming language \cite{yang2011finding}. The first version of the compiler \cite{kumar2014cakeml} was implemented as an interactive read-eval-print loop (REPL) in x86-64 machine code, the second version \cite{yang2011finding} runs as a non-interactive application without dynamic compilation, but produces notably more efficient code for several 32-bit and 64-bit architectures. CakeML is a subset of Standard ML, including datatypes and pattern-matching, higher-order functions, references, exceptions, polymorphism, modules, and signatures. CakeML integers have arbitrary precision. The main feature of Standard ML not included in CakeML is functors (similar to generic classes in object-oriented languages).

\subsubsection{Components}
The verification (in the form of a correctness theorem) ensures that the implementation prints only those results permitted by the semantics of CakeML \cite{kumar2014cakeml}. This includes lexing, parsing, type checking, incremental and dynamic compilation, garbage collection, arbitrary-precision arithmetic, and compiler bootstrapping. With the last point, the verified compiler applies to itself to produce a verified machine-code implementation of the compiler.

\subsubsection{Verified properties}
The correctness of the x86-64 machine-code implementation of the CakeML is stated with respect to the machine model for x86-64 \cite{kumar2014cakeml}. The main theorem states that the machine loaded with the CakeML code and provided with a finite input character stream does one of the following:
\begin{itemize}
    \item terminates successfully with an output stream, expected by the CakeML semantics, and ending with a termination indicator;
    \item diverges after producing a finite output stream, matching the one expected by the CakeML semantics, and terminating with a divergence indicator;
    \item terminates with an out-of-memory error reported in the output stream.
\end{itemize}
The project verifies not only the compiler, but also the language properties, namely, type system soundness \cite{tan2015verified} as well as soundness and completeness of the type inference algorithm, and the run-time part that is not generated by the compiler: a generational garbage collector\cite{ericsson2019verified}.

\subsubsection{Project context}
Apart from being a verified compiler, the CakeML compiler is built with 2 other objectives in mind \cite{tan2019verified}: 1) to be as realistic as is practically possible; 2) to be simple and tidy enough to be used as a platform for future research and student projects.

\subsubsection{Decisions}
The compiler is written as a function in higher-order logic that synthesizes CakeML. Then the compiler is applied to itself to get a machine-code implementation inside the logic. This avoids a tedious manual refinement proof relating the compilation algorithm to its implementation. The compiler correctness theorem   thereby applies to the machine-code implementation of the compiler.
In contrast to CompCert (section~\ref{system:compcert}) that comes without a verified parser, the compiler uses one of two verified parsers. The first one synthesizes a CakeML abstract syntax tree (AST) from monadic functions in HOL \cite{abrahamsson2020proof}. Monadic effects are used to capture non-mathematical properties like I/O and exceptions while using shallow embedding of the functions to simplify proofs in HOL. This parser is used to bootstrap the compiler from the source code in HOL. The second parser is written manually using parsing expression grammar (PEG) \cite{tan2019verified} and comes with both soundness and completeness theorem that ensure that, if the input is correct, a single parse tree is found, as well as that this holds for any valid parse tree. This parser is used to compile programs written in CakeML rather than HOL.

The design decisions related to the backend of the compiler include \cite{tan2019verified}:
\begin{itemize}
 \item Sufficiently many intermediate languages (IL) for ease of compiler implementation and verification: it resulted in 12 ILs (pretty close to 11 ILs used by CompCert (section~\ref{system:compcert}));
\item Direct-style compilation (as  opposed to continuation-passing style (CPS)), and the call stack separate from the heap: quite a few standard ML implementations use CPS, but it is more sophisticated and goes against the objective to be as simple as possible;
\item Curried functions, compilation to multi-argument functions, and good closure representations: Standard ML supports only single-argument functions, but a decent implementation should optimize for the cases when a function is provided several arguments at once (f (a) (b)` is replaced by `g (a, b)` with exactly the same semantics);
\item A clean and simple first-order functional intermediate language: code is referenced as any other data by a pointer;
\item Structured  control-flow  (e.g.  if-statements  as  opposed to gotos) for nearly all the intermediate  languages: as opposed to control flow graphs, this simplifies reasoning such as liveness analysis;
\item One place for concretization of data representations into finite memory: it is delayed to the later stage to enable low-level optimizations (CakeML comes with an infinite-precision integer numbers, therefore fixing a representation at an earlier stage could impact performance);
\item Concrete machine code targets for several architectures: the compiler generated little-endian and big-endian 32-bit and 64-bit code.

\end{itemize}

\subsubsection{Tool stack}
The whole development is performed in HOL \cite{tan2019verified}. Standard ML and, therefore, CakeML are designed to perform proofs and fit HOL naturally. This makes it possible to bootstrap the compiler,(compile it with itself). The second version of the compiler requires an external assembler and linker, whereas the first one comes with these tools built-in and verified.

\subsubsection{Style}
Originally, the proof relied on big-step operational semantics specification for the whole program and small-step semantics for expression evaluation to handle the cases of divergence \cite{kumar2014cakeml}. Both semantics were shown to be equivalent. Later, the specification was done using only big-step semantics \cite{tan2019verified} with an additional ``fuel'' parameter that is decreased with every recursive call and allows detecting diverging behavior.
Hoare’s triples are insufficient to prove the correctness of the CakeML compiler because they capture the relation between the initial state in which a program starts and the final state in which the program terminates. This approach does not work for diverging programs. For such programs the state should be replaced with a trace of states. This is achieved by using characteristic formulae that are also mechanically verified in HOL \cite{gueneau2017verified}. These formulae
play the same role to prove the relation between the initial and the final state/trace of a given program like the weakest precondition play for the classical Hoare’s triples:
\begin{itemize}
    \item if `P $\implies$ wp (e, Q)` then `\{P\} e \{Q\}` -- Reasoning with Hoare’s triples
\item if `cf e P Q` then `\{P\} e \{Q\}` -- Reasoning with characteristic formulae
\end{itemize}

\subsubsection{Software characteristics}
The whole development takes about 100 KLoC in HOL \cite{tan2019verified}. In addition to the verified compiler with two front-ends to enable bootstrap, and with back-ends producing code for different 32-bit and 64-bit CPUs, a verified type system and a generational GC, the project triggered several bachelor’s and master’s theses, similar, but not directly connected projects (such as a minimalistic verified bootstrapped compiler \cite{myreen2021minimalistic}, several verified utilities like ``grep'', a verified certificate checker for finite-precision error bounds \cite{bard2019formally}, and others).

\subsubsection{Project characteristics}
The project was started in 2012 \cite{tan2019verified} and open-sourced on GitHub\footnote{\url{https://github.com/CakeML/cakeml}}, with a verified Lisp compiler as a predecessor. The development still continues, and so far it involved 53 contributors.

\subsubsection{Lessons}
Among the most notable language features, CakeML does not have functors  \cite{tan2019verified}, available in Standard ML.
The compiler comes with some optimizations, but their list is quite limited, for example, the straightforward peephole and common sub-expression elimination optimizations are still missing. As a result, the generated code is sub-optimal. The bootstrap is not fast either: it requires roughly 9 hours for the translation step, and between 24 and 30 hours for evaluation in the logic.
Finally, authors consider the project as a research one, without the focus on industrial usage.

The project proved the feasibility of the end-to-end completely verified bootstrapping compiler for a non-trivial functional language. It required, however, quite a bit of theoretical research to get to the current state, thus ruling out the possibility to carry out the verification by an ordinary programmer.
}{}

\iftoggle{inBook}{
\subsection{Vellvm}

\begin{table*}[htbp]
 \scriptsize
  \centering
  {\renewcommand{\arraystretch}{1.3}
        \begin{tabular}{|p{40pt}|p{45pt}|p{55pt}|p{55pt}|p{45pt}|p{43pt}|p{40pt}|}
    \hline
     Nature & {Properties} & {Programming language(s)} & {Programming style} & {Verification framework} &{ Specification language(s)} & {Number of LOC} \\
    \hline
    Framework for reasoning about programs
 & Semantics preservation &  Coq, C, OCaml & Functional & small-step and big-step operational semantics & Coq & 32k  
    \\    \hline \hline
      {Spec/Imp}  & {Proof Engine} & {Team Background} & {Institution(s)} & {Development period} & {Effort}  & {Size of user community} \\
    \hline
    Inf (specification and proofs are also the implementation) & Coq  & Researchers  & University of Pennsylvania & 2012 - present (ongoing) & N/A (12 contributors during 9 years) &  10 contributors, 23 watchers, 246 stars, 18 forks
    \\ \hline
    \end{tabular}%
    }
  \label{vellvm_id}%
\end{table*}%

\subsubsection{Scope}
The Vellvm project builds a framework for reasoning about programs expressed in LLVM's intermediate representation (IR) and transformations. It provides a mechanized formal semantics of LLVM's intermediate representation, its type system, and properties of its SSA form  (Static Single Assignment is a property of IR to assign every variable exactly once). It includes multiple operational semantics and proves relations among them to facilitate different reasoning styles and proof techniques.

\subsubsection{Components}
An interpreter extracted from the formalized LLVM specification can be used instead of the reference LLVM implementation. The project also enables reasoning about transformations on LLVM IR, in particular verification of SSA-based optimizations \cite{zhao2012mechanized}.

\subsubsection{Verified properties}
Similar to the approach used to verify soundness of a programming language type system, Vellvm comes with two theorems about the formalized LLVM IR \cite{zhao2012formalizing}:
Preservation: from a well-formed configuration (which includes a set of invariants such as each variable is well-typed, types are well-formed, there are no cycles in the type definitions), the next execution step leads to another well-formed configuration.
Progress: if code starts in a well-formed state, then it either terminates successfully, gets stuck, or can make another execution step. The stuck state correspond to fatal errors in the code, and include calls to non-function pointers, double memory deallocation and others.

Being a verification framework rather than a single-purpose tool, Vellvm was the base of verification of several algorithms and transformations, involving LLVM:
\begin{itemize}
    \item SoftBound — the program transformation that hardens C programs against spatial memory safety violations (e.g., buffer overflows, array indexing errors and pointer arithmetic errors) \cite{zhao2012formalizing}.
    \item mem2reg — a critical optimization pass producing a minimal pruned SSA program from a non-optimized SSA form \cite{zhao2013formal}.
    \item block-level dominance algorithms — the project establishes not only soundness, but also completeness \cite{zhao2012mechanized}.
    \item SSA-based optimization — the project demonstrates well-formedness preservation \cite{zhao2012mechanized}.
\end{itemize}

\subsubsection{Project context}
Vellvm addresses the question about validity of transformations of intermediate representations used by compilers. Rigorously proving properties about these IR transformations requires that the IR itself has a well-defined formal semantics, including memory model. Vellvm provides a Coq infrastructure to reason about such transformations, with mechanically-checked relation between static (compile-time) and dynamic (run-time) semantics in the form of classical preservation and progress theorems.

\subsubsection{Decisions}
LLVM IR supports uninitialized variables. The compilers are free to pick arbitrary bit patterns for such occurrences to apply aggressive optimizations and program transformations. This makes formal LLVM semantics inherently nondeterministic. Another source of uncertainty comes from under-specified LLVM memory model where some operations do not have well-defined semantics (e.g., loading with improper alignment, loading from memory with incompatible type). In order to provide a well-founded basis for formal reasoning, Vellvm chooses a sufficiently concrete memory model similar to the one used by CompCert.

Several operational semantics refine the most general non-deterministic small-step one. The deterministic semantics is used to extract a formally verified interpreter of LLVM code,  e.g. all unknown values are replaced by zeros \cite{zhao2012formalizing}.
The big-step operational semantics is used to reason about transformations that do not preserve small-step semantics, such as instruction reordering or invariant code motion. Such optimizations and transformations do not affect the intended program behavior but may change the low-level sequence of intermediate program states.
In order to reduce the number of cases to specify and reason about, the input LLVM code is pre-processed before use, e.g. ``switch'' statements are replaced by the code with regular branch instructions.

\subsubsection{Tool stack}
The formalization as well as proofs about specific transformations and optimizations are performed in Coq. Coq extractor enables generation of OCaml code for the verified LLVM IR interpreter and verified transformations.

\subsubsection{Style}
Vellvm uses operational semantics of both kinds: small-step and big-step.

\subsubsection{Software characteristics}
The syntax and static semantics constitute about 2500 lines of Coq definitions and proof scripts. \mbox{Vellvm’s} memory model implementation extends CompCert’s with approximately 5000 lines of code to support integers with arbitrary precision, padding, and an experimental treatment of casts. In total, the development represents approximately 32,000 lines of Coq code \cite{zhao2012formalizing}.

An additional library consisting of about 5200 lines of OCaml-LLVM bindings supports conversion between the LLVM bitcode representation and the extracted OCaml representation, as well as is used by the extracted interpreter.

Application-specific details:
\begin{itemize}
\item The verified version of mem2reg \cite{zhao2013formal} speeds up various benchmarks by 77\% on average whereas the original unverified version speeds up the same benchmark by 81\% on average. The difference is explained by not considering the standard library `alloca' calls as intrinsic, and not pruning all unnecessary $\phi$-nodes (the nodes combining assignment to the same variable from different basic blocks of an SSA form).
\item The verified version of SoftBound \cite{zhao2012formalizing} demonstrates the same effectiveness in detecting spatial memory safety violations as the implementation in C++, as well as run-time overhead of the instrumented benchmark programs. The verification discovered 2 bugs in the original implementation of SoftBound that could lead to false negatives, thus not reporting violations in the programs under check.
\end{itemize}

\subsubsection{Project characteristics}
The initial project effort took about 1.5 years of 4 researchers to formalize LLVM IR and prove correctness of some transformations in Vellvm. We estimate the overall effort on Vellvm only (the team performed other research tasks involving LLVM and compilation in general) as 10 person-years in total.

\subsubsection{Lessons}
The verified version of `mem2reg' increases the compilation time significantly compared to the unverified version. This happens because the extractor from Coq generates OCaml rather than C, used by the original `mem2reg'. Moreover, the verified version relies on easier-to-reason functional data structures compared to the efficient imperative implementation in the original.
The Coq extraction mechanism is not mechanically checked, so any problems in it could result in misbehavior of the software using the extracted pieces of code.

Any real-life programming language infrastructure — here the LLVM IR — includes many details that usually go beyond the reach of a research project either because of the tricky impact on the selected model or because of the noticeable expansion of the proof effort. The reconciliation of these mutually opposite forces is achieved by restricting the supported input for one set of constructs and by transforming some of the unsupported constructs into an equivalent combination of the supported ones.
The code resulting from application of verified translation tools exhibits performance similar to the one of the same unverified translation. The translation step, however, may take much longer. In order to get the best of both worlds, authors of the tools recommend using unverified 
tools during the development, and verified tools when building the release.
}{}

\iftoggle{inBook}{
\subsection{Vericert} 

\begin{table*}[htbp]
 \scriptsize
  \centering
  {\renewcommand{\arraystretch}{1.3}
        \begin{tabular}{|p{40pt}|p{45pt}|p{55pt}|p{55pt}|p{45pt}|p{43pt}|p{40pt}|}
    \hline
     Nature & {Properties} & {Programming language(s)} & {Programming style} & {Verification framework} &{ Specification language(s)} & {Number of LOC} \\
    \hline
     Compiler
 & Semantics preservation &  Coq, OCaml & Functional & HOL & Coq & 2496   
    \\    \hline \hline
      {Spec/Imp}  & {Proof Engine} & {Team Background} & {Institution(s)} & {Development period} & {Effort}  & {Size of user community} \\
    \hline
   3.63 & Coq  & Researchers  & Imperial College London & 2020 - present & 1.5 & 3 contributors, 8 watchers, 47 forks, 3 stars
    \\ \hline
    \end{tabular}%
    }
  \label{vericert_id}%
\end{table*}%

\subsubsection{Scope}
Vericert is a mechanically verified high-level synthesis (HLS) tool, which automatically compiles software into hardware designs written in Verilog.
It extends a verified C compiler, CompCert, with a new hardware-oriented intermediate language and a Verilog back end, and has been proven correct in Coq.

\subsubsection{Components}
Vericert automatically generates hardware designs from software written in a high-level language like C.
Vericert supports most C constructs, including integer operations, function calls (inlined), local arrays, structs, unions, and general control-flow statements. 

\subsubsection{Verified properties}
The verified property of Vericert is the equivalence between input software and output hardware designs.

\subsubsection{Project context}
Existing HLS tools cannot always guarantee that the hardware designs they produce are equivalent to the software they were given, which undermines any reasoning conducted at the software level.

\subsubsection{Decisions}
C was chosen as the source language as it remains the most common source language amongst production-quality HLS tools.
Verilog was chosen as the output language for Vericert because it is one of the most popular HDLs (hardware design languages) and there already exist a few formal semantics for it that could be used as a target. 

\subsubsection{Tool stack}
The project uses Coq for verification.



\subsubsection{Project characteristics}
It took about 1.5 person-years to build Vericert. About three person-months were spent on implementation and 15 person-months on proofs.

\subsubsection{Lessons}
The authors evaluate the correctness and performance of Vericert based on the PolyBench/C benchmark suite [Pouchet 2020 PolyBench], and compare the
performance of our generated hardware against an existing, unverified HOL tool called LegUp [Canis 2011]. The results show that the hardware generated 
by Vericert is 27 times slower and 1.1 times larger than that generated by LegUp. This is largely attributed to Vericert's lack of support for instruction-level parallelism and the absence of an efficient, piplined division operator.
To alleviate the limitation, the authors intend to introduce and verify their own HLS optimizations, such as scheduling and memory analysis.

}{}

\iftoggle{inBook}{
\subsection{Vermillion}

\begin{table*}[htbp]
 \scriptsize
  \centering
  {\renewcommand{\arraystretch}{1.3}
        \begin{tabular}{|p{40pt}|p{45pt}|p{55pt}|p{55pt}|p{45pt}|p{43pt}|p{40pt}|}
    \hline
     Nature & {Properties} & {Programming language(s)} & {Programming style} & {Verification framework} &{ Specification language(s)} & {Number of LOC} \\
    \hline
      Parser generator
 & Functional correctness &  Gallina & Functional & HOL & Gallina & 8k     
    \\    \hline \hline
      {Spec/Imp}  & {Proof Engine} & {Team Background} & {Institution(s)} & {Development period} & {Effort}  & {Size of user community} \\
    \hline
    -- & Coq  & Researchers and engineers  & Tufts university  & -- & 9 person-months & 5 watchers, 38 stars
    \\ \hline
    \end{tabular}%
    }
  \label{vermillion_id}%
\end{table*}%

\subsubsection{Scope}

Vermillion is a formally verified LL(1) parser generator (using LL-style parsing algorithms) for a wide range of data formats.

\subsubsection{Components}
This tool consists of two main components, including a parse table generator and an LL(1) algorithm implementation. The generator produces a correct LL(1) parse table for any grammar if such a table exists. The algorithm operates on a restricted class of grammars and provides linear-time execution. The tool is dedicated to generating a specific parser regarding the input grammar. 

\subsubsection{Verified properties}
The verified properties involve two aspects, including the completeness and soundness of the parse table generator and parser, and the termination of the parse table generator and parser on both valid and invalid inputs without the use of fuel-like parameters. These properties are verified by the Coq proof assistant.

\subsubsection{Project context}
Using unsafe parsing tools in high-profile software leads to severe consequences. For example, a faulty parser was exploited in a web application framework to obtain a large amount of consumers' data \cite{goodin2018failure}, \cite{CVE-2017-5638}. An HTML parser vulnerability resulted in private user data leaked from several popular online services \cite{cloudflare}. Since parsers mediate between the outside world and application internals, the parsers of strong correctness guarantees are likely to improve the security of applications.

\subsubsection{Decisions}
Functions with multiple parameters should be proved that they terminate. Coq's \textsf{Function} and \textsf{Program} commands can often ease the burden of defining functions with subtle termination conditions. Unfortunately, \textsf{Function} and \textsf{Program} do not support mutually recursive functions that are defined with a well-founded measure. Therefore, the authors implement well-founded recursion ``by hand'', imitating the process that these commands perform automatically.

\subsubsection{Tool stack}
The authors used the Coq Proof Assistant to verify that the generator and the parsers that it produces are sound and complete, and that they terminate on all inputs without using fuel parameters.

\subsubsection{Style}
Let us take the parser completeness as an example. The specification of the parser completeness is illustrated as follows:
\begin{lstlisting}
Theorem parse_complete :
forall (g : grammar) (tbl : parse_table)
(s : symbol) (w r : list token)
(v : symbol_semty s),
parse_table_correct tbl g
-> sym_derives_prefix g s w v r
-> parse tbl s (w ++ r) = inr (v, r).
\end{lstlisting}

This property implies if a grammar symbol derives a semantic value for a prefix of a token sequence, then the parser produces the same value for the same prefix. The proposition parse\_table\_correct tbl g denotes that tbl contains all and only the lookahead facts about g. Relation sym\_derives\_prefix uses a grammar g and a symbol s to derive a semantic value for a prefix of token sequence w. Definition parse that uses an LL(1) parse table tbl and a symbol s to build a semantic value for a prefix of the token sequence w++r. The append function (++) to specify exactly how the function divides its input sequence into a parsed prefix and an unparsed suffix. As for values inr (v, r), v is a semantic value for a prefix of the input tokens and remainder r is the unparsed suffix, indicating a successful parse. Additionally, type symbol\_semty s is the type of semantic values for symbol s.

\subsubsection{Software characteristics}

The algorithm for constructing an LL(1) parse table from a grammar is characterised by:
\begin{itemize}
\item using an ``iterate until convergence'' strategy to perform a data flow analysis over the grammar. Such an algorithm is difficult to implement in a total language because it has no obvious (syntactic) termination metric.
\item computing NULLABLE, FIRST, and FOLLOW simultaneously, so a proof of the function’s correctness must simultaneously deal with the correctness of all three sets. 
\end{itemize}
It has been proved that the LL(1) parser generated from the parser generator is sound and complete. The parser terminates on valid and invalid inputs without using fuel (a fuel parameter can produce ``out of fuel'' return values that do not clearly indicate success or failure).

\subsubsection{Project characteristics}
The verification took 9 person-months \cite{verve_materials}.

\subsubsection{Lessons}
Lessons are mainly learnt from the experiences of verification. The parse table generator conducts dataflow analyses with non-obvious termination metrics over context-free grammars. Defining recursive functions with well-founded measures in Coq allows us to implement and verify these analyses. The use of well-founded recursion on a non-syntactic measure leads to the initial implementation performing an expensive runtime computation to terminate provably. The final version of the parser uses dependent types to avoid this penalty while still proving termination. Two issues are mentioned in this work. One is concerned with the parser’s branches that represent error states. These branches unfortunately slow down the resulting code, although it has been proved that the algorithm never reaches them when applied to a correct LL(1) parse table. The other involves the parsers’ tables. The tables are treated as finite maps and the map lookups are performed at decision points, which is a likely source of inefficiency.

\subsection{Q*cert}
\begin{table*}[htbp]
 \scriptsize
  \centering
  {\renewcommand{\arraystretch}{1.3}
        \begin{tabular}{|p{40pt}|p{45pt}|p{55pt}|p{55pt}|p{45pt}|p{43pt}|p{40pt}|}
    \hline
     Nature & {Properties} & {Programming language(s)} & {Programming style} & {Verification framework} &{ Specification language(s)} & {Number of LOC} \\
    \hline
    Query compiler& Functional correctness  & Coq, Ocaml & Functional programming 
 & HOL & Coq
 & 17080 entries (each entry consists of specification and proofs)
    \\    \hline \hline
      {Spec/Imp}  & {Proof Engine} & {Team Background} & {Institution(s)} & {Development period} & {Effort}  & {Size of user community} \\
    \hline
    -- 
 & Coq
 & Engineers & IBM research & 2016 - present & NA (7 contributors during 5 years) & 6 watchers, 47 stars, 9 forks  
    \\ \hline
    \end{tabular}%
    }
  \label{qcert_id}%
\end{table*}%

\subsubsection{Scope}
Q*cert is a platform aiming at the specification, verification, and implementation of query compilers with the help of the Coq proof assistant.

\subsubsection{Components}
This platform is open source and partially supports SQL and OQL (Object Query Language). It also enables the code generation from Coq to SPARK and Cloudant \cite{cloudant}. It allows users to implement a compiler or develop new optimizations that can be fully verified as well. In addition, a web-based interface is provided to allow users to explore various compilation and optimization strategies.

\subsubsection{Verified properties}
The correctness of most of the query compiler components including the query optimizer is proved using the Coq proof assistant. The compiler specification and proofs are done over a simple data model of complex values with bags and records. 

\subsubsection{Project context}

There is some recent development around query languages and query processing outside traditional database management systems, e.g., language-integrated queries, large-scale distributed processing infrastructure, NoSQL database, or domain specific languages. Understanding and guaranteeing correctness properties for those new data processing capabilities can be important when dealing with business critical or personal data. Thus, they propose to leverage modern theorem proving technology as a foundation for building well-specified and formally verified query compilers, in order to focus on the new features of the compiler.

\subsubsection{Decisions}
The design decision principally includes how to reason scopes when variables are involved as part of the optimization rules. The authors choose $NRA^e$, an extension to a combinator-based nested relational algebra with native support for environments. It avoids blow-ups in query plan size while facilitating correctness reasoning. 

\subsubsection{Tool stack}
Q*cert was implemented in the Coq proof assistant. The compiler implementation is automatically generated from the mechanized specification using Coq's extraction mechanism.
\subsubsection{Style}
As an example, the distinct elimination law
\begin{lstlisting}
Lemma tdup_elim (q: cnraenv)q : nodupA q -> #distinct(q) $\Rightarrow$ q.
\end{lstlisting}
is specified by lemma $tdup\_elim$ with a parameter $q$ of type cnraenv. Predicate $nodupA q$ holds when query plan $q$ always returns a collection without duplicates. Note that $\#distinct()$ and $\Rightarrow$ are notations in the Coq specification.

\subsubsection{Software characteristics}
NRA$^e$, the Nested Relational Algebra with Environments, was introduced in this query compiler to realize explicit support for environment in a way that facilitated the specification and verification of algebraic rewrites. 

A large part of the query compilation pipeline was specified and mechanically checked for correctness, which resulted in strong correctness guarantees for the final compiler.

\subsubsection{Project characteristics}
The development of the compiler took around 4 person-years.

\subsubsection{Lessons}
The optimizers of the compilers reached their performance bottleneck because they could lead to an exponential number of iterations over the tree. Memoization techniques were supposed to be helpful in improving performance \cite{braibant2014implementing}. The future work included further improvements to the compiler structure, especially for the optimizer and backend so that the compiler could support code-generation for distributed query plans.

Some lessons are learnt from the use of Coq. Coq supports functional programming that fits well for writing compilers. Coq allows one to write programs and prove their correctness at the same time. The Coq extraction feature also makes it possible to build certified and relatively efficient functional programs (e.g., OCaml, Haskell or Scheme). Another thing to be addressed is the versatility of Coq. It seems to be useful in tackling a range of database problems. For example, similar techniques could be applied to: the specification of a query language after the fact or during development (e.g., documentation and prototyping), ensuring the correctness of complex algorithms, and building an end-to-end verified compiler.
}{}

\subsection{EiffelBase 2}

\iftoggle{inBook}{
\begin{table*}[htbp]
 \scriptsize
  \centering
  {\renewcommand{\arraystretch}{1.3}
        \begin{tabular}{|p{40pt}|p{45pt}|p{55pt}|p{55pt}|p{45pt}|p{43pt}|p{40pt}|}
    \hline
     Nature & {Properties} & {Programming language(s)} & {Programming style} & {Verification framework} &{ Specification language(s)} & {Number of LOC} \\
    \hline
     Library & Functional correctness & Eiffel & Object-oriented & Axiomatic (Hoare) logic + ``semantic collaboration'' for invariants + frame analysis & Eiffel (assertions integrated in code) & 9372
    \\    \hline \hline
      {Spec/Imp}  & {Proof Engine} & {Team Background} & {Institution(s)} & {Development period} & {Effort}  & {Size of user community} \\
    \hline
   check & AutoProof [YY] + Boogie [XX]  & Researchers  & ETH Zurich & 2010 - 2015 & 6 [check] & N/A 
    \\ \hline
    \end{tabular}%
    }
  \label{eb2_id}%
\end{table*}%
}{}

\subsubsection{Scope}
EiffelBase 2 is a library of fundamental data structure implementations. The project described here provided a proof of full functional correctness.

\subsubsection{Components}
EiffelBase 2 covers core data structures of everyday programming such as arrays, lists, queues and stacks. It is a replacement for the EiffelBase library (hereafter EiffelBase 1 to avoid confusion) widely used by Eiffel programmers; both roughly cover the same scope as the Java Collections and .NET Collections library in respectively the Java\slash JVM and C\#\slash .NET environments. 

\subsubsection{Verified properties}
The verification covers full functional correctness in the following sense: it applies to Eiffel programs equipped with ``contracts'' (assertions expressing preconditions and postconditions of routines and invariants of loops and classes), and verifies that the code satisfies the specification expressed by the contracts. Hence, it verifies that the code fulfills its stated purpose, not just that it meets specific consistency properties.

\subsubsection{Project context}

EiffelBase 1 
was already designed with correctness concerns in mind; it may have been the first general-purpose library equipped with extensive contracts. These contracts are widely used for dynamic verification, that is to say testing; preconditions, in particular, express conditions imposed on the caller and as a result make it possible, when run-time assertion monitoring is turned on, to find errors in application code using the library. 

Dynamic assertion monitoring in a library that is richly equipped with contracts provides an effective way to find bugs, both in client code 
and in the library itself (in the case of postcondition and invariant violations). It falls short, however, of guaranteeing correctness, which requires static proofs of full functional correctness and is the goal of the EiffelBase 2 verification project.

\subsubsection{Decisions}
The initial intent of the EiffelBase 2 project was to extend the assertions of EiffelBase 1 to permit static mechanical verification. 
EiffelBase 2 was devised as a new implementation of the same structures, with a simpler inheritance hierarchy. 
While not identical, the API and the general style of the new library remain close to those of EiffelBase 1.
\vspace{-0.1cm}
\subsubsection{Tool stack}
\begin{figure}[htbp!]
  \centering
  \includegraphics[width=4cm]{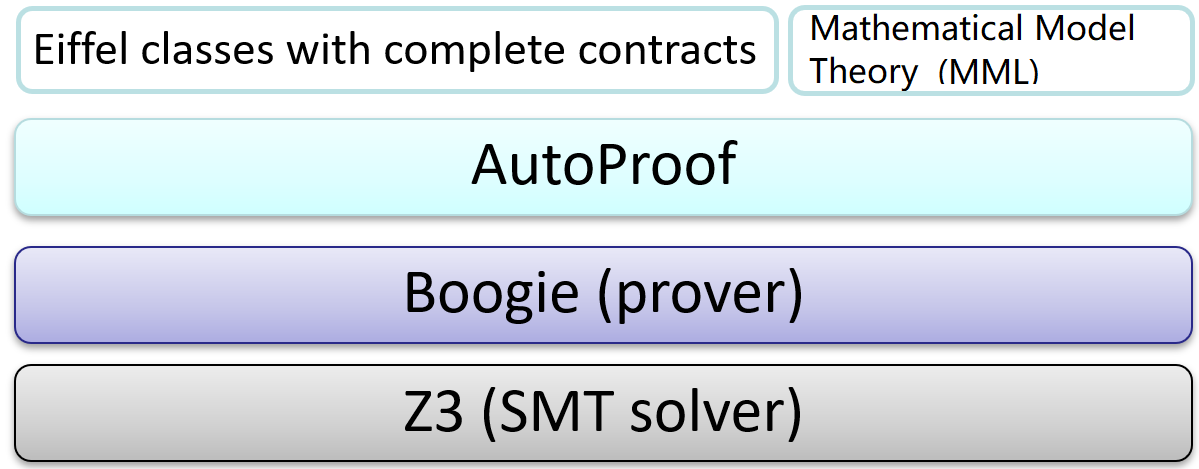}
  \caption{Verification tool stack of EiffelBase 2}
  \label{fig:eb21}
\end{figure}

The verification uses the AutoProof verification environment for Eiffel, relying on the Boogie proof engine from Microsoft Research, itself based on the Z3 SMT solver (Fig. \ref{fig:eb21}).
\vspace{-0.7cm}
\subsubsection{Style}
The need to extend the assertions of EiffelBase 1 arose for two reasons: ensuring complete functional specification; specifying frame properties; and addressing the ``dependent delegate dilemma'' of class invariants.

\noindent \textbf{Full functional specification}: a typical postcondition of EiffelBase 1, for the remove routine, assuming the structure is a hash table and k is the argument representing the key of the element to be removed, would read
\begin{lstlisting}
ensure
	not has (k) 						-- There is no longer any element of key k.
	(old has (k)) implies (count = old count - 1)		-- Number of elements decreased (by 1)
	(not old has (k)) implies (count = old count)	-- if and only if key k was present
\end{lstlisting}
This specification is not complete since it does not say anything about elements associated with keys other than k. An implementation could modify some of these elements and still meet the precondition. EiffelBase 2 classes make the specifications complete by using mathematical models; for example the mathematical model for a hash table is a map (a partial function from keys to elements), so that the postcondition can be given as \e{ensure model = (old model).deprived (k)},
where \e{deprived} yields a map whose domain has been deprived of \e{k}. The original postcondition clauses are consequences of this new one. Models serve specification purposes only and are provided by an auxiliary library, MML (Mathematical Model Library), with classes representing such mathematical objects as sets, functions (including maps) and sequences. They are purely functional in style (for example, no procedure deprive that modifies a map, but a function deprived that yields a new map with the key removed).

\noindent \textbf{Frame conditions}:  postconditions describe how properties of objects change. For practical proofs, it is also necessary to specify what does not change (and prove that it does not). ``Modify'' clauses, added to the contracts, specify the frame of a routine: the objects that it is allowed to change.

\noindent \textbf{Dependent delegate dilemma}: a delicate issue of the semantics of object-oriented programs is the precise role of class invariants, which are only expected to hold on routine entry and exit but can be violated in-between, making the underlying object ``inconsistent''. The ``Dependent Delegate Dilemma'' is the case of a routine r calling another s in a state where the source object O is consistent (as permitted), but s calling back into a routine t of O, whose correctness requires a consistent object on entry. To address this issue, AutoProof and EiffelBase 2 rely on a semantic collaboration model, which requires further annotations.

\vspace{-0.1cm}
\subsubsection{Software characteristics}

While EiffelBase 2 is fully functional, the majority of Eiffel users continue to use EiffelBase 1 for reasons of compatibility with existing code.

\vspace{-0.1cm}
\subsubsection{Project characteristics}
The experience of the EiffelBase 2 project shows that a naive view, ``write fully specified code, then run it through the verifier'', does not reflect the reality of proving program correctness. The process is an iterative one: write code with contracts, attempt to verify, see which parts do not verify, fix the problem (by modifying either the code or the contracts), repeat. It can be compared to the traditional process of debugging a program, except that the debugging is static (running the prover) rather than dynamic (running the program). The process requires significant effort; the principal author of EiffelBase 2 spent 6 months after the completion of her PhD thesis finishing the EiffelBase 2 verification, applying the process described.

The verification required about 5 person-years of work. It was part of the more general AutoProof project, consuming around 10 person-years. Project members were researchers. 

\subsubsection{Lessons}
Libraries are keys to modern programming, with its emphasis on reuse; their correctness is paramount since so many applications will rely on a given library. The verification of EiffelBase 2 is, consequently, a milestone.

The major challenges in generalizing the technology pioneered by the verification of EiffelBase 2 are the level of competence required (project members all obtained PhDs in computer science) and the extensive annotation effort required for frame specification and, particularly, semantic collaboration, the hardest to teach to newcomers. Research efforts are in progress to address both of these needs by analyzing the programs to produce the corresponding specifications automatically.

\subsection{HACL*}
\label{system:hacl*}

\iftoggle{inBook}{

\begin{table*}[htbp]
 \scriptsize
  \centering
  {\renewcommand{\arraystretch}{1.3}
        \begin{tabular}{|p{40pt}|p{45pt}|p{55pt}|p{55pt}|p{45pt}|p{43pt}|p{40pt}|}
    \hline
     Nature & {Properties} & {Programming language(s)} & {Programming style} & {Verification framework} & {Specification language(s)} & {Number of LOC} \\
    \hline
     Cryptographic library & Memory safety, secrets independence, functional correctness & F*, Clight & Functional, Procedural programming & Hoare logic & F* & 7225    
    \\    \hline \hline
      {Spec/Imp}  & {Proof Engine} & {Team Background} & {Institution(s)} & {Development period} & {Effort}  & {Size of user community} \\
    \hline
   3.3 &  Z3  & Researchers  & INRIA, Microsoft Research & -- & -- & --
    \\ \hline
    \end{tabular}%
    }
  \label{hacl_id}%
\end{table*}%

}{}

\subsubsection{Scope}
HACL* \cite{zinzindohoue2017hacl}  is a verified portable C library of cryptographic primitives for a new mandatory cipher suite in TLS 1.3 \cite{rescorla2018transport}. It is also intended as the main cryptographic provider for the miTLS \cite{mitls} verified implementation and integrated in Mozilla NSS cryptographic library.


\subsubsection{Components}

All cryptographic algorithms in HACL* are correct by construction.

\subsubsection{Verified properties}
The developers of HACL* have verified the following properties:
\begin{itemize}
    \item Memory safety.
    The code never reads or writes memory at invalid locations, such as null or freed pointers, unallocated memory, or out-of-bounds of allocated memory.
    In addition, any locally allocated memory is eventually freed (exactly once).
    \item Functional correctness.
    The code of each primitive conforms to its published standard specification on all inputs.
    \item Mitigations against Side-Channel Attacks.
    The code does not reveal any secret inputs to an adversary, even if the adversary can observe low-level runtime behavior such as branching, memory access patterns, cache hits and misses or power consumption.
    \item Cryptographic security.
   The code of each cryptographic construct implemented by the library is (with high probability) indistinguishable from a standard security definition.
\end{itemize}

\subsubsection{Project context}
The project's primary goal was to build a reference implementation in C and to prove that it conforms to computations described in the corresponding RFC standards (Request For Comments --- a publication in a series, usually for the Internet).

\subsubsection{Decisions}

The main implementation vehicles for HACL* are the F* functional language and Low*, an integration into F* of a safe subset of C. HACL* code never allocates memory on the heap; all temporary states are stored on the stack to simplify proofs of memory safety and avoid explicit memory management.
The Low* source code is broken into many small functions to improve readability, modularity and code sharing and reduce the complexity of each proof.
Consequently, the default translation of this code to C would result in a set of small C functions, which can be verbose and hurts runtime performance with some compilers like CompCert (Section \ref{system:compcert}).
For better control of the generated code, program annotations direct the KreMLin compiler to bring certain functions online and unroll certain loops. 
A large chunk of the bignum verified code is shared across three different encryption algorithms: Poly1305, Curve25519 and Ed25519, meaning that this code is verified once but used in three different ways.
The sharing has no impact on the quality of the generated code because KreMLin puts the generic code online and specializes it for one particular set of bignum parameters.
The result is that Poly1305 and Curve25519 contain separate, specialized versions of the original Low* bignum library.

\subsubsection{Tool stack}
The developers first write a high-level specification of the primitive in a higher-order purely functional subset of F* (Pure F*).
They then write an optimized implementation in Low*, a low-level subset of F* that can be efficiently compiled to C. They then verify the code using the F* Z3-based type checker, for conformance to the specification and to the logical preconditions and type abstractions required by the F* standard library.
If type checking fails, there may be a bug in the code, or the type checker may require more annotations.
Finally, KreML translates the Low* code to C. The translation is actually to Clight, a subset of C that can be compiled with CompCert (Section \ref{system:compcert}), although in practice this last step uses GCC for performance reasons.


\subsubsection{Style}
Fig. \ref{fig:f_star_abstraction} contains an example of a pure F* executable specification. 
To implement prime field arithmetic for the Poly1305 algorithm on 64-bit platforms, one strategy is to represent each 130-bit field element as an array of three 64-bit limbs, where each limb uses 42 or 44 bits and so has room to grow.
When adding two such field elements, one can simply add the arrays point-wise, ignoring carries; the \textit{fsum} function performs this task.
Fig. \ref{fig:low_star_contract} contains a contract of the future implementation (fadd) of the abstraction (fsum) expressed in pure F* (Fig. \ref{fig:f_star_abstraction}). 

\begin{figure}[ht]
\vspace{-0.5cm}
    \begin{subfigure}{.49\textwidth}
\begin{lstlisting}
type limbs = b:buffer uint64_s{length b = 3}
let fsum (a:limbs) (b:limbs) =
a.(0ul) $\leftarrow$ a.(0ul) + b.(0ul);
a.(1ul) $\leftarrow$ a.(1ul) + b.(1ul);
a.(2ul) $\leftarrow$ a.(2ul) + b.(2ul)
\end{lstlisting}
        \caption{The pure F* abstraction.}
        \label{fig:f_star_abstraction}
    \end{subfigure}
\hfill
     \begin{subfigure}{.49\textwidth}
\begin{lstlisting}
val fsum: a:limbs $\rightarrow$ b:limbs $\rightarrow$ Stack unit
(requires ($\lambda$ h0 $\rightarrow$ live h0 a $\wedge$ live h0 b
$\wedge$ disjoint a b
$\wedge$ index h0.[a] 0 + index h0.[b] 0 < MAX_UINT64
$\wedge$ index h0.[a] 1 + index h0.[b] 1 < MAX_UINT64
$\wedge$ index h0.[a] 2 + index h0.[b] 2 < MAX_UINT64))
(ensures ($\lambda$ h0 _h1 $\rightarrow$ modifies_1 a h0 h1
$\wedge$ eval h1 a = fadd (eval h0 a) (eval h0 b)))
\end{lstlisting}
        \caption{F* contract}
        \label{fig:low_star_contract}
     \end{subfigure}
\vspace{-0.3cm}
\caption{The specification style of HACL*.}
\vspace{-0.3cm}
\end{figure}

\noindent The contract is proven to correctly implement the F* abstraction; then the Low* implementation is proven to correctly implement the contract.
After that, the KreMLin tool converts Low* to Clight.

\subsubsection{Software characteristics}
The library's code base totals 801 lines of pure F* code (specification), 22926 lines of Low* code (code and proofs), and 7225 lines of C code (translated from Low*) \cite{zinzindohoue2017hacl}, and gets verified in 9127 seconds.


\subsubsection{Project characteristics}
Symmetric algorithms like Chacha20 and SHA2 do not involve sophisticated mathematics, and are relatively easy to prove.
The proof-to-code ratio hovers around 2, and each primitive took around one person-week.
Code that involves bignums requires more advanced reasoning.
While the cost of proving the shared bignum code is constant, each new primitive requires a fresh verification effort.
The proof-to-code ratio is up to 6, and verifying Poly1305, X25519 and Ed25519 took several person-months.
High-level APIs like AEAD and SecretBox have comparably little proof substance, and took a few person-days.

\subsubsection{Lessons}
Although formal verification can significantly improve confidence in a cryptographic library, it relies on a large trusted computing base.
The semantics of F* has been formalized and the translation to C has been proven to be correct on paper, but the results still rely on the correctness of the F* type checker, the Z3 SMT solver, the KreMLin compiler, and the C compiler (when GCC is used instead of CompCert).

One goal of the design of such systems is to achieve "secret independence", a necessary but not complete mitigation of side-channel attacks. Secret independence provably rules out certain classes of timing side-channel leaks that rely on branching and memory accesses, but does not account for others such as power analysis.
The current model only protects secrets at the level of machine integers; it treats lengths and indexes of buffers as public values, so that it is impossible to verify implementations that rely on constant-time table access or length-hiding constructions.

The verification tool chain stops with verified C code.
The research team did not consider the problem of compiling C to verified assembly, which is an important but independent challenge.

HACL* performs well for symmetric algorithms on 32-bit platforms, and suffers a penalty for algorithms that rely on 128-bit integers because they are represented as pairs of 64-bit integers.
If CompCert (Section \ref{system:compcert}) supports 128-bit integers in the future, the penalty will disappear.

Verification may help optimize code.
In the process of verifying HACL*, the researchers found that some algorithms are too conservative.
They optimized them and then successfully verified the correctness of the optimization.

Taking an RFC and writing a specification for it in F* is straightforward, as well as translating existing C algorithms to Low*.
Proving that the Low* code is memory safe, secret independent, and that it implements the RFC specification is the bulk of the work.

\iftoggle{inBook}{
\subsection{Amazon s2n}

\begin{table*}[htbp]
 \scriptsize
  \centering
  {\renewcommand{\arraystretch}{1.3}
        \begin{tabular}{|p{40pt}|p{45pt}|p{55pt}|p{55pt}|p{45pt}|p{43pt}|p{40pt}|}
    \hline
     Nature & {Properties} & {Programming language(s)} & {Programming style} & {Verification framework} &{ Specification language(s)} & {Number of LOC} \\
    \hline
     Cryptogra- phy libraries & Functional correctness & C & Procedural & HOL & Cryptol, Coq & 700     
    \\    \hline \hline
      {Spec/Imp}  & {Proof Engine} & {Team Background} & {Institution(s)} & {Development period} & {Effort}  & {Size of user community} \\
    \hline
   0.86 & Coq proof assistant, SAW (software analysis workbench)  & Researchers and engineers  & Galois, Amazon Web Services, University of Washington, University College London & 2016 - present (deployed in 2016) & More than 3 (2 people full time over 1.5 years) & 141 contributors; over 4k stars, 567 forks 
    \\ \hline
    \end{tabular}%
    }
  \label{s2n_id}%
\end{table*}%

\subsubsection{Scope}

Amazon s2n is a cryptography library that implements the TLS (Transport Layer Security) encryption protocol for privacy and authentication of online communication. s2n has been applied in numerous Amazon services. A continuous verification approach was developed to maintain the functional correctness in s2n automatically and efficiently.

\subsubsection{Components}
s2n is an open source TLS implementation, which has been used by many Amazon and Amazon Web Services (AWS) products to secure various user online activities, e.g., web browsing and connection to cloud resources. It was developed for both high performance and low power as required by a diverse range of AWS products. The implementation of s2n includes HMAC (keyed-Hash Message Authentication Code) algorithm, DRBG (deterministic random big generator) and the TLS handshake \cite{chudnov2018continuous}. HMAC ensures that both the sender and recipient have exchanged the correct secrets before a TLS connection can proceed to the data transmission phase.

\subsubsection{Verified properties}
The verification focuses on both high-level security properties and low-level functional correctness of HMAC, DRBG and TLS handshake. The main security property for HMAC is its indistinguishability, meaning that it should be indistinguishable from a function returning random bits. Specifically, there is no effective strategy by which an attacker that is viewing the output of the cryptographic function HMAC and a real random output can distinguish the two \cite{beringer2015verified}. The verification also aims to prove that 1) s2n implements a subset of TLS 1.2 (the transport layer security protocol version 1.2) [TLS1.2] and 2) the socket corking API that optimizes how data is split into packets is used correctly. 

\subsubsection{Project context}
The s2n library was introduced by Amazon in 2015 and has been applied in many AWS services, e.g., Elastic Load Balancing (ELB), Amazon CloudFront, Amazon S3. As many customers (especially those from financial services or government) use AWS to store or transmit security-critical data, it is essential to provide the customers with concrete information about how security is established. On the other hand, to facilitate rapid and cost-efficient development, this project also seeks to develop an automatic and continuous formal verification approach, in which each time the implementation changes, the proof can be updated and executed automatically or with low effort. 

\subsubsection{Decisions}
The s2n library was designed to prioritize simplicity: it avoids implementing rarely used options and extensions, which is favorable for code auditing and testing. The simplicity can further contribute to proof automation and produce proof which can be integrated into the development workflow of s2n. 

\subsubsection{Tool stack}
The verification started with constructing a high-level specification of HMAC functions in Cryptol language \cite{lewis2003cryptol}, which is a domain-specific language that Galois has created for the high-level specification of cryptographic algorithms. The operational semantics of Cryptol specification was then mapped into the Coq language and proved using the Coq proof assistant. Based on the Cryptol specification, the verification of C implementation of HMAC was conducted by using SAW (Software Analysis Workbench) \cite{saw} to establish the equivalence of the implementation and specification. In the verification, SAW firstly uses bounded symbolic execution to translate Cryptol and C programs into logical expressions, and then proves properties by leveraging automated SAT and SMT solvers (e.g., Z3, Yices, CVC4). Furthermore, to enable the continuous verification of s2n, Travis CI \cite{travisci} platform was applied to integrate the automatic proof procedure into development workflow of s2n: when there is a push to the s2n repository \cite{s2nrepo}, the SAW proof will be executed automatically. 

\subsubsection{Style}
The high-level descriptions of the s2n functions were given in Cryptol. For example, the monolithic definition of HMAC (given in FIPS standard [FIPS]) can be specified in Cryptol as the hmac function: \emph{hmac key message = H((key $^\wedge$ opad) \# H((key $^\wedge$ ipad) \# message))}, where key and message are inputs arguments, H is any hash function, $^\wedge$ is bitwise xor, and \# is concatenation, opad and ipad are relevant constants.

The equivalence between C implementation and Cryptol specification was expressed in SAWScript. Fig. \ref{fig:s2n1} shows an example of SAWScript specification of HMAC update function (denoted hmac\_update\_spec), which incrementally updates the state of function with chunks of the message. The two arguments of the function, msg\_size and cfg, represent the message size and configuration parameters (e.g., the input and output sizes of the hash block function). 

 \begin{figure}[htbp]
  \centering
  \includegraphics[width=6cm]{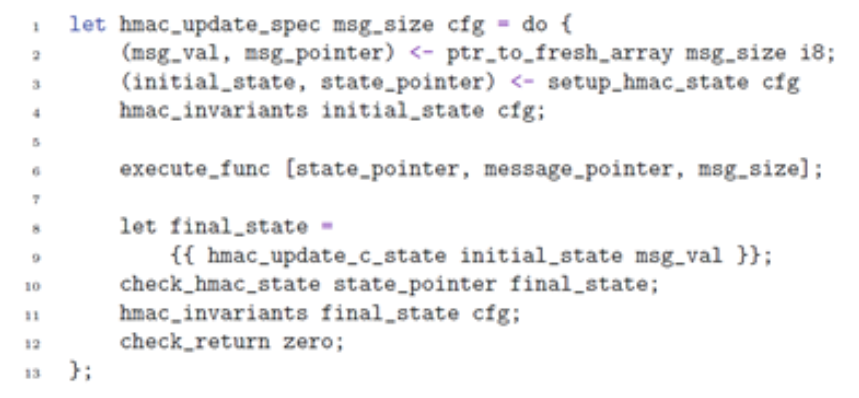}
  \caption{Specification for the HMAC update function in SAWScript (excerpted from \cite{chudnov2018continuous})}
  \label{fig:s2n1}
\end{figure}

The specification represents a Hoare triple, where the precondition and postcondition are separated by the body, execute\_func command (line 6). Lines 2-3 set up the values and pointers that match the necessary types. Line 8-10 specifies that the final state resulting from executing the C function should be equivalent to the state produced by evaluating the Cryptol specification. Specifically, lines 8-9 capture the final state of Crytol specification (embraced by double curly braces) and line 10 asserts that the final state should match the state of the C program stored in memory state\_pointer. The parameters (e.g., msg\_val, initial\_state) are either referenced in the C implementation or the Cryptol specification while pointers (e.g., msg\_pointer, state\_pointers) are only referenced in the C implementation.

\subsubsection{Software characteristics}

The fact that the structures of Cryptol specification and the C implementation are similar, coupled with the usage of automated verification tools (SMT solvers), enables a proof that continues to work through a variety of code changes \cite{chudnov2018continuous}. In particular, changes to the code in function bodies often require no changes in the proof scripts, and changes in the API (e.g., function arguments) require minor changes of proof scripts. This allows the maintainers of s2n to quickly determine the correctness of submitted changes with little to no effort. In one year of code changes only three manual updates to the proof were required.

\subsubsection{Verification effort}
HMAC and DRBG each took roughly 3 months of engineering effort, and the TLS handshake verification took 8 months. The effort includes understanding the C code, writing the Cryptol specifications, developing the code-spec proofs using SAW, the CI implementation work, and the process of merging the proof artifacts into the upstream code-base.

\subsubsection{Lessons}
SAW has limited support for reasoning about unbounded data. Since SAW does not support induction, the properties were proved for only finite message sizes. While this can still provide higher state space coverage than testing, full proof for all message sizes is not provided. Moreover, as symbolic execution is the key technique in SAW for translating imperative programs into formal models, formal models can be successfully generated only in cases where symbolic execution terminates \cite{dockins2016constructing}. Symbolic execution is guaranteed to terminate when control flow only depends on concrete values (not symbolic ones) and may fail to terminate in other cases.

In this verification project, a continuous verification approach was proposed and integrated into the development chain of s2n. At each change to the code, proofs can be automatically reestablished with little to no effort from the developers. This nearly eliminates the need for developers to understand or modify the proof following modifications to the code \cite{chudnov2018continuous}.
}{}

\subsection{seL4}
\label{system:seL4}

\iftoggle{inBook}{

\begin{table*}[htbp]
 \scriptsize
  \centering
  {\renewcommand{\arraystretch}{1.3}
        \begin{tabular}{|p{40pt}|p{45pt}|p{55pt}|p{55pt}|p{45pt}|p{43pt}|p{40pt}|}
    \hline
     Nature & {Properties} & {Programming language(s)} & {Programming style} & {Verification framework} &{ Specification language(s)} & {Number of LOC} \\
    \hline
     Microkernel & Functional correctness & C & Procedural & HOL & HOL & Over 10k     
    \\    \hline \hline
      {Spec/Imp}  & {Proof Engine} & {Team Background} & {Institution(s)} & {Development period} & {Effort}  & {Size of user community} \\
    \hline
  100 & Isabelle, Z3, Sonolar SMT solver  & Researchers and experienced engineers  & Trustworthy systems group in Data61 & 2004 - present (first verified version in 2009) & 31.2 (development and verification) & 10 members (institutes and companies) in seL4 Foundation
    \\ \hline
    \end{tabular}%
    }
  \label{sel4_id}%
\end{table*}%

}{}

\subsubsection{Scope}
In an operating system (OS), a kernel is a central component that runs in a privileged mode to manage the communication between software and hardware. seL4 \cite{seL4-whitepaper} is a microkernel  that provides essential kernel services. 

\subsubsection{Verified System}
As a microkernel, seL4 only undertakes the minimum tasks required in privileged mode, leaving others to user mode. Core OS functions include isolation through sandboxes, in which a program can execute without interference from others, and inter-process communication (IPC) to transport function inputs and outputs between programs. 
seL4 is also a hypervisor that creates and executes virtual machines (VMs) running mainstream OSes (such as Linux). This enables provisioning system services, borrowing services from guest OSes running in seL4's VMs. 


\subsubsection{Verified properties}
The verification of seL4 covers the following aspects \cite{blackham2011timing,klein2014comprehensive,sewell2017high,sewell2013translation}: 
\begin{itemize}
    \item Functional correctness: the C implementation of seL4 is free of defects (e.g., system crashes, unsafe operations).
\item Security properties: confidentiality, integrity and availability.
\item Worst-case execution-time (WCET) analysis by obtaining hard upper bounds for all system call latencies.
\item Semantics equivalence: the binary is a correct translation of the verified C implementation.
\end{itemize}

\subsubsection{Project context}
A fault in an OS kernel can compromise the safety and correctness of the rest of the OS. In other words, a kernel is a trusted computing base (TCB) that must operate correctly for the OS to be secure. SeL4 is one of the L4 microkernels \cite{L4_family} whose TCB is minimized to reduce its exposure to attacks. They aim to achieve greater confidence and better performance. The objectives of the project include: 1) having its implementation proved correct; 2) solving problems on resource management in microkernels and OSes in general; 3) making the system suitable for real-world use.

\subsubsection{Decisions}
The design of seL4 conforms to the principle of least authorities (POLA) \cite{POLA} ---the rights given to a component are restricted to an absolute minimum.
To perform access control in line with POLA, seL4 uses capabilities, which are access tokens that delegate to components the rights (read or write) to use certain objects. 

\subsubsection{Tool stack}
The proof tool is Isabelle/HOL \cite{nipkow2002isabelle}. Two SMT solvers --- Z3 and Sonolar --- are used to validate the semantic equivalence between C code and binary. 
The WCET analysis was performed using the Chronos tool \cite{blackham2011timing}. 

\subsubsection{Style}
\begin{figure}[htbp]
  \centering
  \includegraphics[width=3.7cm]{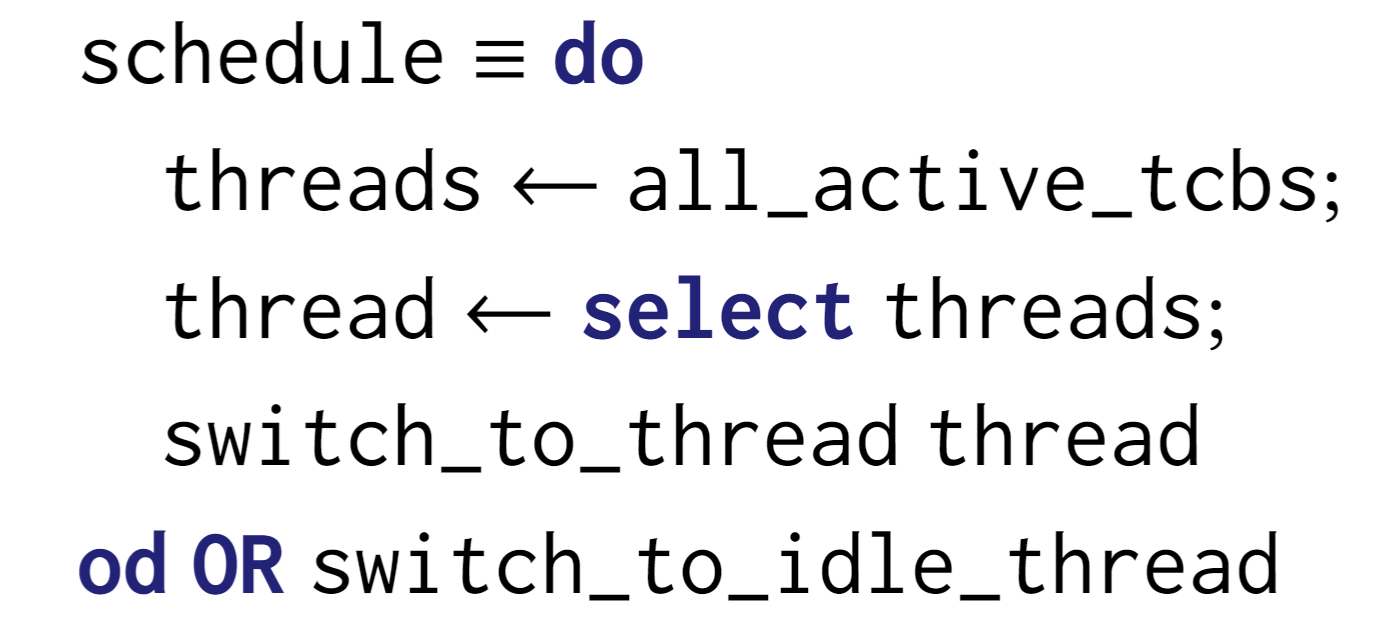}\\
 \vspace{-.2cm}
 \caption{HOL specification for scheduler}
  \label{fig:seL42}
\end{figure}

Functional specifications of seL4 are given in HOL. Fig. \ref{fig:seL42} depicts an example of HOL specification for scheduling: the scheduler is modelled as a function picking one \e{thread} from all active runnable \e{threads} in the OS or from the idle threads; the function $all\_active\_tcbs$ returns the abstract set of all runnable threads in the system; the \e{select} statement picks any element from the set \e{threads}. The \e{OR} operator indicates non-deterministic choice between the first block (embraced by \e{do} and \ekw{od}) and \e{switch\_to\_idle\_thread}.

\vspace{-0.6 cm}
\subsubsection{Software characteristics}
seL4 documentation reports exceptionally good performance, complementing the guarantees of safety and security. IPC (interprocess communication) performance is 2 to 10 times faster than some existing microkernels \cite{seL4-whitepaper} \cite{mi2019skybridge}: seL4 is more than five times faster than CertiKOS for a round-trip IPC operation, over twice as fast as Fiasco.OC and 9 times faster than Zircon. An example of practical deployment of seL4 is the DARPA High-Assurance Cyber Military Systems (HACMS) \cite{fisher2017hacms}, where seL4 guarantees non-interference between components and protects from tampering the computer in the Boeing ULB autonomous helicopter.

\subsubsection{Project characteristics.}
The total development efforts is 31.2 person/years; the effort for the seL4 functional correctness proof, about 20.5; for the security proof 4.1 and the binary verification 2. Re-verification of full functional correctness is feasible \cite{klein2014comprehensive}; adding a significant feature  (for example, a new data structure) to the kernel would cost 1.5 to 2 p/y.

\subsubsection{Lessons}

The cost for development of seL4 and its functional correctness proof is \$362/LOC \cite{klein2014comprehensive} --- inexpensive compared to the industry's widely accepted baseline of \$1000/LOC for the EAL6 low level (``Common Criteria" standard \cite{Common_Criteria}), while the assurance level of seL4 is higher than EAL6. These results suggest that formally-verified software can be less expensive than traditionally engineered high-assurance software, while providing stronger assurance \cite{klein2014comprehensive}. 

\iftoggle{inBook}{
\subsection{ProvenCore}
\label{system:ProvenCore}

\begin{table*}[htbp]
 \scriptsize
  \centering
  {\renewcommand{\arraystretch}{1.3}
        \begin{tabular}{|p{40pt}|p{45pt}|p{55pt}|p{55pt}|p{45pt}|p{43pt}|p{40pt}|}
    \hline
     Nature & {Properties} & {Programming language(s)} & {Programming style} & {Verification framework} &{ Specification language(s)} & {Number of LOC} \\
    \hline
     Microkernel & Security & C & Procedural & Hoare logic & Smart & --     
    \\    \hline \hline
      {Spec/Imp}  & {Proof Engine} & {Team Background} & {Institution(s)} & {Development period} & {Effort}  & {Size of user community} \\
    \hline
  Inf (specification is implementation) & ProvenTools  & Researchers & ProvenRun & 2009 - & 3 & 6 official partners
    \\ \hline
    \end{tabular}%
    }
  \label{provencore_id}%
\end{table*}%

\subsubsection{Scope}
ProvenCore is an off-the-shelf microkernel developed by ProvenRun \cite{provenrun}. It has been proven to be secure in the sense that it guarantees the integrity and confidentiality of processes running on the kernel. ProvenCore is particularly designed for running in the secure part of ARM Cortex-A TrustZone microprocessor architecture [TrustZone] (see Sect. \ref{provencore_context}), where ProvenCore can be used as a trusted execution environment (an isolated processing environment where various applications can be securely executed) \cite{sabt2015trusted}.

\subsubsection{Components}
As a microkernel, ProvenCore only incorporates the critical services running in privileged mode, which include: 1) creation and deletion of processes; 2) execution of some programs, e.g., memory management, virtual address space arrangement and thread control; 3) synchronous message-passing inter-process communications (IPCs) with timeouts; 4) asynchronous notifications; 5) process-to-process data copies using dedicated system calls; 6) a shared memory system to allow potential zero-copy transfers. 

\subsubsection{Verified properties}
The main property of ProvenCore is the isolation property, which can be divided into two properties, integrity and confidentiality. Integrity ensures that the resources of a process (code, data, registers) cannot be tampered by unauthorized processes. Confidentiality requires that the resources of a process cannot be observed by other unauthorized processes.

\subsubsection{Project context}
\label{provencore_context}
Nowadays, embedded systems, such as mobile devices, usually encompass sensitive applications, e.g. e-banking and digital rights management. Secure environments are required for running those applications. To address this, many processor designs adopt a dual OS architecture where the execution platform is divided into two sections: the rich section (that runs the normal user applications) and the secure section (that runs the security-sensitive services). In this architecture, a conventional rich OS (Android, iOS) runs alongside with a secure OS (ProvenCore), which allows users to execute security-critical services without any modifications on the conventional OS side. This technique is called the ARM TrustZone technology \cite{trustzone}. ProvenCore is designed to provide a trusted execution environment for such secure sections in the TrustZone architecture. Formal verification was conducted on ProvenCore to achieve the EAL7 (highest level of Common Criteria \cite{Common_Criteria}) certification of the kernel.

\subsubsection{Decisions}
ProvenCore is designed for ARM architectures and uses Memory Management Unit (MMU) to manage virtual address spaces. It also enforces a number of security policies related to resource allocation, access control and information-flow. For example, the information-flow policy requires that synchronous IPCs can only go ``downward'', from a  privileged kernel-space task to a less privileged task, which prevents the kernel from being stuck waiting for an less privileged process. 

\subsubsection{Tool stack}
This project uses the dedicated tool-chain developed by ProvenRun, in the form of a set of Eclipse plugins, which provides support for editing specifications, browsing and proving proof obligations and generating C code. The proof engine in the tool-chain is called ProvenTools \cite{lescuyer2015provencore}, which can generate proof obligations and prove them semi-interactively (both automatic and interactive modes are enabled). ProvenTools allows users to see all proof steps in detail, which makes reviews of the proofs possible, unlike other verification tools using black-box back-end solvers. 

\subsubsection{Style}
ProvenCore is specified and implemented in the Smart language \cite{lescuyer2015provencore}, which is a proprietary specification language developed by ProvenRun. Smart is a strongly-typed functional language with algebraic datatypes (structures and variants). It also supports logical specifications like pre- and postconditions, local assertions and loop invariant.

As an example, here is a simple auxiliary lemma on one of the abstract models of ProvenCore:

 \begin{figure}[htbp]
  \centering
  \includegraphics[width=6cm]{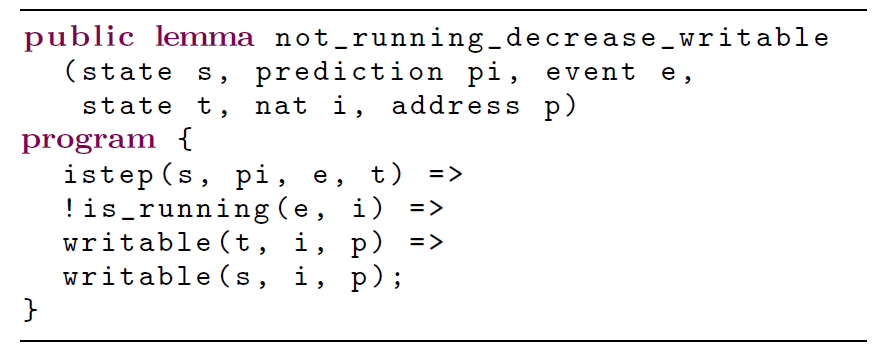}\\
  \label{fig:provencore1}
\end{figure}

This lemma expresses that given a transition of the system
from a state \emph{s} to a state \emph{t}, summarized by a trace \emph{e}, and
provided that \emph{i} was not the index of a process running on
the system during that transition, then the set of addresses
in \emph{i}’s address space that are writable by some other process
can only decrease. 

\subsubsection{Software characteristics}

The proof proceeds by establishing refinements between successive models, with the most abstract models being used for proving high-level security properties and the most concrete models for generating C code. Different levels of abstraction can provide some separation of concerns in the overall proof, with the lower-level proofs cluttered with low-level properties and invariants and the higher-level proofs centered on functional specifications. Incidentally, a high-level Common Criteria evaluation also requires different levels of descriptions of the system being evaluated \cite{lescuyer2015provencore}.

Reuse of the abstract models is possible. For example, the abstract models that specify security policies or functional properties are flexible enough to account for different implementations. Furthermore, ProvenCore's highest abstraction model is very general and could be a suitable specification for many separation kernels.

ProvenCore has been used as a trusted execution environment on the Zynq UltraScale+ devices developed by XLINX \cite{xlinx-whitepaper}. It reduces the attack surface of security-critical applications by isolating those applications from the rest of the system and keeping all confidential data out of the non-secure Linux server.

\subsubsection{Project characteristics}
In an earlier proof of concept, the project team performed an isolation proof on Minix 3.1 \cite{tanenbaum1997operating}, which took about 3 person-years. They estimated the effort of proving the refinement in ProvenCore to be roughly the same \cite{lescuyer2015provencore}. 

\subsubsection{Lessons}
ProvenCore is a TrustZone-based security solution and it has obtained the EAL7 certification.

\subsection{Verve}
\label{system:Verve}

\begin{table*}[htbp]
 \scriptsize
  \centering
  {\renewcommand{\arraystretch}{1.3}
        \begin{tabular}{|p{40pt}|p{45pt}|p{55pt}|p{55pt}|p{45pt}|p{43pt}|p{40pt}|}
    \hline
     Nature & {Properties} & {Programming language(s)} & {Programming style} & {Verification framework} &{ Specification language(s)} & {Number of LOC} \\
    \hline
     OS Microkernel for research & Safety & Beat, C\#, Boogie, TAL & Object-oriented & Hoare logic & Boogie & 7552     
    \\    \hline \hline
      {Spec/Imp}  & {Proof Engine} & {Team Background} & {Institution(s)} & {Development period} & {Effort}  & {Size of user community} \\
    \hline
  3 & Boogie, Z3  & Researchers & Microsoft Research & 2009 - 2010 & 0.75 & --
    \\ \hline
    \end{tabular}%
    }
  \label{verve_id}%
\end{table*}%

\subsubsection{Scope}
Verve is an operating system and run-time system developed by Massachusetts Institute of Technology and Microsoft Research. This system consists of a Nucleus, a kernel and one or more applications. It serves as reliable and secure lower-level software layer for high-level computer applications.


\subsubsection{Components}
Verve is a microkernel based operating system mechanically verified to guarantee type and memory safety. It has a ``Nucleus'' providing primitive access to hardware and memory, a kernel building services on the Nucleus, and applications running on the kernel. The Nucleus implemented allocation, garbage collection, multiple stacks, interrupt handling, and device access, which was written in verified assembly language. The kernel built higher-level services, such as preemptive threads on the Nucleus, which was written in C\# and compiled to TAL. The applications were written in C\# as well.

\subsubsection{Verified properties}
The safety and correctness of Verve are both verified at the assembly level. The safety refers to type and memory safety. More specifically, the safety of the kernel and applications, and the safety and correctness of the Nucleus have been mechanically verified. 

\subsubsection{Project context}
Low-level software still suffered from a steady stream of bugs, making computers vulnerable to attacks. Safe languages were utilized to ensure type safety and memory safety of low-level systems. Unfortunately, this safety cannot ensure the absence of many common bugs, such as buffer overflow vulnerabilities. The implementations of the language’s underlying run-time system might have bugs as well. Hence, Verve was designed to realize the formal verification of low-level operating systems and the run-time system code.

\subsubsection{Decisions}
Two complementary verification technologies, TAL and automated theorem proving, drove Verve’s design. On one hand, the compiler could automatically transform C\# code into TAL code, which enabled TAL to scale easily to large amounts of code. On the other hand, automated theorem provers were able to verify deeper logical properties about the code than a typical TAL type system could express. To make full use of the tradeoff between TAL and automated theorem proving, authors decided to divide the Verve operating system code into two parts: a Nucleus, verified with automated theorem proving, and a kernel, verified with TAL. The balance between the two parts was based on the difficulty of theorem proving. Only the functionality of Kernel that TAL could not verify as safe goes to Nucleus.

\subsubsection{Tool stack}
Fig. \ref{fig:verve1} shows multiple tools were used to build the Verve system. Three compilers were involved, including Beat (a small extension of Boogie) compiler, C\# compiler, and Bartok compiler \cite{chen2008type}. The Beat compiler transformed verifiable Beat expressions and statements into verifiable Boogie assembly language instructions. The Bartok compiler transformed the C\# code into the TAL code. Thus, Boogie was used to verify the safety and correctness of the Nucleus. TAL checker was also employed to verify the safety of the kernel and applications.

The run-time systems and drivers of Verve have been mainly verified by Boogie. Fig. \ref{fig:verve2} shows the Verve structure. The Nucleus interacts with five components: memory, hardware devices, the boot loader, interrupt handling, and TAL code (kernel and application code). Memory and hardware devices export functionality to the Nucleus, such as the ability to read memory locations and write hardware registers. The verifications for run-time systems and drivers ensure that the Nucleus satisfies the preconditions to each operation on memory and hardware. 

 \begin{figure}[htbp]
  \centering
  \includegraphics[width=6cm]{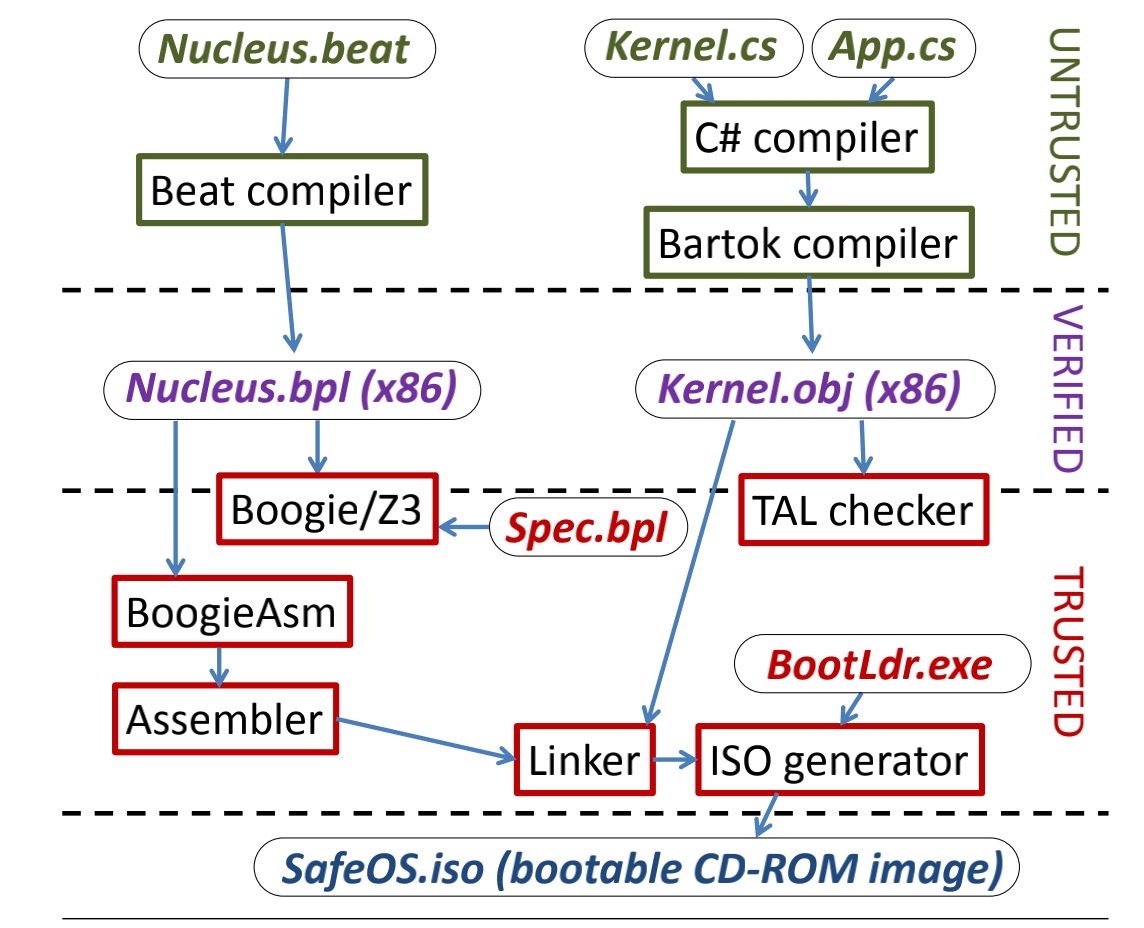}\\
  \caption{Building the Verve system: trusted, untrusted components}
  \label{fig:verve1}
\end{figure}

 \begin{figure}[htbp]
  \centering
  \includegraphics[width=6cm]{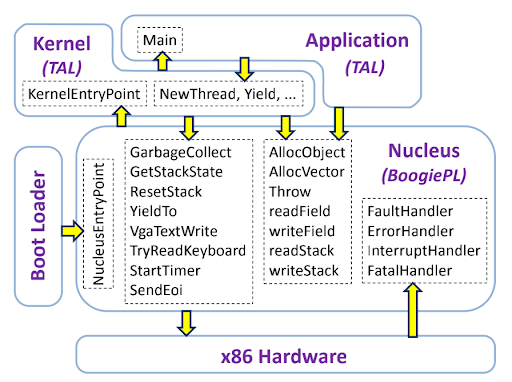}\\
  \caption{Verve structure, showing all 20 functions exported by the Nucleus}
  \label{fig:verve2}
\end{figure}

\subsubsection{Style}
The specifications written in Boogie are used to verify that the Nucleus behaves correctly. Boogie procedures have pre- and postconditions, written as Boogie logical formulas, shown as follows:
\begin{lstlisting}
var a:int, b:int;
procedure P(x:int, y:int)
  requires a < b && x < y;
  modifies a, b;
  ensures a < b && a == x + old(a);
  {
    a := a + x;
    b := b + y;
  } 
\end{lstlisting}
This example shows procedure P can only be called when global variable a is less than global variable b; and parameter x is less than parameter y. These preconditions are implied in the requires statement. Upon exit, the procedure’s postconditions ensure that a is still less than b, and that a is equal to x plus the value of old version a (before P executed). Note that the modifications of all the global variables must be explicitly revealed in the procedure, so that the callers will be aware of the modifications.
No concrete example of specifications of TAL verification is found so far.

\subsubsection{Software characteristics}

TAL and automated theorem proving are combined to perform the verification of Verve. A TAL checker verifies the safety of the kernel and applications. A Hoare-style verifier with an automated theorem prover verifies both the safety and correctness of the Nucleus. Every instruction of every piece of software except the boot loader has been mechanically verified for safety at the assembly level.


This verification architecture excludes high-level language compilers from the trusted computing base. It also resulted in small annotation-to-code ratios (2-3, depending on certain design decisions).
The trusted computing base of Verve includes:
\begin{itemize}
\item The assembler.
\item The linker.
\item The ISO CD-ROM image generator.
\item The boot loader.
\item The specification of correctness for the Boogie code.
\end{itemize}

Because of the guaranteed type safety, Verve keeps the initial boot-loader-supplied address space rather than creating virtual memory address spaces to protect applications from each other.
The verified type-safety also guarantees type-safe behavior of applications running on Verve (which is not the case for seL4, for example).

The authors report the ability to quickly redesign and re-verify as a key benefit of the verification approach. This is an important benefit --- some verification approaches subdue the developer into architectural decisions that simply let the verification process somehow terminate. Instead, the authors of Verve were able to try out and fully re-verify several architectures within just 9 person-months before retaining the system as it is right now.

The authors also point out several commonalities between the architectures of seL4 and Verve:
\begin{itemize}
\item Both work on uni-processors, with preemption over interrupts.
\item Both rely on non-blocking system calls.
\item Both provide memory protection (differently, however).
\item These architectural decisions might have been made due to the need to develop a fully verified system, which existed in both cases.
\end{itemize}

\subsubsection{Project characteristics}
Verve design, implementation, verification took just 9 person-months, spread between two people.

\subsubsection{Lessons}
In its current implementation, Verve suffers from a number of limitations. It does not verify termination. Its does not cover the full C\# language, library and environment; among the features not supported are exception handling, the standard .NET class library (because of the unsafe code it contains), and dynamic loading. It lacks a comprehensive isolation mechanism between applications, such as Java Isolates, C\# AppDomains or Singularity SIPs \cite{fahndrich2006language}. To make verification possible, it uses a verified but non-incremental garbage collector \cite{hawblitzel2009automated}, with interrupts disabled.  These decisions, while necessary for Verve's approach to verification, would make the tool unrealistic for production applications.


Some lessons were learned from the debugger and the linker. The debugger stub allowed a remote debugger to connect to the kernel, so that the memory could be examined and any inadvertent traps that might occur could be debugged. However, the debugger stub turned out to cause more bugs than it fixed: it required the memory, the interrupt handling, and the thread context definition. As a result, without the debugger, it seemed that Verve's invariants were more secure. After dropping the debugger, everything worked for the first time, except when the garbage collector ran. This caused the GC and the TAL checker to fail to operate normally. The cause finally found was that the mismatch of the relocation information for the Bartok-generated objects. Thus, more attention should be paid to the relocation information of objects.
}{}

\subsection{mCertiKOS}
\label{system:mCertikOS}

\iftoggle{inBook}{

\begin{table*}[htbp]
 \scriptsize
  \centering
  {\renewcommand{\arraystretch}{1.3}
        \begin{tabular}{|p{40pt}|p{45pt}|p{55pt}|p{55pt}|p{45pt}|p{43pt}|p{40pt}|}
    \hline
     Nature & {Properties} & {Programming language(s)} & {Programming style} & {Verification framework} &{ Specification language(s)} & {Number of LOC} \\
    \hline
      Hypervisor & Security, strong behavioral correctness, strong contextual correctness & ClightX, LAsm & Procedural & HOL & Coq & 3000     
    \\    \hline \hline
      {Spec/Imp}  & {Proof Engine} & {Team Background} & {Institution(s)} & {Development period} & {Effort}  & {Size of user community} \\
    \hline
   6 & Coq  & Researchers & Yale University, University of Science and Technology of China & 2011-2014 & 1  & --
    \\ \hline
    \end{tabular}%
    }
  \label{mCertiKOS_id}%
\end{table*}%

}{}

\subsubsection{Scope}
mCertiKOS is a fully verified single-core operating system kernel that can serve as a hypervisor capable of booting unmodified Linux guests. 
It is deployed on the Landshark Unmanned Ground Vehicle (UGV) and other American-built car platforms \cite{mcertikos}. 

\subsubsection{Components}
The project included the development and verification of several variations of the mCertiKOS kernel.
An OS kernel is the portion of the OS code that is always resident in memory, facilitating interactions between hardware and software components \cite{wiki2021kernel}.


\subsubsection{Verified properties}
The research team has verified security \cite{costanzo2016end}, as well as behavioral and strong contextual correctness properties (described below) \cite{gu2015deep}.

\subsubsection{Project context}
Modern computer systems consist of abstraction layers, such as OS kernels, device drivers, network protocols.
Each layer defines an interface that hides the implementation details of a particular functionality level.
A client program built on top of a given layer should be able to understand the layer functionality solely from the layer's interface, independently of its implementation.
The mCertiKOS project formalizes this methodology through an \emph{abstraction layer calculus}, relying on \emph{deep specifications}, to specify, implement and verify several OS kernels.
A deep specification is supposed to capture the precise functionality of the underlying implementation.

\subsubsection{Decisions}

The researchers wanted deep specifications to satisfy the \emph{implementation independence} property: two distinct implementations $M_1$ and $M_2$ of the same deep specification $L$ should have \emph{contextually equivalent} behaviors.
Specifically, given any \emph{whole-program} client $P$ built on top of $L$, running $P \oplus M_1$ ($P$ linked with $M_1$) should lead to the same observable result as $P \oplus M_2$. This requirement influenced the following project decisions:


\begin{itemize}
    \item Focus on languages whose semantics are \emph{deterministic} relative to external events \cite{Jaroslav2013CompCertTSO}.
    \item Consider only interfaces whose primitives have deterministic specifications.
    \item Prohibit a client program from calling primitives defined in two different interfaces that reflect conflicting abstract states of the same abstract state. 
\end{itemize}

\subsubsection{Tool stack}
The project relies on the following technologies:
\begin{itemize}
    \item ClightX, a variant of the CompCert (Section \ref{system:compcert}) Clight language \cite{costanzo2016end}.
    \item LAsm, an x86 assembly language. 
    \item Coq for development and refinement of abstraction layers' interfaces.
    \item CompCertX --- a verified compiler for compiling ClightX abstraction layers in LAsm layers. 
\end{itemize}

\noindent Outside the verified kernels, there are 300 lines of C and 170 lines of x86 assembly code that are not verified yet: the preinit procedure, the ELF loader used by user process creation, and functions such as \texttt{memcpy} which currently cannot be verified due to a limitation of the CompCert memory model.
Device drivers are not verified because LAsm lacks device models for expressing the correctness statement.
The CompCert assembler for converting LAsm into machine code remains unverified. LAsm does not cover the full x86 instruction set and many other hardware features.

\subsubsection{Style}
\begin{figure}[htbp!]
  \centering
  \includegraphics[width=7cm]{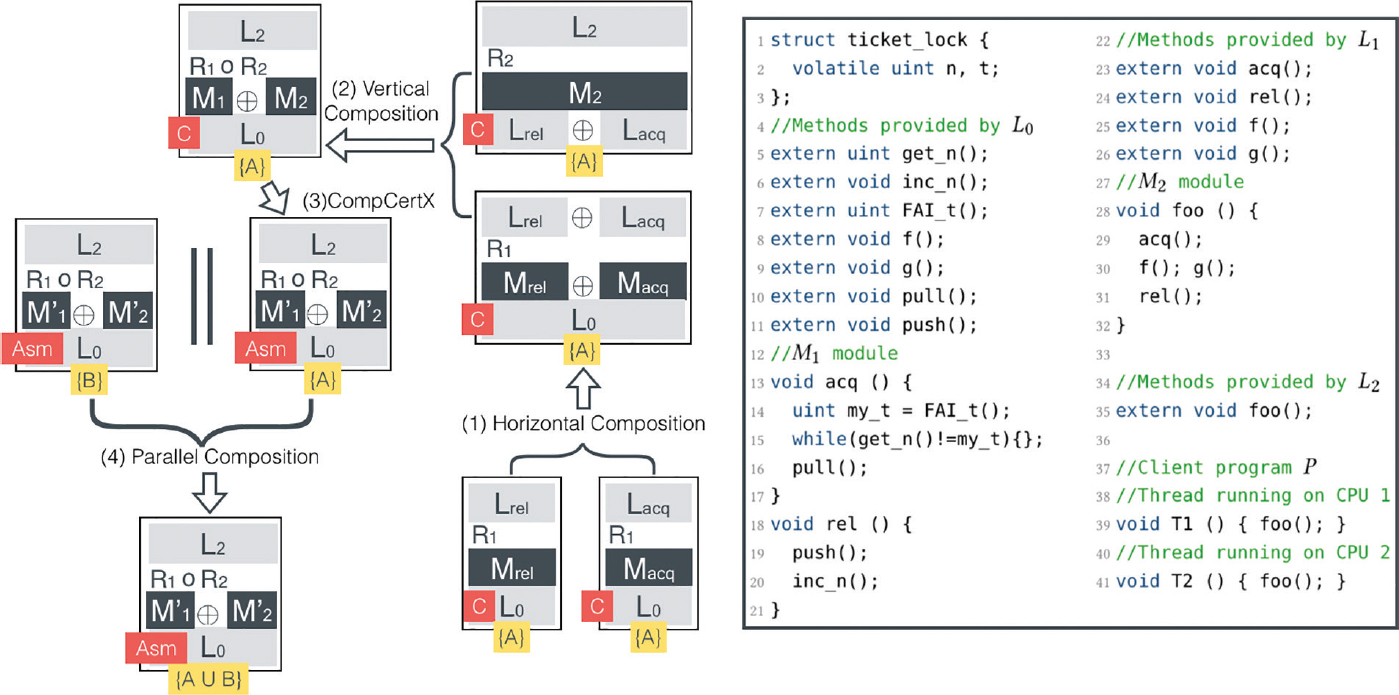}\\
  \vspace{-0.3cm}
  \caption{The calculus of abstraction layers.
  The ticket acquire/release example.}
  \label{fig:mcertikos3}
\end{figure}

The certified abstraction layer, $L_0 \vdash_{R_{1}} M_{acq}:L_{acq}$ (Fig. \ref{fig:mcertikos3}), is a predicate plus a mechanized proof showing that the implementation $M_{acq}$ running on the underlay interface $L_0$, written in C, correctly implements the desirable overlay interface $L_{acq}$, written in Coq.
The implementation relation is denoted by $R_1$.
The layer can be (1) horizontally composed with another layer (such as the lock release operation) if they have identical state views (with the same $R_1$) and are based on the same $L_0$.
The composed layer can also be (2) vertically composed with another that relies on its overlay interface.
The Coq code extraction engine produces an OCaml program containing CompCertX, the abstract syntax trees of the kernels, and the driver functions that invoke (3) CompCertX on pieces of ClightX code and generate the full assembly file.
In a concurrent setting, these layers can also be (4) composed in parallel.\

\subsubsection{Software characteristics}

The project resulted in a \textbf{calculus of abstraction layers} and a series of operating system kernels developed with the help of this calculus:
\begin{itemize}
    \item The baseline version of the mCertiKOS kernel.
    \item The mCertiKOS-hyp kernel adds core primitives to build user-level hypervisors by supporting the AMD SVM hardware virtualization technology.
    \item mCertiKOS-rz augments mCertiKOS-hyp with support of ring 0 processes.
    \item mCertiKOS-emb is a minimal operating system suitable for embedded environments.
\end{itemize}

\noindent The research team reports 3000 lines of ClightX and LAsm code for mCertiKOS-hyp, calling it ``the most realistic kernel''.
The authors do not share the exact number of lines of Coq code, but we estimate around 18500 lines based on the information distributed throughout the papers.

\subsubsection{Project characteristics}
The planning and development of mCertiKOS took 9.5 person-months plus 2 person-months on linking and code extraction. It took additional 1.5 person-months to finish mCertiKOS-hyp, after which mCertiKOS-rz and mCertiKOS-emb took half a person-month each.

\subsubsection{Lessons} \label{mcertikoslessons}

The layered architecture facilitates reuse: the certified abstraction layers developed for the base version of the mCertiKOS kernel were heavily reused for the three variations. The authors, however, report up to 2x slowdown of the mCertiKOS-hyp kernel in the lmbench benchmarks \cite{lmbench}.
The certified abstraction layers approach has demonstrated high applicability for verifying strong properties, such as termination and contextual correctness. The specification/implementation ratio of 6, however, looks high compared to other approaches in this survey.

\iftoggle{inBook}{
\subsection{mC2}
\label{system:mC2}
\begin{table*}[htbp]
 \scriptsize
  \centering
  {\renewcommand{\arraystretch}{1.3}
        \begin{tabular}{|p{40pt}|p{45pt}|p{55pt}|p{55pt}|p{45pt}|p{43pt}|p{40pt}|}
    \hline
     Nature & {Properties} & {Programming language(s)} & {Programming style} & {Verification framework} &{ Specification language(s)} & {Number of LOC} \\
    \hline
     Concurrent OS kernel, hypervisor & Security, strong behavioral correctness, strong contextual correctness, concurrency specific properties & ClightX, LAsm & Procedural & HOL & Coq & 6500     
    \\    \hline \hline
      {Spec/Imp}  & {Proof Engine} & {Team Background} & {Institution(s)} & {Development period} & {Effort}  & {Size of user community} \\
    \hline
   17 & Coq  & Researchers & Yale University, University of Science and Technology of China & 2014-2016 & 2 & --
    \\ \hline
    \end{tabular}%
    }
  \label{mC2_id}%
\end{table*}%

\subsubsection{Scope}
mC2 is an operating system kernel resulting from extending the mCertiKOS hypervisor to support concurrent execution on multicore machines.
mC2 is a fully verified concurrent operating system kernel with fine-grained locking that can also serve as a hypervisor.

\subsubsection{Components}
The project included development and verification of mC2, a concurrent OS kernel with few simple device drivers that doubles as a hypervisor and can boot three Ubuntu Linux systems as guests. The mC2 kernel runs on an Unmanned Ground Vehicle (UGV) with a multicore Intel Core i7 machine. To develop mC2, the researchers added several features on top of the already developed mCertiKOS kernel: dynamic memory management, container support for controlling resource consumption, Intel hardware virtualization support, shared memory IPC, single-copy synchronous IPC, ticket and MCS locks, new schedulers, condition variables etc. 

\subsubsection{Verified properties}
In addition to the properties verified for the mCertiKOS kernel, concurrency specific properties have been verified for the mC2 kernel: safety, liveness, linearizability, starvation freedom, deadlock freedom, data-race freedom. The researchers also prove a strong contextual refinement property for every kernel function, which states that the implementation of each such function will behave like its specification under any kernel/user context with any valid interleaving.
The researchers prove compositionality of the certified abstraction layers by proving a chain of strong contextual refinement lemmas \cite{gu2016certikos} (Fig. \ref{fig:mc21}):

 \begin{figure}[htbp]
  \centering
  \includegraphics[width=6cm]{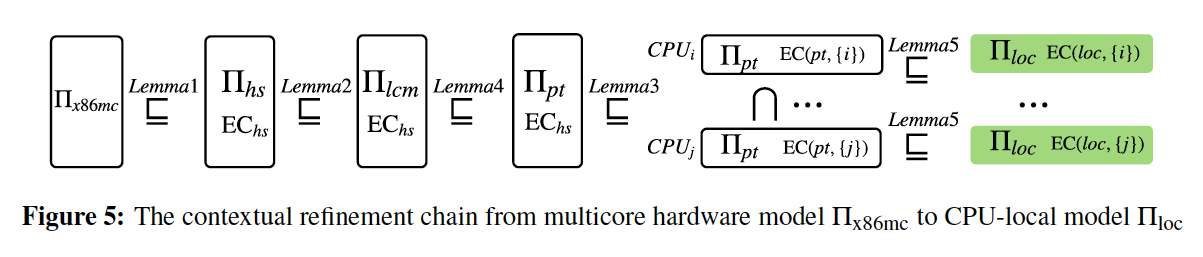}\\
  \caption{}
  \label{fig:mc21}
\end{figure}
\begin{itemize}
\item Lemma 1 asserts correctness of the deterministic hardware scheduler model;
\item Lemma 2 describes correctness of the local copy model;
\item Lemma 3 denotes composition of partial machine models
\item Lemma 4 addresses correctness of the composed total machine
\item Lemma 5 writes correctness of CPU-local machine model.
\end{itemize}

Practical implementation of this model heavily employs spinlocks, and the researchers have proven certain properties of their implementation of spinlocks:

\noindent \textbf {Invariant 1} (Invariants for ticket lock). An environment context that holds the lock i (1) never acquires lock i again before releasing it; and (2) always releases lock i within k steps (for some k).

\noindent \textbf{Lemma 6} (Starvation-freedom of ticket lock). Acquiring ticket-lock in the mC2 kernel eventually succeeds.

They also prove a security invariant of the virtual memory management implementation:

\noindent \textbf{Invariant 2}. (1) paging is enabled only after all the page maps are initialized; (2) pages that store kernel-specific data must have the kernel-only permission in all page maps; (3) the kernel page map is an identity map; and (4) non-shared parts of user processes’ memory are isolated.

\subsubsection{Project context}
The mC2 project context matches that of the mCertiKOS project.
On top of what mCertiKOS offers, mC2 supports multicore processors, can host up to three Ubuntu Linux guests (mCertiKOS can only host one), and contains few simple device drivers.

\subsubsection{Decisions}
The mC2 kernel keeps the layered architecture of mCertiKOS. Other architectural decisions have been introduced, however, to support multiprocessor and multithreaded executions on mC2. The key design decisions introduced in mC2 are (1) the use of fine-grained spinlocks to avoid explicit synchronization on shared data between different threads and (2) a deterministic hardware scheduler. The absence of explicit synchronization is a critical condition for being able to reuse existing certified abstraction layers for concurrent composition.

\subsubsection{Tool stack}
The mC2 kernel development and verification tool stack matches that of the mCertiKOS kernel.

\subsubsection{Style}
mC2 matches the specification style of mCertiKOS.

\subsubsection{Software characteristics}

The authors do not explicitly share the start and end dates of the project, but they state that mC2 is an extension of mCertiKOS, which was first reported in October 2014. The mC2 kernel itself was first reported in November 2016. The paper \cite{gu2016certikos} states the following contributions other than the mC2 kernel:
\begin{itemize}
    \item A framework for composing concurrent OS kernels.
    \item An environment context based approach to applying techniques for verifying sequential programs to verify concurrent programs.
\end{itemize}

The specification base in Coq has significantly grown as compared to that of mCertiKOS. The authors report 5249 lines of additional specifications for the various kernel functions; about 40K lines of code used to define auxiliary definitions, lemmas, theorems, and invariants; 50K lines of proof scripts for proving the newly added concurrency features. The researchers state that at least one third of these auxiliary definitions and proof scripts are redundant and semi-automatically generated, and that they are currently working on a new layer calculus to minimize the amount of redundancy.

\subsubsection{Verification effort}
The new concurrency framework and features took about 2 person-years to verify.

\subsubsection{Lessons}
The assembly machine of mC2 assumes strong sequential consistency for all atomic instructions. More formalization work is needed to turn proofs over sequential-consistent machines into those over Total State Ordering (TSO) \cite{owens2009better} machines. The mC2 assembly machine also does not model transaction lookaside buffers (TLB), and any code for addressing TLB shootdown cannot be verified. The mC2 kernel currently lacks a certified storage system. The mC2 assembly machine only covers a small part of the full x86 instruction set, so the contextual correctness results only apply to programs in this subset. The CompCertX assembler for converting assembly into machine code is unverified. The mC2 kernel also relies on a bootloader, a PreInit module (which initializes the CPUs and the devices), and an ELF loader. Their verification is left for future work.

The mC2 case shows feasibility and practicality of building certified concurrent kernels. The numbers, however,  speak for themselves: the specification/implementation ratio estimate has grown from the already-high 6 for mCertiKOS to 17 for mC2. According to the original report \cite{gu2016certikos},  most of the additional specification overhead comes from the added support for concurrency. The researchers also had to apply strong assumptions and limitations in addition to those applied during the development of mCertiKOS. Notably, none of the assumptions and limitations applied in the mCertiKOS project were relaxed since the 2 year period of developing mC2. 

\subsection{Hyper-V}
\label{system:HyperV}

\begin{table*}[htbp]
 \scriptsize
  \centering
  {\renewcommand{\arraystretch}{1.3}
        \begin{tabular}{|p{40pt}|p{45pt}|p{55pt}|p{55pt}|p{45pt}|p{43pt}|p{40pt}|}
    \hline
     Nature & {Properties} & {Programming language(s)} & {Programming style} & {Verification framework} &{ Specification language(s)} & {Number of LOC} \\
    \hline
     Native hypervisor & Functional correctness, safety, deadlock freedom, lock protectiveness & C, assembly & Procedural & Hoare logic & VCC annotations & 105000  (34500)   
    \\    \hline \hline
      {Spec/Imp}  & {Proof Engine} & {Team Background} & {Institution(s)} & {Development period} & {Effort}  & {Size of user community} \\
    \hline
   5 & Boogie/Z3  & Researchers & German Research Center for Artificial Intelligence, European Microsoft Innovation Center, Microsoft Research, Saarland University & 2007-2010 & 1.5 & $\geq$ 60M
    \\ \hline
    \end{tabular}%
    }
  \label{hyperv_id}%
\end{table*}%

\subsubsection{Scope}
Microsoft Hyper-V Hypervisor is a thin layer of software between hardware and operating system.
It turns an x64 multi-processor machine with virtualization extensions (host) into a set of virtual multi-processor x64 machines (guests). Hyper-V ships since March 2009. The Hyper-V verification project included the following goals:
\begin{itemize}
\item Functional correctness, in the sense of correct virtualization.
\item Memory safety, race freedom, and other concurrent aspects of correctness.
\item Code level verification for full blown C, as opposed to abstract algorithms.
\item Integrating verification into the existing development process.
\end{itemize}


\subsubsection{Components}
The software to be verified was the Hyper-V hypervisor -- a thin software layer between the OS layer and the underlying hardware layer that turns an x64 multi-processor machine with virtualization extensions into a set of virtual multiprocessor x64 machines.
These machines have no virtualization extensions but feature an additional layer of address translation and additional instructions --- hypercalls --- that constitute the calling mechanism for guest operating systems.
A hypercall is to a syscall what a hypervisor is to an OS.
Alternatively, a hypercall is to a hypervisor what a syscall is to a kernel.

 \begin{figure}[htbp]
  \centering
  \includegraphics[width=3cm]{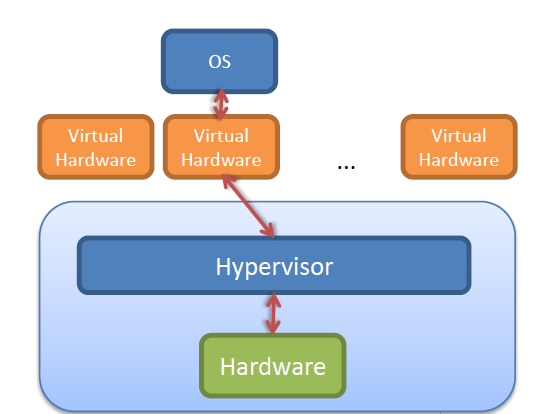}\\
  \caption{Key components of Hyper-V.}
  \label{fig:hyperv1}
\end{figure}

\subsubsection{Verified properties}
The verification project targeted the following properties of Hyper-V:
\begin{itemize}
\item Functional correctness (correct virtualization).
\item Cross-cutting concerns, such as memory safety and absence of race conditions.
\item Concurrency related properties, such as lock freedom (if a shared resource is not used, then it is not locked) and lock safety (if a lock is acquired, it will actually protect the shared resource).
\end{itemize}

\subsubsection{Project context}
The Hypervisor Verification Project was undertaken in the period of 2007-2010 by five parties:
\begin{itemize}
\item European Microsoft Innovation Center
\item German Research Center for Artificial Intelligence
\item Microsoft Research
\item Microsoft’s Windows division
\item Saarland University
\end{itemize}
The project is part of a larger project, Verisoft XT. The parent project assumed verification of two industrially used systems without changing their code -- the Hyper-V Hypervisor and PikeOS micro kernel. The main verification technology, VCC, was being developed simultaneously with verifying the two systems.

\subsubsection{Decisions}

The key challenge of the verification endeavor was the inability to change the codebase. That is to say, the Hyper-V product was not developed with verification in mind, and it was heavily optimized to meet the stringent performance expectations. The verification team was only allowed to introduce verification annotations that would not affect the runtime behavior of Hyper-V. Another challenge was the mix of C and assembly code in the implementation. It was also necessary to realistically model the x64 architecture, including caches, TLBs, store buffers, and the weak memory model. Also, Hyper-V runs in translated mode and maintains its own page tables.

\subsubsection{Tool stack}
The hardware is modeled through two-state invariants over ghost variables. The implementation is equipped with Design by Contract (DbC) style contracts linked with the ghost state through the coupling invariant. The specified code is fed into the VCC prover that translates the implementation into Boogie verification conditions. The Boogie tool then translates the Boogie verification conditions into Z3 proof obligations and runs on them the Z3 tool for the actual verification. To handle flexible invariants (invariants that may be temporarily violated), the VCC tool applies the ownership model. For failed verification attempts, VCC produces execution counterexamples violating the property being verified and displays these counterexamples in the Model Viewer utility. VCC relies on the weakest precondition calculus for program proving.
The following diagram illustrates this process:

 \begin{figure}[htbp]
  \centering
  \includegraphics[width=4cm]{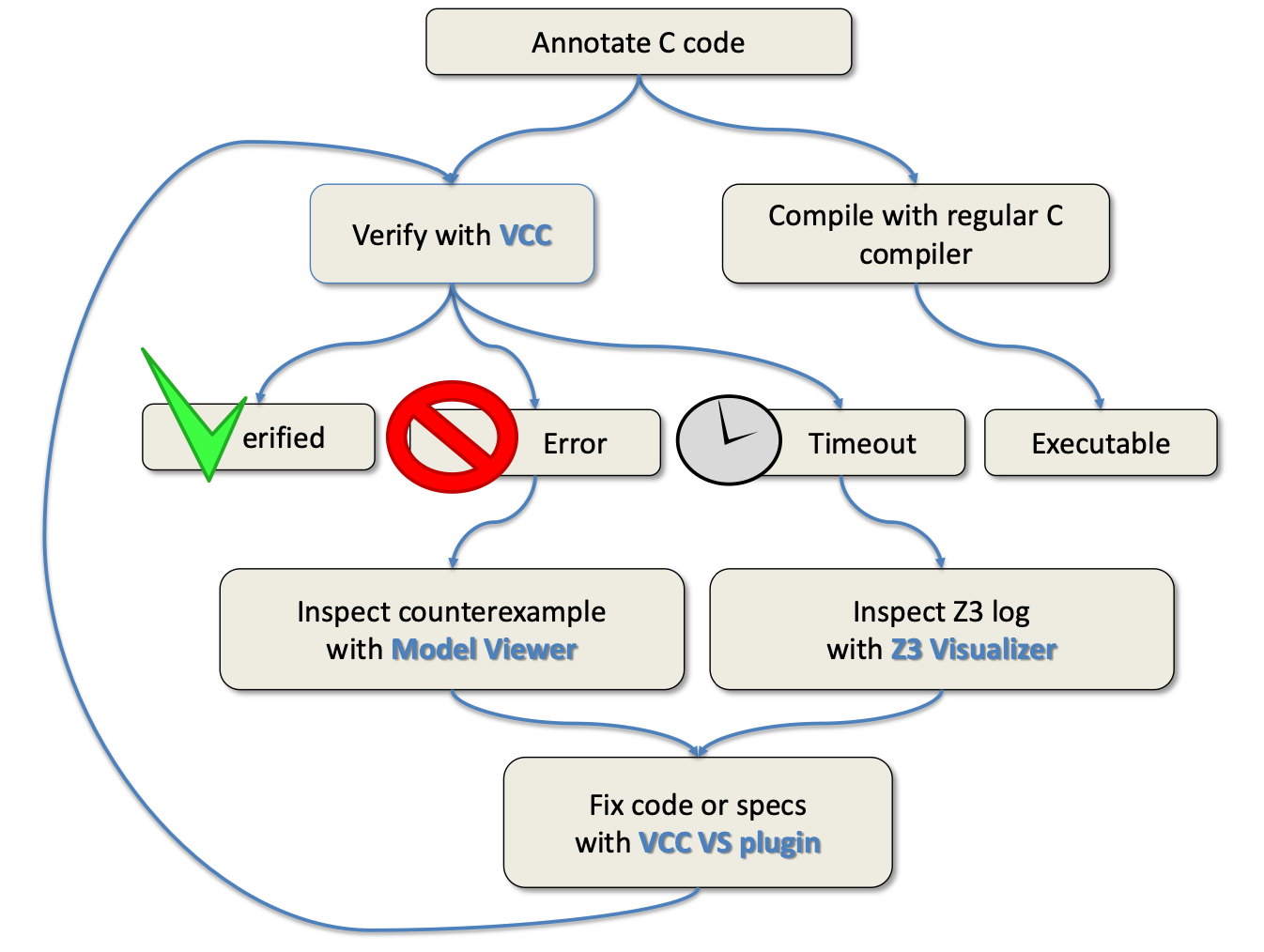}\\
\end{figure}

The annotated C program is passed to the VCC tool, which generates a Boogie program. The Boogie program is then passed to the Boogie tool, which generates verification conditions. These conditions are finally passed to the Z3 tool that generates its verdict regarding the input conditions’ (in)correctness.

\subsubsection{Style}
The following code represents a specified implementation of acquiring a spin lock:
\begin{lstlisting}
void Acquire (struct Lock *l)
  _(maintains \wrapped (l))
  _(ensures l-> locked == 1)
  _(ensures \ wrapped (l-> protected_obj) && \ fresh (l-> protected_obj))
{ 
  int stop = 0;
  do {
    _(atomic l) {
      stop = InterlockedCompareExchange (&l->locked , 1, 0) == 0;
      _(ghost if (stop) l->\ owns -= l-> protected_obj)
    }
  } while (!stop);
}
\end{lstlisting}
Acquiring a spinlock requires that the spinlock stays closed during the function call, not just at call time. A thread cannot assert, however, that an object is closed if it does not own it. The researchers used so-called claims to solve this problem. A claim:

\begin{itemize}
\item is a first class object, which references other objects (e.g. locks);
\item can be owned and used by arbitrary threads or objects;
\item can state property (e.g. lock stays closed while closed handles exist);
\item is just syntactic sugar.
\end{itemize}

Claims are passed as ghost arguments to functions. They serve as stronger preconditions, which can constrain volatile shared state, and hold until the claim is destroyed.

The following diagram illustrates the process of translating a C program annotated with VCC style specifications down to the Z3 level:

 \begin{figure}[ht]
  \centering
  \includegraphics[width=4cm]{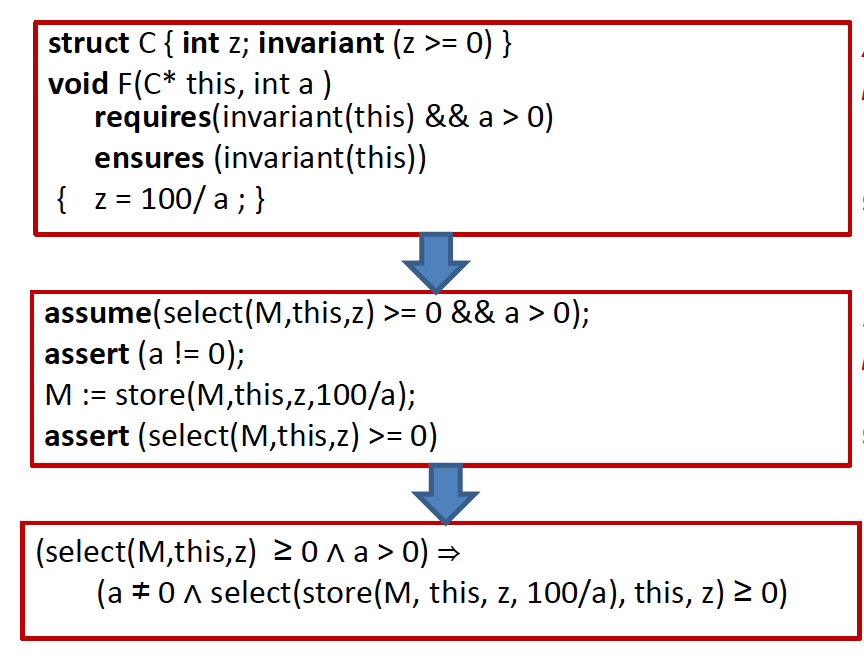}\\
\end{figure}

\subsubsection{Software characteristics}
\label{hyper_v:characteristics}

The verification team found several concurrency and design bugs that testing had not uncovered. The authors of the report say these bugs are virtually impossible to trace back from field failure data. For successful verification attempts, the authors report the typical verification time to be in the interval of 0.5-500 seconds, with the average time of 25 seconds. The biggest time it took to successfully verify an individual feature was 2000 seconds. The biggest time it took to successfully verify the whole codebase was 50000 seconds; this number decreased to 1000 seconds with the upgrade from Z3 to Z3v2. The authors report failing verification attempts to often take much longer than the final successful verification of the whole system. 

\subsubsection{Project characteristics}
The annotations have approximately doubled the codebase as compared to the executable baseline. On average, 3 functions were verified each day during 1 year of work; 4500 (out of 5000) lines of assembler and 30000 (out of 100000) lines of C were verified. The estimated person effort is around 1.5 person-years, including VCC learning period \cite{cohen2009vcc}.

\subsubsection{Lessons}
The researchers report poor performance as the most annoying issue of the verification approach (Section \ref{hyper_v:characteristics} gives the numbers). With the acceptable time for interactive work not exceeding 30 seconds, the current situation looks unacceptable performance-wise.

Hyper-V is a moderately large piece of software consisting of more than 100KLOC that was developed with no concern for verification. The specification language thus should be concise yet powerful enough to specify the properties of interest with an acceptable annotation burden. The complexity and size of the system also puts expectations on the verification technology. Because failed proof attempts often take much longer than the finally successful verification, the prover should map such failures back to the source code level. It should be possible to do profiling of the prover in the event it hits resource constraints. For long running proofs, it should be possible to perform live monitoring of the prover`s work. Summing up these lessons, the specification and verification technology should keep turnaround times for verify \& fix cycles acceptable.

\subsection{PikeOS}
\label{system:PikeOS}

\begin{table*}[htbp]
 \scriptsize
  \centering
  {\renewcommand{\arraystretch}{1.3}
        \begin{tabular}{|p{40pt}|p{45pt}|p{55pt}|p{55pt}|p{45pt}|p{43pt}|p{40pt}|}
    \hline
     Nature & {Properties} & {Programming language(s)} & {Programming style} & {Verification framework} &{ Specification language(s)} & {Number of LOC} \\
    \hline
     Hard real-time operating system, hypervisor & Functional correctness, security & C, assembly & Procedural & Hoare logic & VCC contracts & -- 
    \\    \hline \hline
      {Spec/Imp}  & {Proof Engine} & {Team Background} & {Institution(s)} & {Development period} & {Effort}  & {Size of user community} \\
    \hline
   5 & Boogie, Z3  & Researchers & German Research Center for Artificial Intelligence, European Microsoft Innovation Center, Microsoft Research, Saarland University & 2007-2010 & 1.5  & No number is available; may be very high because the OS is used intensively in embedded systems.
    \\ \hline
    \end{tabular}%
    }
  \label{pikeos_id}%
\end{table*}%

\subsubsection{Scope}
The goal of the Verisoft XT Avionics project was to formally verify an existing operating system which was not designed for deductive code verification. 
PikeOS is one of them.

\subsubsection{Components}
The verified software is PikeOS -- a commercial closed-source hard real-time operating system kernel that can double as a hypervisor developed by SYSGO AG. PikeOS runs mission-critical systems and supports different hardware platforms. It is used in development of software systems that require industrial certifications, such as life-critical avionics software. PikeOS supports ARM, PowerPC, x86, and SPARC hardware architectures.

\subsubsection{Verified properties}
The researchers have verified functional correctness and security properties, where security is understood as non-interference -- an important property that guarantees safe memory separation between concurrently running threads \cite{baumann2011proving}.
Although PikeOS is a hard real-time operating system, we could not find any reported results of verifying the real-time properties. 

\subsubsection{Project context}
One consequence of using an existing and widely deployed system as verification target is the impossibility to modify the code base to make it more amenable to verification \cite{baumann2012lessons}. In addition, PikeOS was not written with deductive verification (as used in VCC) in mind. On the other hand, because of the context of the application domains of PikeOS (especially avionics), the implementation avoids overly complex (and possibly more efficient) implementations in favor of maintainable, adaptable, and robust code. PikeOS as a component of avionics systems has successfully passed DO-178B evaluations \cite{johnson1998178b}.

\subsubsection{Decisions}
  The following issues influenced the project decisions: \cite{baumann2012lessons}: 
\begin{itemize}
\item Implicit behavior in informal specifications. End-user documentation and requirement engineering documentation for PikeOS provide little input for constructing annotations for functional verification. The researchers have overcome this problem by analyzing test cases, industrial certification requirements, and the source code itself to guess the developers’ intentions.
\item No syntactic distinction between different kinds of annotations. The VCC verification approach provides no mechanism to syntactically distinguish between specification- and verification-related annotations. In certification processes, it is important to have a very clear understanding of which annotations form the requirement specification that has been verified. Also, managing annotations in case of software evolution is more difficult when no clear distinction is given: while auxiliary annotations may be changed or removed at will, requirements have to stay unchanged. As a consequence, without this distinction, maintainability of the annotations suffers, e.g., by having to separate requirements from auxiliary annotations in such a case. The researchers solve this problem by keeping requirement annotations in the header file and auxiliary annotations in the source file. Where no such separation is possible, they use keywords (in the style of visibility modifiers public and private).

\item Entangled specifications.
Function specifications may have strong dependencies on each other.
Finding the right annotations for a single function requires the verification engineer to consider several functions at once, due to these dependencies.
Good feedback of the verification tool in case of a failed verification attempt is essential.
Feedback given by annotation-based verification tools so far only focuses on the function currently being verified.
This allows the user to pinpoint and fix bugs in the specification or program respectively to change auxiliary annotations, e.g., loop invariants or contracts for called functions.
For the verification of single functions, this tool support is sufficient.
However, current specification methodologies provide limited help in the presence of interdependent specifications.
The researchers solve this issue by combining software bounded model-checking (SBMC) with deductive verification done by VCC \cite{beckert2011integration}.
Annotations written in VCC's specification language are translated into assertions that can be checked by the LLBMC tool \cite{sinz2010precise}.
The SBMC procedure then allows for checking these assertions without the need to provide any additional auxiliary annotations as in the case with VCC, e.g., functions are inlined and loops unrolled instead of modularized using annotations.
However, in contrast to deductive verification, the number of loop iterations and the function invocation depth are bounded.
If no assertion is violated within the given bounds, this does not imply that the program adheres to its specification, as a bug may still occur, e.g., in a loop iteration outside the given bound.
That is, LLBMC does not provide a full proof (that is left to VCC) but a quick check that can point the user to problems in the annotations early on and thus avoid unnecessary VCC proof attempts.

\item Module granularity.
VCC simply treats each function as a module of verification.
Conversely, modules in PikeOS are not restricted to one function.
The researchers solve this problem by inlining function calls.
This would help to make function contracts obsolete for those functions of the system that are not a module on its own, but only part of some larger functionality.
This would already moderate the issue of imposing fixed module boundaries for functions in some cases.

\item No language support for abstraction.
Abstraction in the VCC tool is mainly achieved by using specification (``ghost'') state together with built-in abstract data types like maps or sets, where the abstract state is related to the concrete program state with the help of coupling invariants.

\item Change management. An issue that arises when verifying an implementation in production use is that it is constantly evolving, for example, to satisfy changing requirements from users of the software. To cope with these changes, when deciding on the frequency of applying code updates, one has to find a balance between costs for re-verification and the benefits of improvements in the system that might simplify specification and verification. The Verisoft XT research team, at the beginning, constantly applied changes occurring in the production code (at the expense of time available for the actual verification progress). At a later point, no more updates were done to the verification target. To support change management, a test harness was used for regression testing to check whether previously completed proofs were still valid or had to be redone.

\end{itemize}

\subsubsection{Tool stack}
The verification tool stack is the same as for the Hyper-V project.
VCC (Verifying C Compiler) is a deductive verification tool for concurrent C programs at the source code level that is used to prove correctness of C implementations against their functional specification. The VCC toolchain allows for modular verification of concurrent C programs using function contracts and invariants over data structures.

 \begin{figure}[htbp]
  \centering
  \includegraphics[width=4cm]{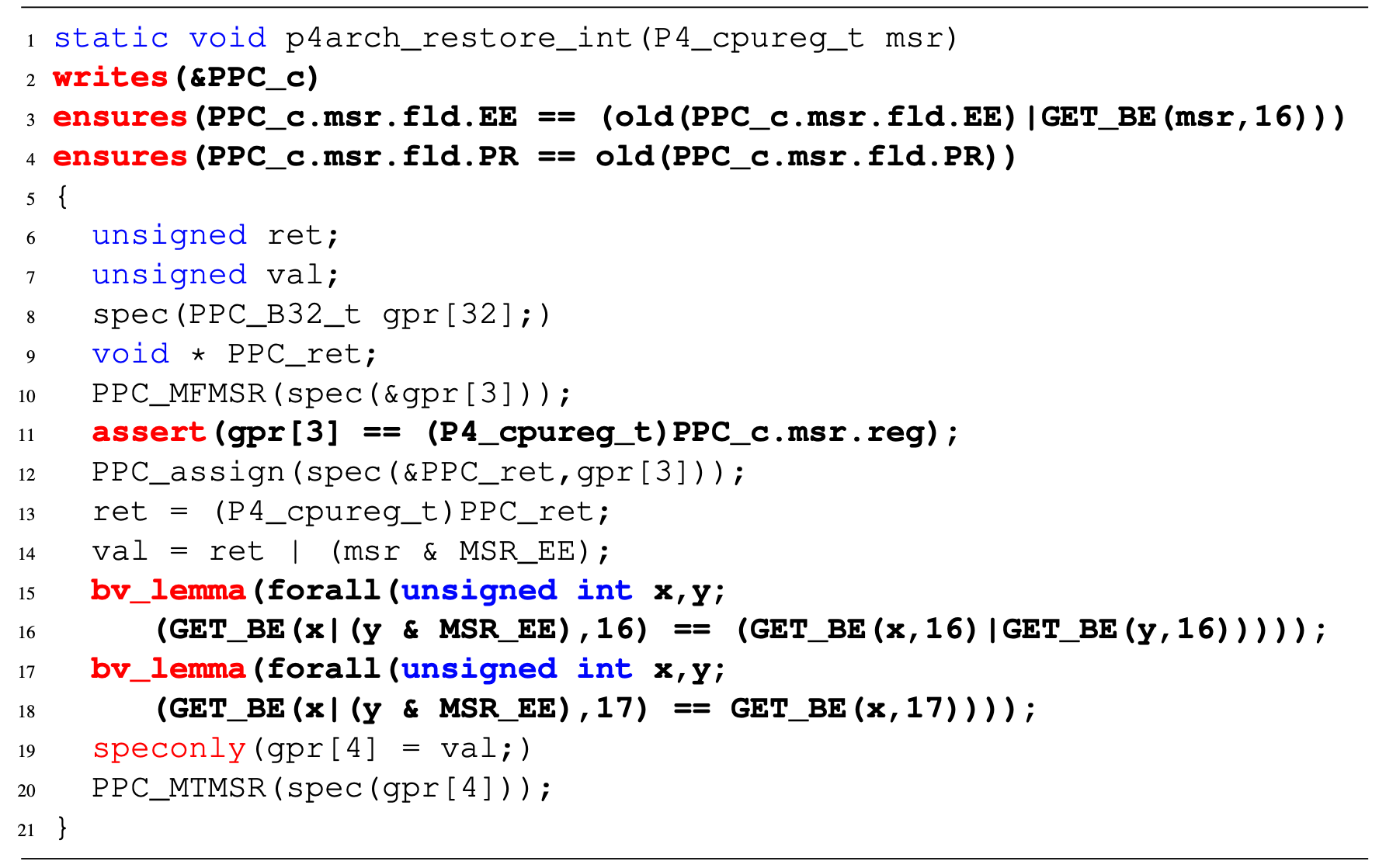}\\
  \caption{}
  \label{fig:pikeos1}
\end{figure}

VCC works using an internal two-stage process. The first stage of the VCC toolchain translates the annotated C code into first-order logic via an intermediate language called Boogie \cite{deline2005Boogiepl}. Boogie is a simple imperative language with embedded assertions. The Boogie executable generates a set of first-order logic formulas stating that the program meets the assertions. These formulas are called verification conditions and stage a verification condition generator (VCG).
In the second stage, the resulting formulas are sent to the Z3 theorem prover \cite{de2008z3} together with a background theory capturing the semantics of C’s built-in operators. The prover checks whether the verification conditions are entailed by the background theory.
Interaction with the VCC tool is only possible before the first step of the toolchain, by providing annotations. Once sufficient annotations have been provided (assuming the program fulfills its specification), the proof is done automatically, hence the term auto-active was coined for systems following the interaction paradigm.
Another tool used for verifying PikeOS was LLBMC \cite{sinz2010precise}, a software bounded model checker. The researchers used it for quickly checking the annotations for possible violations before trying to verify them with VCC. The (b) part of the following picture illustrates the overall verification process:

 \begin{figure}[htbp]
  \centering
  \includegraphics[width=4cm]{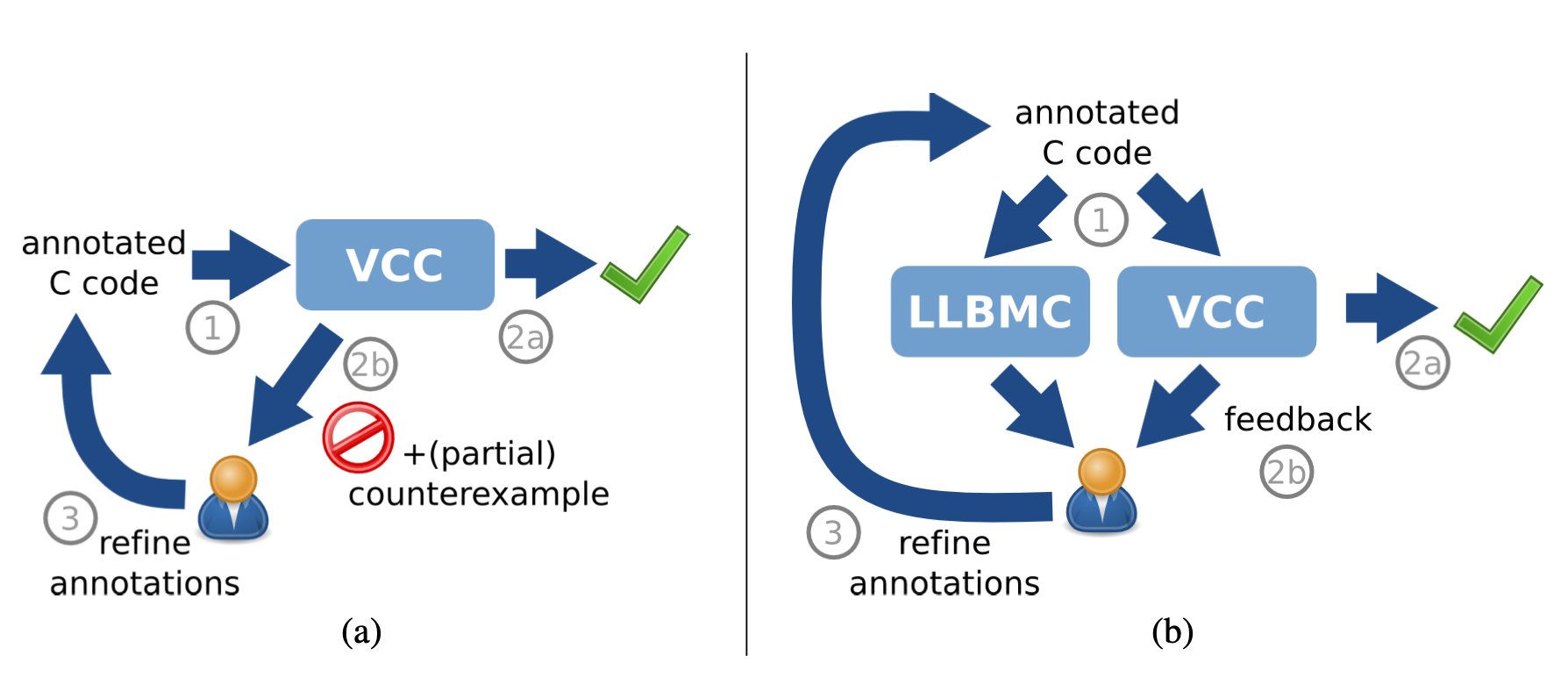}\\
  \caption{}
  \label{fig:pikeos2}
\end{figure}

As opposed to the standard VCC process, the new process starts with finding possible violations of the annotations with the help of the LLBMC model checker. If it does not find any problems within the allotted time limit, the verification engineer tries to verify the source code against the annotations.

\subsubsection{Style}
The specification style is the same as for the Hyper-V project. The following example of the annotated PikeOS code reflects this style:

 \begin{figure}[htbp]
  \centering
  \includegraphics[width=4cm]{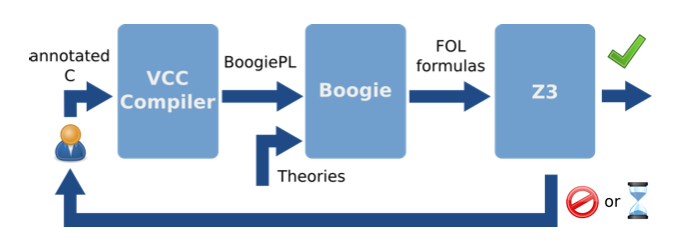}\\
  \caption{}
  \label{fig:pikeos3}
\end{figure}

The ``writes'' and ``ensures'' clauses outside the body of the method form its contract, while the auxiliary annotations inside the body (in red) assist the verification process.

\subsubsection{Software characteristics}
We could not find explicitly stated characteristics of the PikeOS verification project. The Verisoft XT project providing the funding was a three-year project. It included, however, verification of Hyper-V and another small microkernel. The authors of the most recent report on PikeOS verification \cite{baumann2012lessons} state that the amount of code to verify in Hyper-V by orders of magnitude exceeds that of PikeOS.

\subsubsection{Lessons}
The researchers report the following issues, in addition to those discussed in the Design decisions section, that have no straightforward solution:
\begin{itemize}
\item Finding the right abstraction. Multiple dependencies between the functions that all operate on (parts of) the same data structure make it hard to find the right abstraction: what properties of a data structure are needed and relied upon in the software depends on its usage in the functions of the system.
\item Support in finding auxiliary annotations. Auxiliary annotations, such as loop (in)variants and lemmas, are not visible at the interface level (unlike pre- and postconditions), but give additional information to the prover so that it can discharge the interface-level proofs. The ability of a verification engineer to come up with right auxiliary annotations in the right places in the code still heavily depends on their experience. The VCC tool does not provide any hints for finding the right auxiliary annotations and the right places to put them.
\item Amount of annotations. For the functional verification of parts of the memory manager in PikeOS \cite{baumann2011proving} the overhead of specification compared to the code compared by lines was still 5:1, using VCC as of Feb. 11, 2011. The high ratio is due to the auxiliary annotations written to ``convince'' VCC that the given implementation is correct.
\end{itemize}

The researchers emphasize the two most important issues: (1) reducing the amount and complexity of annotations to improve their maintainability and (2) simplifying the process of constructing auxiliary annotations required to make the prover accept correct programs.
To reduce the amount of annotations, the researchers propose utilizing specification patterns and specification refactoring. To reduce the complexity of annotations, they propose to apply specialized mathematical formalisms, such as the Common Algebraic Specification Language (CASL), Communicating Sequential Processes (CSP) or Abstract State Machines (ASM).
To simplify the process of constructing auxiliary annotations, the researchers propose to delegate the proof task altogether to proof assistants, so that the main verifier (VCC) could reuse their results, kept in dedicated objects, to complete verification attempts. As an example of such a toolchain, they mention the HOL-Boogie tool that has been developed to interactively perform complex proofs that VCC cannot handle.
}{}

\subsection{ExpressOS}
\label{system:ExpressOS}

\iftoggle{inBook}{

\begin{table*}[htbp]
 \scriptsize
  \centering
  {\renewcommand{\arraystretch}{1.3}
        \begin{tabular}{|p{40pt}|p{45pt}|p{55pt}|p{55pt}|p{45pt}|p{43pt}|p{40pt}|}
    \hline
     Nature & {Properties} & {Programming language(s)} & {Programming style} & {Verification framework} &{ Specification language(s)} & {Number of LOC} \\
    \hline
     mobile phone operating system & security invariants & C\# / Dafny & Object-oriented
 & Hoare logic, abstract interpretation & Code contracts \& Dafny annotations & N/A
    \\    \hline \hline
      {Spec/Imp}  & {Proof Engine} & {Team Background} & {Institution(s)} & {Development period} & {Effort}  & {Size of user community} \\
    \hline
   2.8\% annotation overhead & Z3 solver & Researchers & University of Illinois & 2013 & --  & --
    \\ \hline
    \end{tabular}%
    }
  \label{expressos_id}%
\end{table*}%

}{}

\subsubsection{Scope}

ExpressOS is a mobile operating system kernel featuring the interface of Android that enforces seven security invariants.
Those proven invariants provide a foundation for applications to isolate their state and run-time events from other applications running on the same device.

\subsubsection{Components}
The scope of verification is limited to the ExpressOS kernel.
The ExpressOS kernel is responsible for managing the underlying hardware resources and providing abstractions to applications running above.

\subsubsection{Verified properties}
The verification effort included checking 7 security invariants that guarantee storage security, memory isolation, UI isolation, and secure IPC.

\subsubsection{Project context}
The distinguishing characteristic of the project was its focus on preserving security invariants, rather than establishing full functional correctness.
It allowed the researchers to verify the resulting system against these invariants, whereas attempts to specify and verify other Unix-based kernels achieved only partial success.
The project was motivated by the lack of formal guarantees that mobile applications do not cross each other's security boundaries which can be reinstalled or removed at any time, on smartphones that retain a near-constant Internet connection.

\subsubsection{Decisions}
ExpressOS relies on four main techniques to simplify verification effort: \cite{Mai2013Verifying}:
\begin{enumerate}
    \item It pushes functionality into microkernel services, reducing the amount of code that needs to be verified.
    \item It deploys end-to-end mechanisms in the kernel to defend against compromised services; for example, the kernel encrypts all data before sending it to the file system service.
    \item It relies on programming language type-safety to isolate control and data flows.
    \item It makes minor changes to the IPC system-call interface to expose explicitly IPC security policy data to the kernel.
\end{enumerate}


\subsubsection{Tool stack}
The ExpressOS kernel is implemented in C\# and Dafny \cite{leino2010dafny}.
The verification uses Boogie proof engine \cite{leino2008boogie} and Z3 SMT solver \cite{de2008z3} from Microsoft Research.
The researchers combined two specification and verification techniques to keep the annotation burden under control: (i) using Code contracts in C\# code for specifying lightweight properties, then verified using abstract interpretation based techniques, (ii) using Dafny for specifying, implementing, and verifying data structures that are less amenable to straightforward verification, such as doubly linked lists.
Both C\# and Dafny code compile to object code for native execution.

\subsubsection{Style}
Fig. \ref{fig:expressos_1} contains a fragment of the C\# method that handles page faults and includes an assertion enforcing the following property, enforcing memory isolation, which illustrates well the specification style of ExpressOS.

\noindent \textbf{Property 5-} \textit{When the page fault handler serves a file-backed page for a process, the file has to be opened by the same process.}

\begin{figure}[h!]
\vspace{-0.3cm}
\begin{lstlisting}[language={[Sharp]C},morekeywords={var}]
static uint HandlePageFault(Process proc, uint faultType, Pointer faultAddress, Pointer faultIP) {
  Contract.Requires(proc.ObjectInvariant());
  ...
  AddressSpace space = proc.Space;
  var region = space.Regions.Find(faultAddress);
  ...
  if (shared_memory_region) {...} else { ...
    if (region.File != null) {
      // Assertion of Property 5
      Contract.Assert(region.File.GhostOwner == proc);
      var r = region.File.Read(...); ... } ... } ... }
\end{lstlisting}
\vspace{-0.3cm}
\caption{Property 5 in code contracts.}
\label{fig:expressos_1}
\vspace{-0.5cm}
\end{figure}

\noindent The invariant of class \emph{MemoryRegion} (Fig. \ref{fig:expressos_2}), implemented in Dafny, ensures the assertion in the C\# client code.
The \emph{GhostOwner} built-in ghost field of an object denotes the object's owner.
In this specific case, it has type \emph{Process} because every memory region is owned by a process.
The invariant of a memory region states that the file backing the region is owned by the same process owning the region, which is what the required property says.\\

\begin{figure}[h!]
\vspace{-0.6cm}
\begin{lstlisting}[language={[Sharp]C},morekeywords={var,function}]
class MemoryRegion { ...
  var File: File; var GhostOwner: Process;
  function ObjectInvariant() : bool ... { File != null ==> File.GhostOwner == GhostOwner ... }}
\end{lstlisting}
\vspace{-0.3cm}
\caption{Dafny code ensuring the C\# assertion of Property 5.}
\label{fig:expressos_2}
\vspace{-0.2cm}
\end{figure}

\noindent The verification results from Dafny are expressed as properties of ghost variables, such as the \emph{GhostOwner} field, exported to code contracts as ground facts with the \emph{Contract.Assume()} statements.
\emph{GhostOwner} fields also appear read-only C\# annotations to ensure soundness.

\subsubsection{Software characteristics}
The resulting system's performance is comparable to that of Android: it only adds 16\% overhead on average to the page load latency time for nine popular web sites.
Focusing on security invariants and combining specification and verification techniques made it possible to achieve around 2.8\% annotation overhead.
To evaluate the security of ExpressOS, the authors built a prototype system and examined 383 vulnerabilities listed in CVE \cite{cve}.
The ExpressOS architecture successfully prevents 364 of them \cite{Mai2013Verifying}.

\subsubsection{Lessons}
During the verification process, the specification using ghost code was sometimes too closely intertwined with the implementation.
The authors believe that an alternate mechanism for writing specifications that is a bit more code-independent, resilient to code changes, and yet facilitates automated proving would make the developers' work more robust and productive.

They also observe that combining code contracts and Dafny reduced the annotation-to-code ratio to about 2.8\%  and that the redesign of the data structures and the implementation of the property using ghost variables made verification easier. 

\iftoggle{inBook}{
\subsection{SeKVM}

\begin{table*}[htbp]
 \scriptsize
  \centering
  {\renewcommand{\arraystretch}{1.3}
        \begin{tabular}{|p{40pt}|p{45pt}|p{55pt}|p{55pt}|p{45pt}|p{43pt}|p{40pt}|}
    \hline
     Nature & {Properties} & {Programming language(s)} & {Programming style} & {Verification framework} &{ Specification language(s)} & {Number of LOC} \\
    \hline
     Hypervisor & Functional correctness, security & C, Assembly& Procedural & HOL & Coq & 15.5k
    \\    \hline \hline
      {Spec/Imp}  & {Proof Engine} & {Team Background} & {Institution(s)} & {Development period} & {Effort}  & {Size of user community} \\
    \hline
    2.1 & Coq & Researchers & Columbia University & 2018 & 3
  & --
    \\ \hline
    \end{tabular}%
    }
  \label{sekvm_id}%
\end{table*}%

\subsubsection{Scope}
SeKVM is a retrofitted and verified version of KVM \cite{li2021secure}, which is a full-featured commodity multiprocessor hypervisor that is integrated with Linux. In this verification project, the microverification (MicroV) approach was introduced, in which the hypervisor was decomposed into a small core and a set of untrusted services such that security properties of the entire hypervisor can be proved by verifying the core alone.

\subsubsection{Components}
In this verification project, the KVM hypervisor was retrofitted into SeKVM, which consists of a small, verifiable core, KCore, and a rich set of untrusted hypervisor services, KServ. KCore mediates all interactions with VMs and enforces access controls to limit KServ access to VM (virtual machine) data, while KServ provides complex virtualization features. KCore protects the security of each VMs’ data from being tampered with by untrusted KServ.

\subsubsection{Verified properties}
The project aimed to prove that SeKVM protects the confidentiality and integrity of VM data. Confidentiality means that adversaries should not be privy to private VM data while integrity specifies that adversaries should not be able to tamper with private VM data.

\subsubsection{Project context}
Formal verification of a commodity hypervisor is largely intractable due to its complexity. Most existing verified systems are specifically designed to be verified, meaning they lack basic features and are far simpler than their commodity counterparts. This suggests that the proof effort potentially required to fully verify a commodity system is far beyond feasible. To address this problem, the authors introduced the MicroV approach to modularize the proofs and thus reduce the proof efforts.

\subsubsection{Decisions}
The functional correctness of the MicroV framework's core was proved, showing that its implementation refines specification. 
The specification was then used to prove security properties of the entire system, (noninterference between VMs and KServ) was proved at the high-level specification. The verification is decomposed into two parts: 1) verification that the refinement relations preserve security properties (the refined implementation preserves the same security properties as the corresponding specification); 2) verify that the top-layer specification preserves security properties. Since the specification is easier to use for higher-level reasoning, it becomes possible to prove security properties that would be intractable if attempted directly on the implementation.

\subsubsection{Tool stack}
All the security-preserving layer specifications, refinement proofs, and noninterference proofs were implemented using Coq and checked by the Coq proof assistant.

\subsubsection{Software characteristics}

A key verified feature of SeKVM is page tables that can be shared across multiple CPUs, which became possible to verify by introducing transparent trace refinement \cite{li2021secure}. This makes it possible to provide verified support for multiprocessor VMs on multiprocessor hardware as well as DMA (dynamic memory access) protection.
The verified KVM performs comparably to stock, unverified KVM. On real application workloads, SeKVM incurs only modest overhead compared to unmodified KVM. In most cases, the overhead for SeKVM is similar to unmodified KVM and less than 10\% compared to native \cite{li2021secure}. The worst overhead for SeKVM versus unmodified KVM is for a hypercall that measures bulk data send performance from the VM to a client.

\subsubsection{Project characteristics}
The SeKVM implementation is based on KVM from mainline Linux 4.18.  In the retrofitting phase, 1.5K LOC (about 10\% of the codebase) were modified in existing KVM code. The entire retrofitting process took one person-year. The Coq development effort took two person-years. The results show that microverification of a commodity hypervisor can be accomplished with modest proof effort.

\subsubsection{Lessons}
This project introduces security-preserving layers to incrementally prove the functional correctness of the core, ensuring the refinement proofs to the specification preserve security guarantees. The verification result show that microverification only required modest KVM modifications and proof effort, yet results in a verified hypervisor that retains KVM’s extensive commodity hypervisor features, including support for running multiple multiprocessor VMs, shared multi-level page tables with huge pages, and standardized virtio I/O virtualization.

\subsection{PROSPER}

\begin{table*}[htbp]
 \scriptsize
  \centering
  {\renewcommand{\arraystretch}{1.3}
        \begin{tabular}{|p{40pt}|p{45pt}|p{55pt}|p{55pt}|p{45pt}|p{43pt}|p{40pt}|}
    \hline
     Nature & {Properties} & {Programming language(s)} & {Programming style} & {Verification framework} &{ Specification language(s)} & {Number of LOC} \\
    \hline
    Separation kernel & Information flow security & C, Assembly& Procedural 
 & Theorem proving, binary code analysis & HOL, BAP Intermediate Language (BIL) & 2717 (c); 942 (Assembly)
    \\    \hline \hline
      {Spec/Imp}  & {Proof Engine} & {Team Background} & {Institution(s)} & {Development period} & {Effort}  & {Size of user community} \\
    \hline
     -- & HOL4, ARMv7 prover, BAP, SMT solver & Researchers and experienced engineers  & KTH Royal Institute of Technology  & 2012 - 2017 (first verified version in 2013) & --  & 3 members (1 institute and 2  companies) in PROSPER website
    \\ \hline
    \end{tabular}%
    }
  \label{prosper_id}%
\end{table*}%

\subsubsection{Scope}
A separation kernel is a low-level software execution platform that allows the simulation of a distributed environment on a single physical machine using partitions that are executed in isolation while controlling communication among them appropriately. PROSPER is a separation kernel for ARMv7 that supports the isolation of component resources, a mechanism for communication between partitions and a scheduling mechanism for allocating shared resources (e.g., processors) among the components. 

\subsubsection{Components}
The PROSPER kernel is designed to handle four minimal tasks, which include separating resources protection into two partitions (sender and receiver), executing two  partitions on a single physical machine, scheduling of partitions and delivering  messages from the sender to the  receiver. Depending on the status of the receiver’s message-box, different message statuses, message contexts and program counters are set. Thus, PROSPER creates an environment where guests are ``isolated'' but still have the possibility of 
communicating with each other via RPCs (Remote Procedure Calls).

To realise PROSPER, a model called ``ideal system'' (the PROSPER specification) which formalizes the Top Level Specification (TLS)  was created. The ideal system models the actual system, which is then transferred to the actual system after being verified against required separation properties. 

\subsubsection{Verified properties}
The verification goal of PROSPER is limited to its information flow security: the partitions have no way to directly or indirectly affect each other, except through the intended (allowed) channels. The verification task thus is to show that, properly set up, the user observable traces of the ideal system are the same as those of the ``real system'', obtained by the execution of the software in different partitions, on top of a real ARMv7-A processor. This is proved with the  bisimulation proof method \cite{dam2013formal}.

\subsubsection{Project context}
To design secure embedded systems, software components that belong to different security domains require adequate isolation from each other, in such a way that only authorized communication between them is allowed. An approach to achieve this is to execute  the components in isolated partitions on a single shared hardware, by using low-level software execution platforms such as separation kernels \cite{rushby1981design,heitmeyer2008applying} or secure hypervisors \cite{gehrmann2011there,mcdermott2012separation,sailer2005shype}. A key requirement is the verification of the tamper resistance of trusted computing base. PROSPER only verifies that the information flow between both partitions is secure.

\subsubsection{Decisions}
PROSPER was implemented in C and Assembly.  It was developed for single-core ARMv5 and ARMv7 architectures. 

\subsubsection{Tool stack}
The overall proof was carried out in the HOL4 theorem prover. Contracts were generated for the parts that depend on kernel code of the ARM chip and the BAP tool (a binary analysis tool that can verify low-level code, such as machine code) was then used to verify them. A number of helper tools were also employed to realise the generation of contracts and the transfer of their verification to BAP. BAP was used to compute weakest preconditions because HOL4 requires a significant engineering effort and also because the BAP \cite{brumley2011bap} toolset provides platform-independent utilities. This BAP toolset reasons on the BIL. Therefore, a front-end for ARMv7 programs was developed to translate the code of the kernel handlers and bootstrap into BIL. In addition to the ARMv7 prover tool implemented in SML/HOL4, various tools and tool components interfacing HOL4 with BAP were also used. Also, a lifter tool that converts ARM assembly to BAP’s input language was used to fit the HOL4 version to the expected BAP input before transfer to BAP binary code analysis tool for verification. An SMT solver was also in the process of binary code verification. 

\subsubsection{Style}
TLS  of PROSPER was formulated to directly formalize the set of computation paths both allowed and required at the implementation level. TLS was formally specified in higher-order logic (HOL) – a mathematical language. 

\subsubsection{Software characteristics}
For the overall verification of parts not dependent on the chip's Kernel code, a monadic HOL4 model of ARMv7 instruction set architecture was used \cite{fox2010trustworthy}. Due to the complexity of memory management, the model needed to be restricted to support only those parts of the Memory Management Unit (MMU) functionality used by the PROSPER kernel. This was achieved through the extension of the ARM model to support MMU functionality in the PROSPER setting. A relational Hoare logic framework is then used to prove the ARM Lemmas. A set of inference rules were implemented in a semi-automatic HOL4 helper tool to aid this process. Pre- and postconditions for the kernel handlers (the parts dependent on kernel code), were generated using HOL4. Then the pre- and postconditions generated by HOL4 are used for the subsequent verification with BAP, whose verification relies on Hoare Logic. 

PROSPER was deployed on Beaglebone, Beagleboard, Beagleboard-xM, NovaThor and the Integrator development board. It was also deployed on emulated platforms within the OVP framework and upon Qemu \cite{dam2013formal}.

\subsubsection{Project characteristics}
The ideal system’s model, the proofs of the theorems and the formalization of the procedures for verification make up 21,000 lines of HOL4 code in total. The tools developed to support the verification of the kernel contracts were completed with a total of 2,000 lines of HOL4 and python code. Sixteen binary contracts formed the basis, with respect to which the binary code was verified. Each of the sixteen contracts consist of approximately 400 lines of assertions that were automatically generated from HOL4. The worst case scenario times for the different processes involved is given below:
\begin{itemize}
\item Verification of one contract using Intel(R) Xeon(R) X3470 core : 30 minutes
\item Generation of the contract: 5 minutes
\item Solving of the indirect jumps: 2 minutes
\item Computation of the weakest precondition: 10 minutes
\item Verification of the resulting condition by the SMT solver: 15 minutes.
\end{itemize}

\subsubsection{Lessons}
The TLS, in specifying the exact set of traces allowed by an implementation, might become too detailed and unwieldy \cite{dam2013formal}. This might be a potential issue, especially for project replication. The approach used in the implementation of the ideal model might also produce differences between the real implementation and the ideal model.  For example, the ideal model was constructed with 2 separated machines, while the actual system is installed on only one hardware.

One of the first major concerns is timing, because the model counts instruction cycles, instead of real clock cycles. This limitation makes it impossible for the current approach to handle access-driven attacks, which requires a more refined analysis of timing behaviour \cite{dam2013formal}. Secondly, there is the issue of unpredictable states, brought about by a limitation in the ARM Architecture. The ARM Architecture Reference Manual \cite{armv7manual} describes that unpredictable behaviour is not allowed to ``perform any function that cannot be performed at the current or lower level of privilege using instructions that are not unpredictable'', which is difficult to accommodate in the framework of PROSPER \cite{dam2013formal}.

Being one in a series of projects to formally verify base kernels, the PROSPER project is the first to apply verification on the information flow security between two partitions in a shared single hardware. The development and verification of the PROSPER project keys into the notion of formally verified systems providing stronger quality assurance in comparison to non-formally verified projects. However, the amount of re-engineering and/or extension of available tools also shows that considerable technical effort and know-how is needed to develop a formally verified system.

\subsection{SHOLIS}

\begin{table*}[htbp]
 \scriptsize
  \centering
  {\renewcommand{\arraystretch}{1.3}
        \begin{tabular}{|p{40pt}|p{45pt}|p{55pt}|p{55pt}|p{45pt}|p{43pt}|p{40pt}|}
    \hline
     Nature & {Properties} & {Programming language(s)} & {Programming style} & {Verification framework} &{ Specification language(s)} & {Number of LOC} \\
    \hline
    Ship-borne information system & Functional correctness; security; timing/memory constraints & SPARK & Procedural
 & Hoare logic & Z notation, SPARK contracts
 & 27k (executable code)
    \\    \hline \hline
      {Spec/Imp}  & {Proof Engine} & {Team Background} & {Institution(s)} & {Development period} & {Effort}  & {Size of user community} \\
    \hline
    3.07 & Simplifier, Proof Checker & Two proof engineers (experienced mathematicians) and two coders & UK Ministry of Defense (MOD)
Altran UK &  1993-1999 & 19  
  & Installed on  UK 23 type Frigates, 160 users (informal guess)
    \\ \hline
    \end{tabular}%
    }
  \label{sholis_id}%
\end{table*}%

\subsubsection{Scope}
Ship/Helicopter Operational Limits Information System (SHOLIS) is a ship-borne computer system that aids the safe operations of helicopters on naval vessels. SHOLIS was developed for the UK Ministry of Defense (MoD) and (intended to be) deployed on the UK Royal Navy and Royal Fleet Auxiliary vessels.

\subsubsection{Components}
 SHOLIS is an information system that gives advice on the safety of helicopter flying operations. It contains a database of Ship Helicopter Operating Limits (SHOLs), which specifies the limits for performing certain operations (e.g., taking off or landing) for different types of helicopters. One major safety-critical function of SHOLIS is to continually make comparisons of sensor information against the SHOL selected by operators. When the current environmental conditions (detected by sensors) exceed the limit, audible and visual alarms will be triggered.

\subsubsection{Verified properties}
The verification of SHOLIS was conducted at requirement, design and implementation level. At the requirement and design level, the analyzed properties include standard properties (e.g., consistency of global variables and constants) and high-level safety requirements (e.g., an alarm must be triggered when the detected sensor value exceeds the currently selected SHOL). At the implementation level, the verified properties include: 1) data- and information-flow properties (e.g., imported/exported data can be read from/written to); 2) functional correctness; 3) absence of run-time errors (invalid index, division by zero, overflow).

\subsubsection{Project context}
In 1991, when the early drafts of UK MoD Defense Standard 00-55 \cite{mod1991procurement} were produced, there was a certain amount of controversy among software suppliers \cite{king1999value}. This was because of the emphasis on use of formal methods in safety-critical applications in the standard, which was deemed to be unrealistic given the technology at the time. The SHOLIS project was started in 1993 and was the first effort to meet the requirements of the Interim version of the standard for critical software. Meanwhile, it was also the first attempt at serious proof of a non-trivial system using SPARK \cite{chapman2014we}.  

\subsubsection{Decisions}
Due to its critical nature, SHOLIS was developed to stringent standards, the MoD Defense Standards (DS) \cite{mod1991procurement}. During the design and verification, the functions of SHOLIS were divided into two categories, safety-critical components and non-critical components. The safety-critical components are the software with the potential to cause catastrophic hazards (errors that would lead to catastrophic damages to the ship/helicopter), and must conform to SIL4 (the highest software integrity level in DS). The non-critical software is less safety-critical but still needs to be subject to a stringent standard roughly equivalent to SIL3. The proof of functional correctness was only performed for SIL4 code. To prevent the incorrect interference of non-SIL4 code with critical data on the same processor, the data- and information-flow analysis was conducted for all code.

\subsubsection{Tool stack}
At the requirement and design level, the behaviors of SHOLIS was specified in Z notation and the proof was carried out manually by a form of rigorous argument, with the assistance of the CADiZ tool \cite{toyn1995cadi}. At the code level, the proof was performed by SPARK commercial verification toolset: the Examiner, Simplifier and Proof Checker. The Examiner performed data- and information-flow analysis and generated verification conditions (VCs) from SPARK annotations. Simplifier then carried out routine simplification using a collection of rules and discharged the VCs automatically. The VCs that cannot be discharged and the Simplifier were discharged by Prover Checker interactively.

\subsubsection{Style}
\label{SPARK_Z_SPEC}
Two specification languages were applied in the project, Z notation and SPARK annotations (SPARK83 version).
Z notation was used to specify the behaviors of SHOLIS in the requirement and design phases. The basic building blocks of Z are schemas, which define variables, constraints and operations in terms of the variables. A schema in Z describes either a state entity or an operation. An example of using schema to describe a state entity Indicator is shown in Fig. \ref{fig:sholis1}. The upper section (signature) declares three variables and their types, while the lower section (predicate) defines logical relationships between entities.


SPARK annotations were used at the code level to specify data- and information-flow properties as well as functional properties. The annotations were generated based on Z specifications. Annotations are encoded in comments (via the prefix --\#). For example, annotation --\# global means that the procedure uses some global variables. Annotation --\#derives specifies that some variable value depends on other variables. Annotations --\#pre and --\#post define preconditions and postconditions of a certain procedure.

\subsubsection{Software characteristics}
 The size of the SHOLIS project is considerable. The system consists of 500 pages of Z proofs and 133000 lines of code. In the verification, 9000 VCs were generated, among which 3100 VCs were for functional and data- and information-flow properties and 5900 VCs for RTC.

About 75\% of the VCs were discharged by the Simplifier successfully and the remaining VCs were discharged by the Proof Checker virtually, with a few cases explored manually through rigorous arguments. The system was certified to Def-Stan 00-55. After 10 years of operational service, the system has a defect density of 0.22 defects per KLOC \cite{white2017formal}.

\subsubsection{Verification project effort }

The proof activity on the project was carried out by four engineers (mathematicians and experienced software engineers). A 5-day training for reading/writing Z specification and a 10-day SPARK training (covering both basic and advanced SPARK) were carried out. The proof costed 19 person-years in total \cite{king1999value}.

\subsubsection{Lessons}
Although formal proof was conducted to provide guarantees for the correctness of source code, the delivered system can still suffer from compiler-introduced errors. To alleviate this, the SHOLIS team applied a commercial and validated Ada compiler to provide some guarantees of quality. Moreover, the optimizer of the compiler was switched off (the only compiler bug found in the project was introduced by the optimizer). However, there is still a lack of rigorous analysis to further eliminate compiler-introduced errors.

SHOLIS has shown both the benefit and practicality of large-scale proof on projects of its kind. A key output was the identification of the need to support state abstraction and refinement in SPARK proofs for reducing the complexity and size of contracts. This was implemented after the SHOLIS project but had a major impact on later projects using SPARK.

The main bottleneck of SHOLIS was the limited computing resources for theorem proving in the late 1990s. Simplification time for a single subprogram was measured in hours or days at the time. Later on a reverification of SHOLIS was carried out on a modern (2013-era) quad-core machine, which took only about 11 minutes and discharged about 95\% VCs \cite{chapman2014we}.

\subsection{Lockheed-Martin C130J}

\begin{table*}[htbp]
 \scriptsize
  \centering
  {\renewcommand{\arraystretch}{1.3}
        \begin{tabular}{|p{40pt}|p{45pt}|p{55pt}|p{55pt}|p{45pt}|p{43pt}|p{40pt}|}
    \hline
     Nature & {Properties} & {Programming language(s)} & {Programming style} & {Verification framework} &{ Specification language(s)} & {Number of LOC} \\
    \hline
    Communi- cation software & Functional correctness & SPARK & Procedural
 & Hoare logic & Z notation, SPARK contracts
 & 350k 
    \\    \hline \hline
      {Spec/Imp}  & {Proof Engine} & {Team Background} & {Institution(s)} & {Development period} & {Effort}  & {Size of user community} \\
    \hline
    -- & Simplifier, Proof Checker & experience in avionic applications development & LASC (Lockheed Aeronautical Systems Company) and Praxis Critical Systems &  1991-1998 & --  
  & 400 delivered aircrafts
    \\ \hline
    \end{tabular}%
    }
  \label{c130j_id}%
\end{table*}%

\subsubsection{Scope}
The Lockheed-Martin C130J is the newest generation of the ``Hercules" military transport aircraft and the only model still in production. Formal verification has been performed on the Operational Flight Program (OFP) software in the Mission Computers (MCs) and the Bus Interface Units (BIUs) of the aircraft.

\subsubsection{Components}
The verified OFP software includes a primary OFP in each of two MCs and a backup OFP in each of two BIUs. The OFPs control the many individual systems on the C130J aircraft to operate and provide information on the integrated aircraft. More specifically, they synchronize and transfer data between various avionic hardware devices, e.g., electronic circuit boards, radar and the fuel system. The OFPs also process and transfer information needed for the flight station hardware (e.g., head-up displays, radio control panels, alert annunciators) to interface with the pilot, copilot, navigator and auxiliary crew members.

\subsubsection{Verified properties}
The verified properties include internal consistency of the software in terms of absence of data- and information-flow errors as well as the functional correctness of implementation regarding  requirements.

\subsubsection{Project context }
In the 1990s, the Lockheed-Martin C-130J (or Hercules II) was updated from C-130 transport aircraft with significantly improved avionics, engines, and propellers. Part of the transformation involved new OFP software, which was required to be certified for both civil and military use. This meant meeting DO-178B for civil applications, as well as the requirements of the UK’s Royal Air Force on static analysis and formal verification.

\subsubsection{Decisions}
The OFP software in MCs and BIUs performs a great multiplicity of functions to transfer information to and from buses, originating and terminating at various types of devices. Those functions were devised to be mutually independent and simple to describe in a precise manner. This exploitation of the ``separability'' of the many functions of the OFPs contributed greatly to the simplicity of the specification of requirements, their implementation and verification.

\subsubsection{Tool stack}
The verification was conducted using the SPARK verification toolset, the Examiner, the Simplifier and the Proof Checker. The SPARK Examiner performed data- and information- analyses, and then generated proof obligations from the SPARK programs. Most of the proof obligations can be discharged automatically by the Simplifier and the rest can be proved interactively using the Proof Checker.

\subsubsection{Style}
The specifications were expressed using the ``Parnas tables'' in CoRE (Cosortium Requirement Engineering), which is a formal requirement modelling method. Fig. \ref{fig:c130j1} illustrates an example of the Parnas table excerpted from \cite{chapman2000industrial}. Each line specifies an element of the overall input-to-output function (a dependency relation), which can be expressed as a clause within the corresponding SPARK postcondition of the relevant operation (as shown in Fig. \ref{fig:c130j2}).

 \begin{figure}[htbp]
  \centering
  \includegraphics[width=6cm]{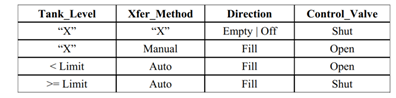}\\
  \caption{An example of Parnas table (``X'' means ``any value'')}
  \label{fig:c130j1}
\end{figure}

 \begin{figure}[htbp]
  \centering
  \includegraphics[width=6cm]{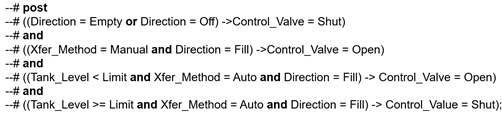}\\
  \caption{Postcondition derived from the Parnas table}
  \label{fig:c130j2}
\end{figure}

\subsubsection{Software characteristics}

When compiling real-time embedded systems written in Ada, run-time checks are usually turned off to reduce code size and improve performance. To provide evidence to justify the ``with checks off'' compilation, proof of the absence of predefined exceptions was conducted. Due to the simplicity of the required input-output functions, the generation and simplification of the proof obligations for SPARK programs was straightforward. Formal verification of the programs was performed before compilation and testing and the expensive testing was performed only once, as confirmation of correctness.

The SPARK verification approach was applied in the C130J project to great effect. Due to the earlier elimination of data- and information-flow errors by the Examiner, the code in modified condition/decision coverage (MC/DC) testing exhibited an unusually low fault density, less than one tenth of the expected industry norm for safety critical software. This contributed to an 80\% saving in the expected budget allocated to MC/DC testing \cite{chapman2000industrial}.

\subsubsection{Lessons}
The Lockheed-Martin C130J project was one of the first projects where the use of formal methods has proved to be advantageous from an economic standpoint. Most of the errors uncovered in the requirements specifications and code by formal verification would have been discovered during integration testing. However, it has been found that detecting such bugs from traditional test results is more expensive than formal analysis of the same code \cite{croxford1995breaking}. Some errors that can be immediately uncovered by formal analysis (e.g., conditional initialization errors) may only emerge after very extensive testing. Although the burden of proof is increased if verification is attempted for code that does not yet meet its specification, the cost savings obtained by catching errors early, and especially before system integration and costly MC/DC testing, are very great in comparison \cite{croxford1995breaking}.

The essential factors of the success of this project include: 1) factorization of the sizeable system into a large number of functionally separate components; 2) formal specification of requirements using CoRE; 3) preservation of function separation in the implementation through the translation from the CoRE requirements into formal specifications of SPARK program components; 4) the technology for generation and discharge of the proof obligations based on the formal definition of SPARK program \cite{croxford1995breaking}.
}{}

\subsection{NATS iFACTS}

\iftoggle{inBook}{

\begin{table*}[htbp]
 \scriptsize
  \centering
  {\renewcommand{\arraystretch}{1.3}
        \begin{tabular}{|p{40pt}|p{45pt}|p{55pt}|p{55pt}|p{45pt}|p{43pt}|p{40pt}|}
    \hline
     Nature & {Properties} & {Programming language(s)} & {Programming style} & {Verification framework} &{ Specification language(s)} & {Number of LOC} \\
    \hline
    Tools for air traffic controller & Absence of run-time exceptions; functional correctness; memory constraints & SPARK & Procedural 
 & Hoare logic & Z notation, SPARK contracts
 & 250k
    \\    \hline \hline
      {Spec/Imp}  & {Proof Engine} & {Team Background} & {Institution(s)} & {Development period} & {Effort}  & {Size of user community} \\
    \hline
    0.296 & Simplifier & Engineers & NATS (national air traffic service)  &  2006 - present (deployed  in 2011)  & $>50$  
  & Deployed by NATS' facility in Swanwick, UK
    \\ \hline
    \end{tabular}%
    }
  \label{ifacts_id}%
\end{table*}%

}{}

\subsubsection{Scope}
With the ever-increasing capacity of air traffic over the UK, the National Air Traffic Service of UK developed a set of electronic tools to assist air-traffic control. The tool set is called interim Future Area Control Tools Support (iFACTS),  which was proven free of run-time errors.

\subsubsection{Components}
iFACTS provides functions to help human controllers to handle air traffic, including electronic management of flight progress strip, trajectory prediction and conflict detection. 
iFACTS enables controllers to look up to 18 minutes ahead. This allows them to test the viability of various options available for a maneuvering aircraft, and thus controllers have more time for decision-making. As the flights progress, the tool updates predictions information (positions, altitude and headings of all aircrafts in their sectors of interest)
and the controllers issue new clearances based on the real-time predictions. 


\subsubsection{Verified properties}
The proof of iFACTS concentrates on type-safety properties: the absence of run-time exceptions such as buffer overflow, arithmetic overflow or division by zero. The verification also covers functional correctness and memory usage constraints \cite{questionnaire}.

\subsubsection{Project context}
The integration of iFACTS into industrial systems requires the system to be certified according to Civil Aviation Authority SW01 standards \cite{authority2001sw01}. iFACTS was the first system developed under these new regulations.


\subsubsection{Tool stack}
Implementation of iFACTS used the SPARK framework; the verification tool set includes the Examiner and Simplifier; the Examiner generates verification conditions (VCs) in FDL language (a typed first-order logic) from input SPARK programs; the Simplifier is a theorem prover for discharging the VCs.

\subsubsection{Style}
In requirement and design phases, iFACTS specifications are in Z: the basic building block is a  schema, which defines variables and associated constraints and operations. 
\vspace{-0,4cm}
\begin{figure}[htbp!]
  \centering
  \includegraphics[width=4cm]{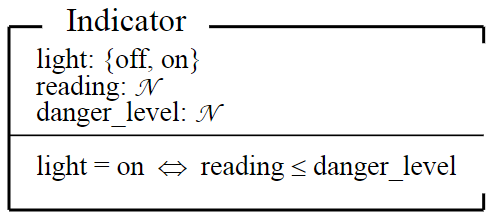}\\
  \vspace{-0.3cm}
  \caption{The graphical layout of Indicator schema}
  \label{fig:sholis1}
\end{figure}

\vspace{-0,3cm}
Fig. \ref{fig:sholis1} shows an example of a schema (from SHOLIS \cite{king1999value}), describing a \emph{state entity} named \gt{Indicator}. The upper section (or \emph{signature}) declares three variables and their types; the lower section (\emph{predicate}) defines logical relationships between the three variables.
In the implementation phase, specifications are given as SPARK annotations, which describe data- and information-flow properties as well as functional properties. Those SPARK annotations were generated from the Z specifications in the previous design phase. Annotations are encoded in comments via the prefix \gt{--\#}. For example, ``\gt{--\#global}'' indicates the use of certain global variables; ``\gt{--\#derives}'' specifies the dependency between the variables; ``\gt{--\#pre and --\#post }'' define pre- and post-conditions of a certain procedure.

\subsubsection{Software characteristics}
The Examiner generates 152927 VCs from the SPARK code; the Simplifier discharged 151026 VCs (98.76\%) automatically and the remaining 1701 VCs were analyzed manually; a single proof run  took 3 hours \cite{chapman2014we}. Owing to the tools' ability to parallelize and  to exploit caching of proof results, re-verification of small changes can take only seconds. In addition, the development also established a `proof-first'  style --- all code changes must be proved correct before being committed.

iFACTS have been in full operation in NATS' facility since November, 2011. The deployment of iFACTS has enhanced the capacity of airspace and the efficiency of air-traffic controllers. It was reported that there was an average 15\% increase in airspace capacity in the UK in 2012 \cite{natsblog2013}.

\subsubsection{Project characteristics}
The research and development team of the project comprised over 140 engineers from NATS and Altran and the project spanned over 10 years; they carried out a one-week formal training, including the training on Z notation and SPARK.

\subsubsection{Lessons}
Verification of iFACTS scaled much better than previous SPARK projects such as SHOLIS and C130J, thanks to the considerable improvement in computing resources and theorem-provers \cite{croxford1995breaking}. The further reduction of time consumption of the overall project, however, is limited due to the rudimentary tool support for Z; as they used MS Word documents to store Z specifications, merging different versions of large documents (over 1000 pages documents) would be difficult and time-consuming.

What makes this verification project stand out from others includes not only the scale of implementation (iFACTS is one of the most ambitious SPARK project to date) but also the application of formal methods throughout the complete development lifecycle. Additionally, the experience showed that training engineers was not a barrier --- engineers of diverse programming backgrounds could easily pick up verification languages/tools \cite{white2017formal}. 

\iftoggle{inBook}{
\subsection{EuroFighter Typhoon Aircraft FCS}

\begin{table*}[htbp]
 \scriptsize
  \centering
  {\renewcommand{\arraystretch}{1.3}
        \begin{tabular}{|p{40pt}|p{45pt}|p{55pt}|p{55pt}|p{45pt}|p{43pt}|p{40pt}|}
    \hline
     Nature & {Properties} & {Programming language(s)} & {Programming style} & {Verification framework} &{ Specification language(s)} & {Number of LOC} \\
    \hline
    Flight control software & Functional correctness & Ada & Procedural
 & HOL & Simulink, Z notation
 & 35k (core functionalities)
    \\    \hline \hline
      {Spec/Imp}  & {Proof Engine} & {Team Background} & {Institution(s)} & {Development period} & {Effort}  & {Size of user community} \\
    \hline
    -- & Supertac, ProofPower & -- & QinetiQ plc  &  2001-2010  & 1.5 
  & More than 550 aircrafts have been delivered, 584000 flying hours with unparalleled operational statistics
    \\ \hline
    \end{tabular}%
    }
  \label{EuroFighter_id}%
\end{table*}%

\subsubsection{Scope}

The Eurofighter Typhoon is a multinational agile combat aircraft controlled by a digital fly-by-wire system, the flight control software (FCS).

\subsubsection{Components}
The FCS is used to control aircraft’s flight surfaces to achieve improved stability and manoeuvrability in the air. It provides carefree handling characteristics to minimize pilot workload and improve flight safety during failure-free flight, when either no warning is alerted or the failure consequences are minor \cite{stich2012clearance}. For instance, if the aircraft decelerates towards its minimum airspeed, the auto-low-speed recovery function takes over and recovers the aircraft at a safe speed before control is handed over to the pilot again. In addition, some emergency features (e.g., activation of alarms, disabling or enabling certain operations) have also been included in the FCS to ensure safety of aircraft operations. 

\subsubsection{Verified properties}
The verification aimed to establish the correctness of a set of control laws incorporated in the FCS, which describe the algorithms of transforming the operations by the pilot into movements of the aircraft control surfaces.

\subsubsection{Tool stack}
The control laws of the FCS were modeled in Simulink \cite{simulink}, a commercial tool of MathWorks for graphical modelling, simulating and analyzing real-time embedded systems. Ada source code can be generated automatically from the Simulink diagram.
Formal verification on the Ada code was carried out using the CLawZ (Control Laws in Z) toolset \cite{vernon2010communication}, which is composed of a set of tools, e.g., Z Producer, RSG (refinement script generator), DAZ, Supertac and ProofPower. Z Producer translates Simulink diagrams into Z specifications. The Z specifications are fed into RSG to generate a refinement script, which specifies how the Ada code correctly implements the Z specifications. Based on the refinement script, DAZ generates verification conditions, which can be discharged by the two theorem provers, Supertac and ProofPower.

\subsubsection{Style}
The control laws of FCS were modeled as Simulink diagrams (or subsystems), which consist of a collection of input/output ports and mathematical/operational blocks, all connected by wires.  Fig. \ref{fig:euro1} shows a Simulink diagram of a differentiator subsystem, namely Diff, in which the output (Deriv) represents the changes of input (Error). The Unit Delay block stores the value of Error for the next iteration of computation: the output of Unit Delay represents the value of Error at the previous step. The Sum block computes the value of Deriv by subtracting the output of the Unit Delay block from Error.

 \begin{figure}[htbp]
  \centering
  \includegraphics[width=4cm]{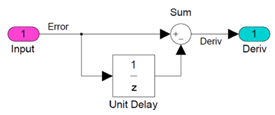}\\
  \caption{An example of Simulink diagram (excerpted from \cite{o2013automated})}
  \label{fig:euro1}
\end{figure}

Based on the Simulink model (stored as an .mdl file), the Ada code and Z specifications of the corresponding subsystem can be generated. The Ada code is then verified against the Z specifications.

For each block in the CLawZ Simulink subset, there is a Z definition that specifies its semantics. The Z schema of a Simulink subsystem can be constructed based on the Z definitions of its constituent blocks. Fig. \ref{fig:euro2} shows the Z specification of the Diff subsystem. The \$``discrete/Diff/Unit Delay'' and \$``discrete/Diff/Sum'' correspond to the Unit Delay and Sum blocks that appear in the subsystem Diff of the discrete system model (a model that uses discrete time-steps). The \$``discrete/Diff'' schema describes the relations between inputs and outputs of the blocks in accordance with the wiring depicted in Fig. \ref{fig:euro2}. state and state’ represent the state of Diff before and after the execution, respectively. 

 \begin{figure}[htbp]
  \centering
  \includegraphics[width=6cm]{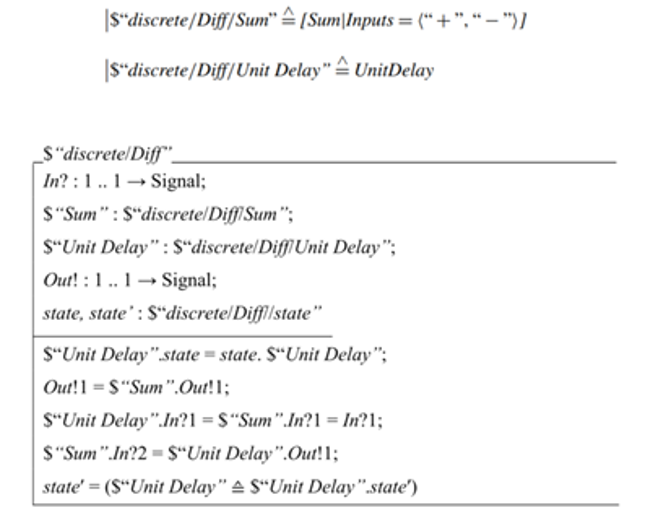}\\
  \caption{Z specifications of Diff subsystem}
  \label{fig:euro2}
\end{figure}
To enable the verification of Ada code of the subsystem, an interface schema is generated. The interface schema specifies the correspondence relation between the variables in the Z specifications and variables in the Ada implementation. It is then used to form the overall postcondition that the code must satisfy.

\subsubsection{Software characteristics}

Due to the predictable automatic code generation (where human creativity was eliminated), the program proof can be automatized, which reduces the overall development cost dramatically. Formal specification of the system was generated automatically from the Simulink model. The refinement script was also automatically generated based on traceability information from the auto-coder.

Furthermore, the verification was cost-effective due to the high level of proof automation and the relatively abstract level of user-interaction. 95\%-98\% of the verification conditions can be proved automatically with Supertac and the remainder is checked by ProofPower interactively. The analysts only need to work on high-level specifications (e.g., modeling control laws in Simulink) and do not need to construct scripts containing low-level details, which reduces the requirement on analysts’ expertise and the amount of user interaction. 

\subsubsection{Verification efforts}
The core control functionality of FCS is implemented in about 35,000 lines of Ada code. Typically, 50-80 LOC can be verified per person-day \cite{adams2005clawz}. The total verification effort is about 538 person-days (1.5 person-years) taking an average verification rate.

\subsubsection{Lessons}
Despite huge improvements to the productivity of the CLawZ toolset, which was ongoing throughout the verification project, the size of the verified systems still pushed the tools to their limits \cite{questionnaire}. Moreover, there is still significant scope for improvement to help further reduce analyst time and expertise \cite{adams2005clawz}, e.g., RSG and Supertac can provide more feedback to the analysts to help in tracking down analyst errors more quickly. Supertac can be enhanced to further automate verification condition proof. Additionally, the CLawZ toolset supports only a well-defined subset of Simulink while mainstream users of Simulink typically want to use features outside the subset.

The CLawZ toolset is dedicated for the specific verification domain where the verification of control system software is specified in Simulink and implemented in Ada. However, the verification approach could equally be used to verify systems implemented in other programming languages, or specified in other design notations. The large scale of the cost reductions brought about by CLawZ opens up the prospect of much wider use of formal verification, including during development, and perhaps even for non safety-critical code. This approach could therefore have a profound impact on the use of formal verification in industrial applications\cite{adams2005clawz}.
}{}

\subsection{Ynot}

\iftoggle{inBook}{
\begin{table*}[htbp]
 \scriptsize
  \centering
  {\renewcommand{\arraystretch}{1.3}
        \begin{tabular}{|p{40pt}|p{45pt}|p{55pt}|p{55pt}|p{45pt}|p{43pt}|p{40pt}|}
    \hline
     Nature & {Properties} & {Programming language(s)} & {Programming style} & {Verification framework} &{ Specification language(s)} & {Number of LOC} \\
    \hline
    Database management system & Functional correctness & Coq, Ocaml& Functional
 & HOL & Coq
 & 3030
    \\    \hline \hline
      {Spec/Imp}  & {Proof Engine} & {Team Background} & {Institution(s)} & {Development period} & {Effort}  & {Size of user community} \\
    \hline
   1 &  Coq & Researchers and engineers & Harvard university  & 2008-2010 &  8 
  & -- 
    \\ \hline
    \end{tabular}%
    }
  \label{ynot_id}%
\end{table*}%

}{}

\subsubsection{Scope}
Relational database management systems (RDBMSs) allow application developers to have a high-level specification of the behavior of the data manager (one of the modules in the RDBMS), which is suitable for formal reasoning about application-level security and correctness properties. It is one of the systems implemented by the team of the Ynot Project.

\subsubsection{Components}
The Ynot RDBMS is a lightweight, fully verified, in-memory database management system. A command-line interface serves to create tables, load tuples into a table, save/restore a table to/from disk, and query tables using an SQL subset.

\subsubsection{Verified properties}
The main verification task focuses on the correctness of executing queries in the RDBMS regarding denotational semantics of SQL and relations.

\subsubsection{Project context}
RDBMSs are omnipresent in modern application software. They are used to store data whose integrity and confidentiality must be maintained. The implementation of the data manager should ensure that a bug cannot bring about accidental corruption or disclosure.

\subsubsection{Decisions}
The design decisions of RDBMSs seek to avoid complicated proofs and reduce computations. The introduction of ptrees and their association with B+ trees \cite{abiteboul1995foundations} as ghost states help in these simplifications. Avoiding disjunctions when something can be readily computed also helps reduce computations.
A ptree is a defined functional model of the tree, simplifying the task of defining predicates that describe when a heap contains a tree with a valid shape.  

\subsubsection{Tool stack}
The verification of the system involved two tools: the formal proof management system Coq and the Ynot extension to Coq. The proof that implementation meets the specification is written and verified in Coq. The Ynot extension to Coq is designed to support writing and proving correct pointer-based code, using a variant of separation logic.

\subsubsection{Style}
The specification language is Coq, which the team used to develop mathematical theories and prove specifications, relying on  OCaml to create Coq plugins for novel tactics or functionality. The following Coq specification example ensures correct usage of semantic optimizations:
\begin{lstlisting}
Definition accurate (m: DbInfo) (G: Ctx) (E: Env G) : Prop :=
forall s I, getRelation s I E = empty $<->$ isEmpty (m s I)
\end{lstlisting}

Here, \e{DbInfo} stands for database information; \e{Ctx} means context; \e{Env} describes environment; \e{s} and \e{I} denote a string and a schema respectively.

\subsubsection{Software characteristics}
The Ynot RDBMS is a proven lightweight memory system, in which the functional specification of system behavior, system implementation, and proof of achieving compliance with the specification were written and verified in Coq.

\subsubsection{Project characteristics}
Four people spent two years finishing this project.

\subsubsection{Lessons}
Successful application for real systems requires a number of tasks. Some need relatively modest extensions to the current implementation; for example, key information can be integrated to enable efficient point and range queries. Reifying the low-level query plan will help fuse operations to avoid materializing transient data (the low-level queries are those executed by the RDBMS using a sequence of operations over imperative finite maps).
Other tasks demand more effort. Realizing the ACID (Atomicity, Consistency, Isolation, Durability) guarantee of concurrency, transactional atomicity and isolation, and fault-tolerant storage is a challenging task. Implementing and verifying the correctness of high-performance, concurrent B+ tree is another challenge.



\subsection{Roissy Shuttle}

\iftoggle{inBook}{
\begin{table*}[htbp]
 \scriptsize
  \centering
  {\renewcommand{\arraystretch}{1.3}
        \begin{tabular}{|p{40pt}|p{45pt}|p{55pt}|p{55pt}|p{45pt}|p{43pt}|p{40pt}|}
    \hline
     Nature & {Properties} & {Programming language(s)} & {Programming style} & {Verification framework} &{ Specification language(s)} & {Number of LOC} \\
    \hline
    Safety critical software  &  safety & B, ADA & Procedural
 & B method & B notation
 & 158612
    \\    \hline \hline
      {Spec/Imp}  & {Proof Engine} & {Team Background} & {Institution(s)} & {Development period} & {Effort}  & {Size of user community} \\
    \hline
   1.16 &  automatic prover of Atelier B & Researchers and experienced engineers & Siemens Transportation Systems, ClearSy  & 2005 & --  & --
    \\ \hline
    \end{tabular}%
    }
  \label{roissyshuttle_id}%
\end{table*}%

}{}

\subsubsection{Scope}
The VAL shuttle  system runs driverless shuttles at CDG airport in Paris. The safety-critical software, developed in this project with the B Method, controls the shuttles' speed.


\subsubsection{Components}
A shuttle is an automatic light train running on the line connecting Roissy terminal 1 to terminal 2. The line is made up of 5 sections. Each section has a wayside control unit (WCU) that controls the speed of the driverless shuttle in the section, composed, in turn,  of a set of blocks. WCUs should localize the shuttle by checking whether a block is occupied. WCUs receive the commands from the traffic control center and drive the shuttles based on the commands. The control logic of WCU is implemented as a safety critical software, called WCU-SCS.


\subsubsection{Verified properties}
The development took place in three phases: abstract model construction, concrete model construction, and code generation. In the first phase, the informal software specifications, given in natural language, were formalized into an abstract model. The next phase builds a concrete model from the non-implementable parts of the abstract model.  The last phase generates the Ada code using both the abstract model and the concrete model. The verified properties include safety requirements for the abstract model, compliance between concrete model and abstract model, compliance between code and models.


\subsubsection{Project context}
The VAL system was developed by Siemens Transportation Systems, with  WCU-SCS subcontracted to ClearSy. The project's objective was to ensure that WCU-SCS satisfied a set of safety properties and that its development was within budget and schedule.



\subsubsection{Decisions}
ClearSy applied the Siemens B Method, a process designed for using B language to build correct software by construction. 
The functionalities of WCU-SCS were firstly modeled in B. After establishing their correctness, the B models were then translated into Ada code. Unit testing was not performed since most of the effort was devoted to the early specification phase, to build correct software directly.


\subsubsection{Tool stack}
Refinement from abstract model to concrete model was performed using two semi-automatic refinement tools, EDiTh B and Bertille. The proof of the B models was made by the automatic prover of Atelier B. Generation of ADA code uses the Digisafe-ADA technology.

\subsubsection{Style}
The high-level properties of the VAL system models are specified using abstract data types such as sets of scalar types, relations, partial and total functions. An example of a property for the abstract model is illustrated below.

\begin{lstlisting}
$\forall$ block (block $\in$ t_block $\land$ (( ctx_block_bs_up [{block}] $\cup$ ctx_block_bs_down [{block}] $\cap$ cut_beam_sensors $\neq$ $\varnothing$ $\lor$ ctx_block_detector [{block}] $\subseteq$ occupied_block_detectors) $\Rightarrow$  block $\in$ occupied_blocks)
\end{lstlisting}

\noindent $t\_block$ represents the block type and $ctx\_block\_xxx[\{block\}]$ denotes abstract variables for a given $block$. The property specifies that a block is regarded as occupied when a beam sensor located at one of the block borders is cut (denoted by $cut\_beam\_sensors \neq \varnothing$) or when the block detector is occupied (denoted by $ctx\_block\_detector[\{block\}] \subseteq occupied\_block\_detectors$).

\subsubsection{Software characteristics}


Automatic refinement can lower the costs of a software development, with an acceptable price of usually 10\% slower code compared to the handwritten version. In addition, although the approach can ensure that the code correctly implements the B model, it is difficult for developers to link the auto-generated code with the corresponding software specification.   

\noindent It was easier to prove the concrete model than the abstract model. The ratio between the effort on the abstract model phase and the effort on the concrete model phase is 2. The abstract model proof consists of complex and lengthy interactive demonstrations, making the proof difficult to reuse. The concrete model proof is broken down into small steps with the same patterns.
 As far as maintenance is concerned, it is guaranteed that the modified version will remain consistent as long as the proof is completely redone. The cost will also be limited since 1) the whole development environment has been set up and 2) the refinement and proof rules are likely to be reused.

\subsubsection{Project characteristics.}
The interactive verification took about 50 person-days.


\subsubsection{Lessons}
The method used in this project is also suitable for other industrial domains. It is suitable for software mainly based on discrete logic description, but not for software based on continuous calculation or on floating point numbers which cannot be considered as decimals. 

While compliance between the code and the model has been established by a mechanical proof, compliance between the model and the natural language specifications is checked manually, which still leaves room for potential errors.

\iftoggle{inBook}{

\subsection{Dutch Tunnel Control System}
\label{system:dutch_tunnel}

\iftoggle{inBook}{

\begin{table*}[htbp]
 \scriptsize
  \centering
  {\renewcommand{\arraystretch}{1.3}
        \begin{tabular}{|p{40pt}|p{45pt}|p{55pt}|p{55pt}|p{45pt}|p{43pt}|p{40pt}|}
    \hline
     Nature & {Properties} & {Programming language(s)} & {Programming style} & {Verification framework} &{ Specification language(s)} & {Number of LOC} \\
    \hline
    safety-critical control system & safety and liveness & mCRL2, Java& Modeling language
 & model-checking and code verification & Process 
algebra $\mu$-calculus
 &  4662 (Java )
700 (Mcrl2 )
~700 (Pvl)
    \\    \hline \hline
      {Spec/Imp}  & {Proof Engine} & {Team Background} & {Institution(s)} & {Development period} & {Effort}  & {Size of user community} \\
    \hline
   0.15 &  mCRL2 model checker, VerCors & Researchers & University of Twente  & 2015-2020  & 8  & --
    \\ \hline
    \end{tabular}%
    }
  \label{dutchtunnel_id}%
\end{table*}%

}{}

\subsubsection{Scope}
The considered  software component is  a safety-critical component of a control system for a traffic tunnel that is currently in use in the Netherlands.
It is responsible for handling emergencies (fires for example).
The Dutch government imposes very high reliability demands on the traffic tunnel control software, and in particular on this emergency component. 

\subsubsection{Components}
This project \cite{DutchTunnelControlSystem} investigates how formal methods can help developers to find potential problems in their specification and (Java) implementation, with realistic effort and preferably at an early stage of development. 
The tunnel was already deployed and the verification was done afterwards. 

\subsubsection{Verified properties}
The main properties of interest concern reliability and recoverability: Does the system always go into the Calamity state in real emergency situations? And is it always possible to recover from calamities, and thereby go back to the Normal operational state?

\subsubsection{Project context}
The tunnel control software is developed by Technolution, a Dutch software and hardware development company which has a big experience in developing safety-critical, industrial software. 
The development process of the traffic tunnel control system is accompanied by an elaborate quality assurance control process to meet high reliability requirements.

\subsubsection{Decisions}
Technolution invested significantly in an extensive design phase to ensure the quality of the control system and to cope with the high reliability demands. During design, which involved domain experts, the intended behavior of the control software was written out in pseudocode. 
The pseudocode specifications have been structured into a finite state machine (FSM). 
The states of this FSM are the operational states of the tunnel system (normal operation, under repair, being evacuated etc.), while the transitions are the pseudocode descriptions of the system behavior. 
The FSM thus illustrates how different behaviors and events of the tunnel system should change its operational state. 

\subsubsection{Tool stack}
A formal algebraic process model of the informal pseudocode and FSM description of the tunnel control software is defined using mCRL2.
This mCRL2 model is then analyzed by state-space exploration and by checking desired $\mu$-calculus properties on the model such as blocking freedom or strong connectivity. 
Finally, the automated verifier VerCors is used to prove that the program refines the mCRL2 model, and thus to verify deductively that the control system correctly implements (in Java) the pseudocode specification \cite{10.1007/978-3-030-61467-6_18}. 

\subsubsection{Style}







\noindent Once the mCRL2 model obtained (which formalizes the specification), desired properties are formulated as $\mu$-calculus formulae, and checked on the reduced model.
Here is an example of such a formula, expressing that the StandBy state can only ever be reached via the Normal state.

\begin{center}
\begin{lstlisting}[xleftmargin=.1\textwidth]
vx.( [($\neg$ enter(Normal)]))$^\ast$ $\cdot$ enter(StandBy)] false $\land$ [true$^\ast$ $\cdot$ enter(StandBy)]x )
\end{lstlisting}
\end{center}

\subsubsection{Software characteristics}
This project aims to establish on the one hand whether the specification is consistent and cannot reach problematic states (e.g. deadlocks in the FSM) and on the other hand whether the implementation of the Java code is consistent with the specification in pseudocode of the intended behavior.
Thanks to this project, an undesired behavior was detected: an internal deadlock due to an intricate combination of timing and events. 

\subsubsection{Project characteristics}
Two weeks of a PhD student specializing in formal methods where  necessary to formally model and analyze the traffic tunnel system. 
Rerunning the verification would require no effort. The step that will probably cost the most is generating the model with MCRL2, and that should take about 4 minutes. After that, verification of each property should be instantaneous. Verification with VerCors should also take less than a minute. 

\subsubsection{Lessons}

During the development phase, significant time and effort were invested in ensuring that the code was correctly implemented against the specification. 
This was done primarily by unit testing and code review. But the unit tests and code review revealed to not prevent from a serious bug in the code, which the current verification work has pointed out. 
This work demonstrated that formal methods can really help to detect undesired behaviors within reasonable time, that would otherwise be hard to find.
Despite this, the actual verification of mCRL2 was done separately from the development and deployment. There was no relationship between the two projects (no feedback on the verification of the actual software implementation was provided). 


}{}


\subsection{Sizewell B}

\iftoggle{inBook}{
\begin{table*}[htbp]
 \scriptsize
  \centering
  {\renewcommand{\arraystretch}{1.3}
        \begin{tabular}{|p{40pt}|p{45pt}|p{55pt}|p{55pt}|p{45pt}|p{43pt}|p{40pt}|}
    \hline
     Nature & {Properties} & {Programming language(s)} & {Programming style} & {Verification framework} &{ Specification language(s)} & {Number of LOC} \\
    \hline
    Safety critical system - Nuclear & Functional correctness & PL/M-86, ASM86, PL/M-51, ASM51 & procedural
 & Hoare logic & IL
 & 150k
    \\    \hline \hline
      {Spec/Imp}  & {Proof Engine} & {Team Background} & {Institution(s)} & {Development period} & {Effort}  & {Size of user community} \\
    \hline
    -- &  MALPAS & Industrialists & TA Consultancy Services (For Nuclear Electrics, UK)  & 1989-1993 & 250
  & --
    \\ \hline
    \end{tabular}%
    }
  \label{sizewell_id}%
\end{table*}%

}{}

\subsubsection{Scope}
Sizewell `B'  is a Westinghouse-designed Nuclear Pressurised Water Reactor (PWR) built in Sizewell, Suffolk in the UK. It possesses two diverse protection systems whose role is to provide an automatic reactor trip when plant conditions reach safety limits and to actuate emergency safeguard features to limit the consequences of a failure condition. 

\subsubsection{Components}
Nuclear Electric plc constructed and operated the Sizewell B PWR. Two systems provides overall protection (Primary Protection System or PPS) and secondary protection (SPS). The PPS uses microprocessor-based logic, while the SPS uses magnetic core logic (laddic) technology.
The design requirement for the PPS was to provide reactor trip and engineered safety feature (ESF) actuation for all design faults. 
The requirement for the SPS was to provide reactor trip and ESF actuation, in parallel with the PPS, for faults with a frequency in excess of about once in 1000 years. The reliability targets of the two systems were therefore 1 in 10,000 for PPS and 1 in 100,000 for SPS. They could have been achieved by hardware, but it was necessary to guaranteeing the same level on the software side.
A full code and compliance analysis was performed with the MALPAS analyzer; it showed that the code fully matched the requirements.

\subsubsection{Verified properties}
The verification ensures that the implementation of complex protective functions is accurate and that there is a high degree of confidence in this. 

\subsubsection{Project context}
The confirmatory assessment of the software using MALPAS (consisting of static and semantic analysis tools for all safety software) was conducted under the TACIS program of the European Union. It took place in a large acceptance process. An extensive program of software verification was carried out on the PPS software to detect and remove any errors generated in the design and coding phases. All of the assessments were carried out by independent companies or independent entities of the Nuclear Electric plc. 

\subsubsection{Decisions}
The analysis of the PPS software was conducted on a procedure-by-procedure basis in a bottom-up manner. 
The analysis commenced with procedures that call no others, and then progresses up the call hierarchy until the top level application code is reached. 
All the MALPAS analysers were run on each procedure and the results verified against two levels of specification: a higher level specification (the  Software
Design Requirements, SDR) and  a lower level (the Software Design Specification, SDS). 
The SDR was the primary document against which the code was verified, with supporting information provided by the SDS.
All the code that can be accessed during on-line operation was subjected to MALPAS analysis.

\subsubsection{Tool stack}
The MALPAS toolset includes a set of analyzers addressing various program properties. 
They process input in the MALPAS Intermediate Language (IL), produced from the original source code by an automated translation tool assisted by an analyst who can make suitable manual changes.
There are five analyzers:
\begin{enumerate}
    \item Control Flow Analyser examines the program structure, and provides a summary report drawing attention to undesirable constructs and an indication of the complexity of the program structure.
    \item Data Use Analyser separates the variables and parameters used by the program into distinct classes and identifies errors.
    \item Information Flow Analyser identifies the data and branch dependencies for each output variable or parameter.
    \item Semantic Analyser reveals the exact functional relationship between all inputs and outputs over all semantically feasible paths through the code.
    \item Compliance Analyser compares the mathematical behavior of the code with its formal IL specification to ensure that the code of each procedure:conforms to the functional aspects of its specification (SDR, supported by SDS), respects the state invariants, performs its specified functions without corrupting the computing environment nor data owned by any other module or procedure, conforms to the language in which it is written. 
  \end{enumerate}  
  
\subsubsection{Style}
We have not found an example of Sizewell B specification in the literature, so this section exceptionally includes no illustration. 
The IL text is fed into MALPAS, which constructs a directed graph and associated semantics for the program.
For compliance with the MALPAS Compliance Analyzer, the IL specification is written as pre- and postconditions for each procedure, plus optional code assertions.
The analyzer determines whether the code meets the specification.

The first part of the Compliance Analysis process is the construction of the mathematical specification from the natural language SDR and SDS. 
This work ensures both that the interpretation of the existing specifications is correct and that the important functionality and properties are modelled in the mathematical specifications.

\subsubsection{Software characteristics}
Sizewell B has required the major effort by Westinghouse, NE, the NNC and others to ensure that complex protective functions are implemented accurately in code and that there is a high degree of confidence in this. The quantity of software involved in the PPS is large, but not too large to have been verified meticulously and in its entirety. 
The PPS software is verified with very demanding reliability requirements (imposed by UK Nuclear Installations Inspectorate (NII)),  that attest to its suitability for reactor protection.
About 2000 comments have been raised during the verification process (One for 30 lines of code). 
40\% of them induced minor specification changes or checks, but no major issue was detected. We found no  representation of this semantics in the literature.

\subsubsection{Project characteristics}
The team grew from 15 persons at the beginning of year 1992, to 95 at the end of the project. 
It took four and a half years to complete the project; The Compliance Analysis was the most important part of the MALPAS analysis of the PPS software and also involved the majority of the effort.

\subsubsection{Lessons}
The analysis does not prove the safety of the software, but does prove formally that it meets its specifications.
It also increased the integrity of the software (through code and documentation modifications that have resulted from detected anomalies) and confidence in its correctness.  
Five main measures were adopted to ensure its accuracy, consistency and reproducibility: (1) all work was conducted under a defined quality system and in accordance with the requirements of ISO9001\footnote{\url{https://www.iso.org/standard/16533.html}}; (2) a detailed ``standards and procedures'' document defined very precisely how the analysis was to be conducted; (3) full recording of all aspects of the analysis was ensured; (4) peer reviews of all analytical work were carried out; (5) strict configuration management of all analytical documents and results was applied, both in hard copy and magnetic media.

The Malpas analysis work conducted in the scope of this project has shown the feasibility of conducting a rigorous retrospective analysis of a large software system.
Even if the costs are high in absolute terms, they are low in relation with any potential software malfunction, for example.
Conducting rigorous static analysis, including Compliance Analysis, has been then considered to be essential for software controlling systems with the level of criticality and potential failure consequences similar to that of the Sizewell `B' PPS.



\iftoggle{inBook}{
\subsection{Ironclad Apps} \label{ironcladapps}

\begin{table*}[htbp]
 \scriptsize
  \centering
  {\renewcommand{\arraystretch}{1.3}
        \begin{tabular}{|p{40pt}|p{45pt}|p{55pt}|p{55pt}|p{45pt}|p{43pt}|p{40pt}|}
    \hline
     Nature & {Properties} & {Programming language(s)} & {Programming style} & {Verification framework} &{ Specification language(s)} & {Number of LOC} \\
    \hline
    Data transmitter  & Functional correctness, security & Dafny & Imperative
 & Hoare logic & Dafny annotation
 & 85286 (43720 SLOC and 41566 LOC)
    \\    \hline \hline
      {Spec/Imp}  & {Proof Engine} & {Team Background} & {Institution(s)} & {Development period} & {Effort}  & {Size of user community} \\
    \hline
    0.5 &  Boogie, SymDiff & Researchers & University of Pennsylvania and Cornell University  & 2014 & 3 
  & deployed in Trusted Platform Module
    \\ \hline
    \end{tabular}%
    }
  \label{ironclad_id}%
\end{table*}%

\subsubsection{Scope}
Ironclad Apps are dedicated to securely transmitting data to a remote machine so that every instruction executed on that machine is conformed with a formal abstract specification of the app's behavior. 

\subsubsection{Components}
Ironclad Apps are verifiably end-to-end secure. The verification involves all code that executes on the servers, including the app, the OS, libraries, and drivers. The verification is conducted at both source code level and the assembly level. Finally, the verification demonstrates remote equivalence: the app’s implementation is indistinguishable from the app’s high-level specifications.

\subsubsection{Verified properties}
The correctness and security of Ironclad Apps were proved, using Boogie and SymDiff (a Boogie-based experimental tool used to prove noninterference properties).

\subsubsection{Project context}
Nowadays, people are worried about the security of their personal data when they ask for web services. They have vague legal guarantees provided by the service’s privacy policy and hope the owner will follow industry best practices. In addition, a vulnerable OS, library, or application may harm the service providers’ best intentions.

\subsubsection{Decisions}
The design decisions are devoted to performing low-level, full-system verification with model effort. Unlike tools used in previous efforts, the authors use Dafny because it supports automated verification via the Z3 SMT solve. In order to enable the verification of the entire system at the assembly level, the authors build automated tools to translate verified Dafny code to BoogieX86. Thus, any bugs in Dafny or in the compiler can be caught at this stage.

\subsubsection{Tool stack}
Three tools were used to develop and verify the system, including Dafny, translator (built by authors), Boogie verifier, and SymDiff. First of all, the specifications and code were written in Dafny. Since Dafny supports automated verification via the Z3 SMT solver, the code in Dafny was proved to be correct. To verify the entire system from the assembly level, an automated tool was developed to translate the Dafny code to BoogieX86. Boogie verifier was then employed to conduct assembly-level verification. The SymDiff was also used to prove noninterference properties, serving as a complement to the system’s security.

\subsubsection{Style}
Ironclad Apps are verified by Floyd\_Hoare reasoning \cite{floyd1993assigning}. The specifications are embedded into the Dafny code. Here is an example illustrated as follows:
\begin{lstlisting}
method Main() {
  var count := 0;
  while(true) 
  invariant even(count) {
    count := count + 2;
    WriteOutput(count);
  } } 
\end{lstlisting}
The loop invariant \emph{even (count)} checks whether variable \emph{count} is even in each iteration of the loop.

\subsubsection{Software characteristics}

This project is characterised by several aspects: (1) The verification does not assume that any piece of server software is correct. Thus, it covers all code, including the app, the OS, libraries, and drivers. (2) The verification assumes that the hardware where the assembly code is executed is correct. (3) Property Remote equivalence should be demonstrated.
The outcome shows all the apps are proved to meet their specifications. The total spec size for the apps is 3,546 lines, which meets the goal of a small computing base.

\subsubsection{Project characteristics}
The verification project effort took 3 person-years.

\subsubsection{Lessons}
Several assumptions have been made. Hardware is supposed to be correct, including the CPU, memory, and chipset. It is assumed that the attackers do not mount physical attacks, such as electrically probing the memory bus. The platform where the apps are deployed is assumed to be trusted. The verification tools used in this project are supposed to be correct. In addition, the hardware is not modeled in enough detail to prove the absence of covert or side channels that may exist due to timing or cache effects. 

Potential future work includes proving liveness of the system, and connecting the guarantees to even higher-level cryptographic protocol correctness proofs.

Lessons are principally learned from the experiences of using modern verification tools in a large-scale systems project. 
\begin{enumerate}
\item Verification automation varies by theory
Automated theorem provers like Z3 support a variety of theories: arithmetic, functions, arrays etc. But Z3’s theory of nonlinear arithmetic is found to be slow and unstable. Small code changes often caused unpredictable verification failures.
\item  Verification needs some manual control
Verification projects often avoid automated tools because such tools could be unstable and/or too slow to scale to large, complex systems. Modular verification could be an efficient alternative solution. More specifically, opaque functions can be a good choice. For example, Z3 may unwrap function definitions too aggressively, leading to timeouts for large code. To avoid this, Dafny is modified to enable programmers to designate a function as opaque. So the verifier will ignore the bode, except in places where the programmer explicitly indicates. The mechanism used in verifiers may lead to instability. For instance, as for a complicated method that includes a nonlinear expression, Z3 has many options for applicable heuristics, leading to instability. Thus, we can choose to prove simple, fundamental lemmas. Then we use those fundamental lemmas to prove it.
\item  Existing tools make simple specs difficult
Dafny’s verifier requires that whenever one function invokes another, the caller meets the callee’s pre-conditions. So, the extra preconditions should be added to the caller’s preconditions. If the types of variables in the caller and callee are different, it will make the specs more complicated. It is because a corresponding proof has to be embedded in the body of the caller.
\item  Systems often use bounded integer types
Dafny only supports integer types int and nat, both representing unbounded-size values. However, almost all of our code concerns bounded-size integers such as bits, bytes, and 32-bit words. This results in many more annotations and proofs than we would have thought.
\item Libraries should start with generality
Conventional software development principle is to start with simple, specific code and generalize only as needed to avoid writing code paths that are not exercised or tested. This suggestion seems to be inefficient. On the contrary, it is often easier to write, prove, and use a more-general statement than the specific subset. 
\item  Spec reviews are productive
Independent spec reviews help us find bugs. For example, three bugs were detected in the segmentation spec. Similarly, two bugs in the SHA-1 spec were detected as well during spec reviewing.
\item  High-level tools have bugs
The idea is to verify the low-level code that will actually run. It seems that bugs could be more easily hidden in the high-level code. For example, the Dafny-to-C\# compiler suppressed calls to methods with only ghost return values. Another example is that the Dafny CC failed to report incorrect code when there was a complex bug in the translation of while loops.

\end{enumerate}


\subsection{Quark}

\begin{table*}[htbp]
 \scriptsize
  \centering
  {\renewcommand{\arraystretch}{1.3}
        \begin{tabular}{|p{40pt}|p{45pt}|p{55pt}|p{55pt}|p{45pt}|p{43pt}|p{40pt}|}
    \hline
     Nature & {Properties} & {Programming language(s)} & {Programming style} & {Verification framework} &{ Specification language(s)} & {Number of LOC} \\
    \hline
    Browser  & security & Coq, OCaml & Functional 
 & Lambda-calculus 
Calculus of constructions & Coq;
Crash Hoare Logic (CHL)
 & Kernel Code 859
Kernel Security Properties 142 
Kernel Proofs 4,383 
Kernel Primitive Specification 143 
Kernel Primitives (Ocaml/C) 538 
Tab Process (Python) 29 
Input Process (Python) 60 
Output Process (Python) 83 
Cookie Process (Python) 135 
Python Message Lib (Python) 334 
WebKit Modifications (C) 250 
WebKit (C/C++) 969109
    \\    \hline \hline
      {Spec/Imp}  & {Proof Engine} & {Team Background} & {Institution(s)} & {Development period} & {Effort}  & {Size of user community} \\
    \hline
     For kernel primitives: 0.2 &  Coq, Ynot & Researchers & University of California, San Diego  & 2012 (6 months) & 0.83
  & --
    \\ \hline
    \end{tabular}%
    }
  \label{quark_id}%
\end{table*}%

\subsubsection{Scope}
Quark is a Web browser with formally verified integrity properties.
Quark is structured similarly to Google Chrome. It consists of a small browser kernel which mediates access to system resources for all other browser components. 

\subsubsection{Components}
Quark is written inside a theorem prover. It exploits formal verification to verify security properties. To achieve this goal, all browser components run in sandboxes which only allow the component to communicate with the kernel.
In this way, Quark is able to make strong guarantees about a million lines of code (e.g., the renderer, JavaScript implementation, JPEG decoders etc.) while only using a proof assistant to reason about a few hundred lines of code of  the Quark kernel. The underlying system is protected from Quark's untrusted components (everything other than the kernel).

\subsubsection{Verified properties}
The following security properties are verified: tab non-interference, cookie integrity and confidentiality, address bar integrity. 

\subsubsection{Project context }
Web browsers mediate access to valuable private data in domains ranging from health care to banking. Despite this critical role, attackers routinely exploit browser vulnerabilities to exfiltrate private data and take over the underlying system.

\subsubsection{Decisions}
QUARK factors a browser into distinct components which run in separate processes. Fig. \ref{fig:quark1} shows how QUARK isolates complex and vulnerability-ridden components in sandboxes, forcing them to access all sensitive resources through QUARK (arrows indicate information flow).

\begin{figure}[htbp]
  \centering
  \includegraphics[width=4cm]{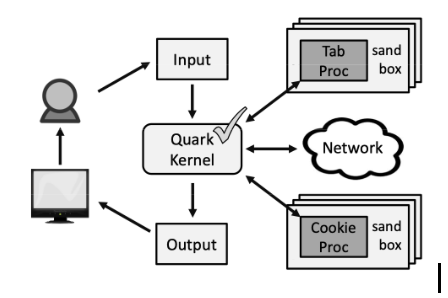}\\
  \caption{QUARK Architecture}
  \label{fig:quark1}
\end{figure}

Security properties are  guaranteed by formally verifying the QUARK Kernel in the Coq proof assistant.

\subsubsection{Tool stack}
\begin{figure}[htbp]
  \centering
  \includegraphics[width=4cm]{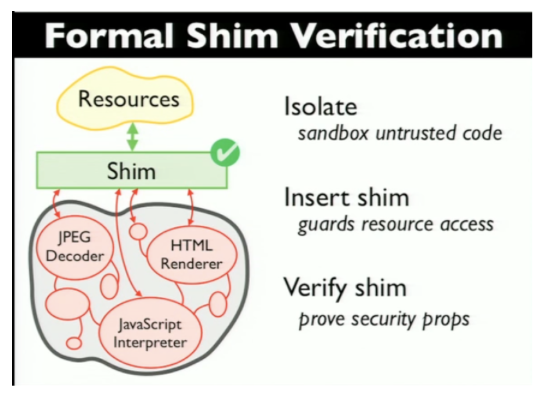}\\
  \caption{Formal Shim verification}
  \label{fig:quark2}
\end{figure}

What the authors of this approach call formal shim verification addresses the formal verification of a large system by reducing the amount of code that must be considered: instead of formalizing and reasoning about all system components, the verification is made at the kernel level. All components communicate through a small shim which ensures the components are restricted to only exhibit allowed behaviors. Formal shim verification only requires one to reason about the shim, proving that  the actual browser implementation is correct. So, security properties have been established by verifying a small amount of code in Coq (about 1000 lines).

First, the correct behavior of the QUARK’s kernel is specified in terms of traces. Second, the proof is made that the kernel satisfies this specification using Ynot’s monads (Ynot library \cite{ynot2008} enables reasoning about imperative programming features, e.g., impure functions like fopen, which are otherwise unavailable in Coq’s pure implementation language). Finally, the proof is made that the kernel specification implies the target security properties. 

\subsubsection{Style}
The Coq code below is an extract of the kernel specification. $step\_correct$ is a predicate over triples containing a past trace, a request trace, and a response trace; it holds when the response is valid for the given request in the context of the past trace. $tcorrect$ defines a correct trace for the kernel to be a sequence of correct steps, the concatenation of valid request and response trace fragments.

\begin{lstlisting}
Inductive tcorrect : Trace -> Prop :=
  | tcorrect_nil: 
tcorrect nil
  | tcorrect_step: forall tr req rsp,
 tcorrect tr ->
      	step_correct tr req rsp ->
     	 tcorrect (rsp ++ req ++ tr)

Inductive step_correct :
  Trace -> Trace -> Trace -> Prop :=
  | step_correct_add_tab: forall tr t,
     step_correct tr
        (MkTab t :: Read stdin "+" :: nil)
        (WroteMsg t Render)
 | step_correct_socket_true: forall tr t host port,
      is_safe_soc host (domain_suffix t) = true ->
      step_correct tr
       (ReadMsg t (GetSoc host port))
       (SentSoc t host port)
  | step_correct_socket_false: forall tr t host port,
     is_safe_soc host (domain_suffix t) <> true ->
      step_correct tr
       (ReadMsg t (GetSoc host port) ++ tr)
        (WroteMsg t Error ++ tr)
\end{lstlisting}

\subsubsection{Software characteristics}
QUARK furnishes a framework to easily experiment with additional web policies without re-engineering an entire browser or formalizing all the details of its behavior from scratch.
The  approach’s performance impact was evaluated by comparing QUARK’s load times for the top 10 Alexa Web sites to WebKit’s load times. QUARK loads these websites with 24\% overhead. QUARK’s overhead is due to factoring the browser into distinct components which run in separate processes and explicitly communicate through a formally verified browser kernel. 

\subsubsection{Project characteristics}
Quarck's development took 6 person-months. 

\subsubsection{Lessons}
The Quark team assumes that an attacker can compromise any QUARK component which is exposed to content from the Internet, except for the kernel which is formally verified.
Besides, it assumes OCaml's correctness (including not running out of memory) in Quark's trusted computing base. Nevertheless,  as Quark's guarantees are safety properties, an out of memory error would crash the kernel, but would not lead to violating any of Quark's guarantees.

The browser is structured so that all target security properties can be ensured by a very small browser kernel. Then reasoning only about that single component can be possible. Leveraging this technique makes guarantees about the behavior of a million of lines of code while reasoning about only a few hundred, in the mechanical proof assistant Coq. (``What you run  is what you proof'' says Quarck’s authors.)
The trusted computing base (Coq’s core calculus and type checker, the formal statement of the security properties, several primitives used in Ynot, several primitives unique to QUARK, the Ocaml compiler and runtime, the underlying Operating System kernel, the used chroot sandbox) consists of all system components assumed to be correct. A bug in the trusted computing base could invalidate the security guarantees.
}{}

\subsection{CoCon}
\label{Cocon}

\iftoggle{inBook}{

\begin{table*}[htbp]
 \scriptsize
  \centering
  {\renewcommand{\arraystretch}{1.3}
        \begin{tabular}{|p{40pt}|p{45pt}|p{55pt}|p{55pt}|p{45pt}|p{43pt}|p{40pt}|}
    \hline
     Nature & {Properties} & {Programming language(s)} & {Programming style} & {Verification framework} &{ Specification language(s)} & {Number of LOC} \\
    \hline
    CMS  & security  & Scala & Functional 
 & Theorem Proving  + Unwinding Proof Method(?) & Isabelle
 & +3000 Scala lines for CoCon’s kernel
-2000 Isabelle lines for CoCon’s specification 
~ 10000 Isabelle lines for CoCon’s verification
~200 Lines dedicated to various safety properties to support the confidentiality proofs
    \\    \hline \hline
      {Spec/Imp}  & {Proof Engine} & {Team Background} & {Institution(s)} & {Development period} & {Effort}  & {Size of user community} \\
    \hline
    2/3 & Isabelle / HOL BD-Security
 & Researchers & Middlesex University /  Global NoticeBoard / German Research Center for Artificial Intelligence (DFKI) Bremen & 2013-2016 2018-2020 & 0.25 (verification)
  & 180
    \\ \hline
    \end{tabular}%
    }
  \label{cocon_id}%
\end{table*}%

}{}

\subsubsection{Scope}
CoCon is a conference management system, providing a web interface. 

\subsubsection{Components}
Cocon involves five types of stakeholders, each with a specific role. 
It supports 45 operations, some open to some stakeholders only. They are divided into 5 categories: creation, update, nondestructive update, reading, listing. 
A conference goes through 7 phases: No-Phase, Setup, Submission, Bidding, Reviewing, Discussion, Notification, Closing.

\subsubsection{Verified properties}
Confidentiality properties of CoCon are verified through a framework, called verification infrastructure for Bounded-Deducibility (BD) security. It is  a general framework for the verification of rich information flow properties of input/output automata. BD security is parameterized by declassification bounds and triggers (in a context of a fixed topology of bounds and triggers). 
Informally, BD Security can be summarized as follows: If trigger T never holds, then attacker Obs can learn nothing about secrets Sec beyond B.
CoCon’s verification also involves a form of traceback properties and several safety properties proofs. All of them are described by Isabelle scripts (A zip archive with the Isabelle formalization is available\footnote{\url{https://www.andreipopescu.uk/papers/Formal_Scripts_CoCon.zip} (accessed in March 2023)}).

\subsubsection{Project context}
A Conference Management System (CMS) presents confidentiality, integrity and security problems. In addition, a CMS must guarantee the selective availability of information.
CoCon addresses information flow security problems of realistic web-based systems by interactive theorem proving—using a proof assistant, Isabelle/HOL. 

\vspace{-0.1cm}
\subsubsection{Decisions}
The architecture of CoCon follows the paradigm of security by design \cite{deogun2019secure} (the guarantees do not apply to the application layer).

\vspace{-0.1cm}
\subsubsection{Tool stack}
Cocon relies on a three layers stack as follows: 
\begin{enumerate}[label=\arabic*)]
\item Formalization and verification of  the kernel of the system in the Isabelle proof assistant.
\item Automatic translation of the formalization obtained in step 1) into Scala.
\item Wrapping the translated program (from 2) in a web application, using trusted components.
\end{enumerate}
\vspace{-0.1cm}
\subsubsection{Style}
To ensure that all these properties are handled, CoCon’s kernel is formalized in Isabelle as an executable input/output automaton (extracted from the Isabelle specification to a Scala program using Isabelle’s code generator).
A state of the CoCon’s I/O Automaton stores the lists of registered conference IDs, user IDs and paper IDs. 
For each ID, the state stores actual conference, user or paper information. 
For user IDs, the state also stores (hashed) passwords. 
In the context of a conference, each user is assigned one or more of the roles described by the following Isabelle datatype:
\e{datatypeRole=Chair | PC| AutPaperID | RevPaperIDNat}.

The initial state of the system, $istate \in State$ (Fig. \ref{fig:cocon2}), is the one with no conference and a single user, Andrey Voronkov, as SuperChair (the superChair is the first user of the system whose  role is to approve new conference requests).
\begin{figure}[ht]
\vspace {-0.45cm}
    \begin{subfigure}{.53\textwidth}
\begin{lstlisting}
                           istate =
                               confIDs = []                        
                               userIDs = ["voronkov"]              
                               user = ($\lambda$ uid.emptyUser)    
                               paperIDs = ($\lambda$ cid.[])   
                               pref = ($\lambda$ uid pid.NoPref)   
                               news = ($\lambda$ cid.[])           
\end{lstlisting}    
    \end{subfigure}
\hfill
     \begin{subfigure}{.45\textwidth}
\begin{lstlisting}

conf = ($\lambda$ cid.emptyConf)
pass = ($\lambda$ uid.emptyPass)
roles = ($\lambda$ cid uid.[])
paper = ($\lambda$ pid.emptyPaper)
voronkov = "voronkov"
phase = ($\lambda$ cid.noPh)
\end{lstlisting}
     \end{subfigure}
\vspace{-0.3cm}
\caption{Specification of the initial state}
\label{fig:cocon2}
\vspace{-0.6cm}
\end{figure}

\subsubsection{Software characteristics}
CoCon has been used so far to manage the submission and review process of two international conferences, TABLES 2015 and ITP 2016, hosting about 70 and 110 users respectively, consisting of PC members and authors.

\vspace{-0.1cm}
\subsubsection{Project characteristics}
Verification took 3 person-months, including development of the reusable proof infrastructure and automation. Developing the web application around the kernel took longer, much of it done incrementally with the system already up and running. Two master students worked on it (6 person-months); one of them  completely changed the implementation.

\vspace{-0.1cm}
\subsubsection{Lessons}
The second application of CoCon evealed a bug that was difficult to catch. This bug made privacy violations possible, which is exactly what CoCon is trying to avoid. In reality, the bug was caused by the web interface and had nothing to do with CoCon's verification. It did highlight the need to verify the API layer and all the trusted components.

The formal guarantees provided in Isabelle must be combined with trust steps applied to the entire system. 
The verification targets only the system’s implementation logic — attacks such as browser-level forging are out of its reach, but are orthogonal issues that should be handled.

\iftoggle{inBook}{

\subsection{CoSMed}

\begin{table*}[htbp]
 \scriptsize
  \centering
  {\renewcommand{\arraystretch}{1.3}
        \begin{tabular}{|p{40pt}|p{45pt}|p{55pt}|p{55pt}|p{45pt}|p{43pt}|p{40pt}|}
    \hline
     Nature & {Properties} & {Programming language(s)} & {Programming style} & {Verification framework} &{ Specification language(s)} & {Number of LOC} \\
    \hline
    social media platform  & security  & Scala & Functional 
 & Theorem Proving  + Unwinding Proof Method & Isabelle
 & +700 l
 Scala for kernel
~ 6500 l 
 Isabelle  for specification and verification 
~200 l for various safety properties to support the confidentiality proofs
    \\    \hline \hline
      {Spec/Imp}  & {Proof Engine} & {Team Background} & {Institution(s)} & {Development period} & {Effort}  & {Size of user community} \\
    \hline
    1 (system’s kernel)
0.33 (web-app) 
 & Isabelle / HOL BD-Security
 & Researchers & Middlesex University /  Global NoticeBoard / German Research Center for Artificial Intelligence (DFKI) Bremen & 2016-2017 & 0.33 
  & 0
    \\ \hline
    \end{tabular}%
    }
  \label{cosmed_id}%
\end{table*}%

\subsubsection{Scope}
CoSMed \cite{Cosmed1} is a social media platform with verified document confidentiality. CoSMed allows users to register and post information. Access to this information is restricted based on friendship relationships established between users.

\subsubsection{Components}
An I/O automaton represents the formally verified CoSMed’s kernel. This kernel is automatically translated to isomorphic Scala code using Isabelle’s code generator.

\subsubsection{Verified properties}
CoSMed offers verified confidentiality guarantees. To support the verification that there are no unintended flows of information to attackers who can observe and influence certain aspects of the system execution, what is specified is
what the capabilities of the attacker are, 
which information is (potentially) confidential, and
which flows are allowed.

The first point is captured by a function O taking a trace and returning the observable part of that trace. Similarly, the second point is captured by a function S taking a trace and returning the sequence of (potential) secrets occurring in that trace. For the third point, we add a parameter B, which is a binary relation on sequences of secrets. It specifies a lower bound on the uncertainty of the observer about the secrets, in other words, an upper bound on these secrets’ declassification. In this context, BD security states that  O cannot learn anything about S beyond B.

\subsubsection{Project context}
A. Popescu’s team addresses information flow security problems of realistic web-based systems by interactive theorem proving (using the proof assistant Isabelle/HOL). A security notion was introduced in CoCon (cf. the description of CoCon \ref{Cocon}). It supports  a very fine-grained specification of what an attacker can observe about the system, what information is to be kept confidential and in which situations. In CoSMed, the ``documents'' of interest are friendship requests, friendship statuses, and posts by the users. The latter consist of title, text, and an optional image. In CoSMed, like in CoCon, the observers are the users of the system, and it is necessary to verify that, by interacting with the system, the observers cannot learn more about confidential information than what has been specified. 
The roles include admin, owner, and friend. Modeling the restrictions on CoSMed’s information flow poses particular challenges due to the dynamic variations of roles. (For example, A user U1 learns nothing about the friend-only posts posted by a user U2 unless U1 becomes a friend of U2. However, this property is too weak—given that U1 may be ``friended'' and ``unfriended'' by U2 multiple times. A stronger confidentiality property would be: U1 learns nothing about U2’s friend-only posts beyond the updates performed while U1 and U2 were friends).

\subsubsection{Decisions}
Like CoCon, CoSMed is based on a three layers design.

\subsubsection{Tool stack}
Architecturally, CoSMed is an I/O automaton formalized in Isabelle, exported as Scala code, and wrapped in a web application (using Scalatra).

\subsubsection{Style}
The system can be viewed as an I/O automaton, having a state and offering some actions through which the user can affect the state and retrieve outputs. 
The state stores information about users, posts, and the relationships between them:
\begin{itemize}
    \item user information: pending new-user requests, the current user IDs and the associated user info, the system’s administrator, the user passwords;
    \item post information: the current post IDs and the posts associated with them, including content and visibility information;
    \item post-user relationships: the post owners;
    \item user-user relationships: the pending friend requests and the friend relationships.
\end{itemize}
The state is represented as an Isabelle record: 
\begin{center}
	  \includegraphics[width=7cm]{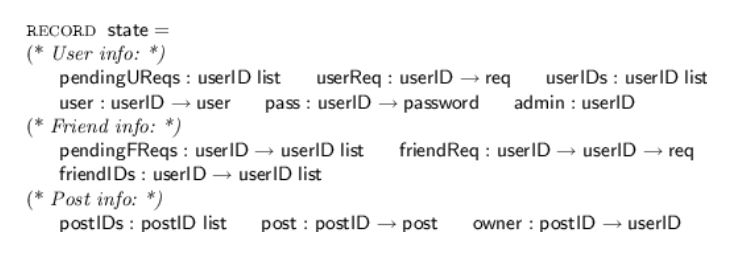}\\
\end{center}
Each pending request (Ureqs for user or Freqs for friend relationship) stores a request information (of type req), which contains a message of the requester for the recipient (the system admin or a given user). The type user contains user names and information. The type post of posts contains tuples (\textit{title}, \textit{text}, \textit{img}, \textit{vis}), where the title and the text are essentially strings, \textit{img} is an (optional) image file, and \textit{vis} $\in$ {FriendV,PublicV} is a visibility status that can be assigned to posts: 
\begin{itemize}
    \item FriendV means visibility to friends only, whereas 
    \item PublicV means visibility to all users
\end{itemize}

The initial state of the system is empty.
Users can interact with the system via six categories of actions: start-up, creation, deletion, update, reading, and listing. These kinds of actions are wrapped in a single datatype through specific constructors:
\begin{center}
	  \includegraphics[width=7cm]{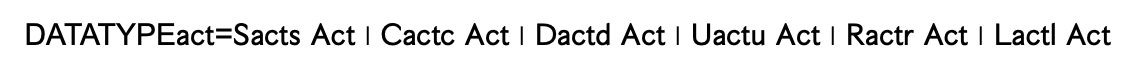}\\
\end{center}

The actions take varying numbers of parameters, indicating the user involved and optionally some data to be loaded into the system. Each action’s behavior is specified by two functions:
An effect function, performing the action, possibly changing the state and returning an output
An enabledness predicate (marked by the prefix ``e''), checking the conditions under which the action should be allowed
When a user issues an action, the system first checks if it is enabled, in which case its effect function is applied and the output is returned to the user. If it is not enabled, then an error message is returned and the state remains unchanged.
The start-up action, 
\begin{center}
	  \includegraphics[width=4cm]{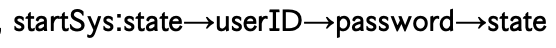}\\
\end{center}
initializes the system with a first user, who becomes the admin:
\begin{center}
	  \includegraphics[width=7cm]{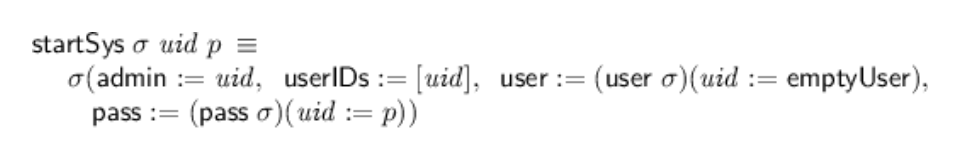}\\
\end{center}

The I/O automaton formalized by the initial state \textit{istate:state}
and the step function \textit{step:state→act→out×state}, represents the formally verified CoSMed’s kernel. The kernel is automatically translated to isomorphic Scala code using Isabelle’s code generator.

CoSMed verification is based on BD security, the framework for information-flow security formalized in Isabelle, developed in the CoCon project (see the description of CoCon above \ref{Cocon})). 
CoSMed’s specific confidentiality needs require a dynamic topology of declassification bounds and triggers, since users’ roles can dynamically vary. 

This time, BD security is parameterized by the following data:
\begin{itemize}
    \item an I/O automaton (state, act, out, istate, step)
    \item a security model, consisting of:
    \begin{itemize}
        \item a secrecy infrastructure (secret,isSec, getSec)
        \item an observation infrastructure (obs, isObs, getObs)
        \item a declassification bound B.
    \end{itemize}
\end{itemize}
Where: 
\begin{itemize}
        \item isSec:trans→bool, filters the transitions that produce secret
        \item getSec:trans→secret, produces a secret out of a transition
        \item isObs:trans→bool, filters the transitions that produce observations
        \item getObs:trans→obs, produces an observation out of a transition
\end{itemize}

The functions O and S are defined by: \begin{center}
	  \includegraphics[width=5cm]{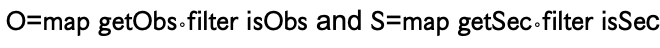}\\
\end{center}

O uses a filter to select the transitions in a trace that are (partially) observable according to 
isObs, and then maps this sequence of transitions to the sequence of their induced observations, via getObs. Similarly, S produces sequences of secrets by filtering with isSec and mapping with getSec.

For example, the following property of post text confidentiality has to be verified: 
A user can learn nothing about the updates to a post content beyond those updates that are performed while one of the following holds: either that user is the post’s owner, or he is a friend of the owner, or the post is marked as public.

Using BO,  an auxiliary predicate to cover the case when the window is open, the bound for post text confidentiality is: 
\begin{center}
	  \includegraphics[width=7cm]{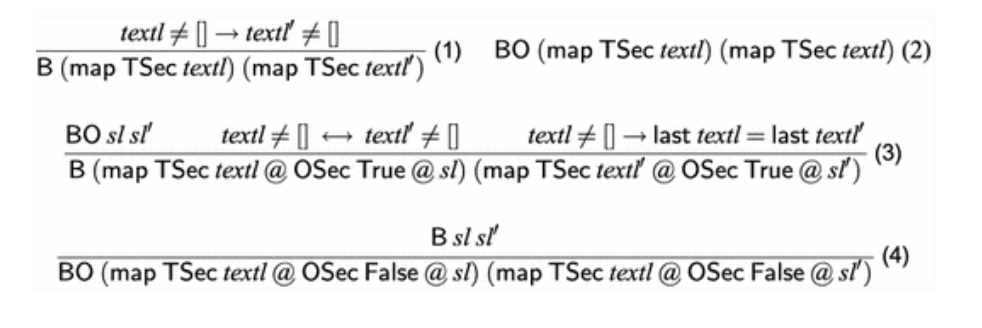}\\
\end{center}

\subsubsection{Software characteristics}
CoSMed has been developed to fulfill the functionality and security needs of a charity organization. The current version is a prototype not yet deployed for charity usage. Both the formalization and the running website are publicly available in the CoSMed website\footnote{\url{https://cosmed.globalnoticeboard.com}(NA!!)}.

\subsubsection{Project characteristics}
The verification took 3 person-months. The web application layer is more basic than for CoCon, so it took about one person-month.
CoSMeDis \cite{Cosmedis} is a distributed multi-node extension of CoSMed. Its verification involved (1) proving a compositionality theorem, (2) slightly strengthening the previously verified properties of single nodes (CoSMed) to fit the hypotheses of the theorem, and finally (3) applying the theorem to lift the single-node guarantees to the entire multi-node system. This took 4 person-months.

\subsubsection{Lessons} 
Although CoSMed’s architecture involves trusted code, the authors believe that the confidentiality guarantees of the kernel also apply to the overall system. They argue that the Scalatra API essentially forwards requests back and forth between the kernel and the outside world and that the web application operates by calling combinations of primitive API operations, without storing any data itself. User authentication, however, is also part of this unverified code. Complementing the secure kernel with a verification that ``nothing goes wrong'' in the outer layer would give stronger guarantees.

CoSMed/CoSMeDis do not have user communities; they are research prototypes of verified running systems that were used to inspire a secure design of a real-world platform, Global Noticeboard.


\subsection{FloVer}
\begin{table*}[htbp]
 \scriptsize
  \centering
  {\renewcommand{\arraystretch}{1.3}
        \begin{tabular}{|p{40pt}|p{45pt}|p{55pt}|p{55pt}|p{45pt}|p{43pt}|p{40pt}|}
    \hline
     Nature & {Properties} & {Programming language(s)} & {Programming style} & {Verification framework} &{ Specification language(s)} & {Number of LOC} \\
    \hline
    certificate checker & roundoff errors 
soundness with respect to standard semantics
  & Coq, HOL4 & Functional 
 & higher-order type theory (calculus of constructions) & Coq, HOL4
 & $<1000$
Connecting the semantics to IEEE-754 is around 2000 lines of specifications  and proofs
    \\    \hline \hline
      {Spec/Imp}  & {Proof Engine} & {Team Background} & {Institution(s)} & {Development period} & {Effort}  & {Size of user community} \\
    \hline
    10 
 & Coq, HOL4
 & Researchers & MPI-SWS, ENS Lyon, Chalmers University of Technology, University of Cambridge & 2016-2018 still evolves & 2 person-years   & 1
    \\ \hline
    \end{tabular}%
    }
  \label{flover_id}%
\end{table*}%

\subsubsection{Scope}
FloVer \footnote{\url{https://gitlab.mpi-sws.org/AVA/FloVer}} is an automated certificate checker for finite-precision roundoff error bounds.

\subsubsection{Components}
FloVer has been implemented and proved correct in both Coq and HOL4. FloVer provides a forward dataflow static analysis to  support both floating-point and fixed-point arithmetics. 
FloVer is built modularly to allow reusability and extensions, and supports dataflow analysis with both interval and affine arithmetic abstract domains.
FloVer has been implemented and verified with two theorem provers, to be able to connect to projects in both provers and thereby be widely applicable. 

\subsubsection{Verified properties}
The certificate checker is developed modularly, and each module is proven sound with respect to its semantics, e.g. for the real-number range inference, the proof has been made that the inferred ranges are upper bounds to the concrete values of any run of the analyzed program. To strengthen their results, the researchers connect the floating-point version of their semantics to standardized IEEE-754 semantics.

\subsubsection{Project context}
Heiko Becker and his team \cite{becker2018verified} have developed a verified certificate checker of the correctness of finite-precision roundoff error bounds encoded in a certificate.
A verified binary of the certificate checker is produced using the CakeML toolchain\footnote{\url{https://cakeml.org/}} (depicted section \ref{system:CakeML}).
To validate an encoded roundoff error analysis results, FloVer runs a set of boolean functions on the encoding and the encoded result is correct if all of these functions return true.

\subsubsection{Decisions}
The researchers used big-step-style functional semantics, and designed the language  similar to an imperative language as they split between commands (let-bindings) and expressions. Further, they have made their checker modular, such that it  step-by-step validates a certificate using boolean functions. Each function comes with a once and for all proven soundness proof. The overall soundness proof is a straightforward combination of those. With the overall soundness theorem, correctness of  an encoded roundoff error analysis follows from the certificate checker returning true when run on this encoding.

\subsubsection{Tool stack}
To perform the proofs and to develop the certificate checking functions, the HOL4 and Coq theorem provers were used. The Daisy static analyser \cite{darulova2018daisy} was used as an example (unverified) of a static analysis tool to generate certificates which were analyzed in an experimental evaluation.
The HOL4 implementation is compiled to machine-code using the proof-producing synthesis procedure of the CakeML compiler.

\subsubsection{Style}
The following extract encodes a certificate for a single addition of a constant to a variable.
Most of the file (up until line 24) encodes the AST and the analysis result, and the theorem in lines 26-29 encodes the actual certificate checking. The preamble from lines 1-6 and line 32 are just HOL4 boilerplate. 

\begin{lstlisting}
1  open  FloverTactics AbbrevsTheory MachineTypeTheory CertificateCheckerTheory            FloverMapTheory;
2  open simpLib realTheory realLib RealArith;
3 
4  val _ = new_theory "certificate_AdditionSimple";
5 
6 
7  val C12_def = Define `C12:(real expr) = Const M64 ((1657)/(5))`;
8  val u0_def = Define `u0:(real expr) = Var 0`;
9  val PlusC12u0 = Define `PlusC12u0:(real expr) = Binop Plus C12 u0`;
10 val RetPlusC12u0_def = Define `RetPlusC12u0 = Ret PlusC12u0`;
11 
12 
13 val defVars_additionSimple_def = Define `
14  defVars_additionSimple : typeMap = 
15 (FloverMapTree_insert (Var 0) (M64) (FloverMapTree_empty))`;
16 
17 
18 val thePrecondition_additionSimple_def = Define ` 
19  thePreconditionadditionSimple (n:num):interval = 
20 if n = 0 then ( (-100)/(1), (100)/(1)) else (0,1)`;
21 
22 val absenv_additionSimple_def = RIGHT_CONV_RULE EVAL (Define `
23  absenv_additionSimple:analysisResult = 
24     FloverMapTree_insert PlusC12u0 (( (1157)/(5), (2157)/(5)), (7771411516990528329)/(81129638414606681695789005144064)) (FloverMapTree_insert u0 (( (-100)/(1), (100)/(1)), (25)/(2251799813685248)) (FloverMapTree_insert C12 (( (1657)/(5), (1657)/(5)), (1657)/(45035996273704960)) (FloverMapTree_empty)))`);
25
26 val _ = store_thm ("ErrorBound_additionSimple_Sound",
27 ``? Gamma. CertificateCheckerCmd RetPlusC12u0 absenv_additionSimple
28 the PreconditionadditionSimple defVars_additionSimple = SOME Gamma``,
29  flover_eval_tac \\ fs[REAL_INV_1OVER]);
30
31
32  val _ = export_theory();
\end{lstlisting}

\subsubsection{Software characteristics}
FloVer is a re-usable, modular tool that can theoretically be used to verify and infer roundoff error bounds for small numerical kernels without loops and conditionals. 
FloVer has been used in several projects, and the overall design of the tool allows simple extensions to validate the results, like adding new operators or new functions for example. 

\subsubsection{Project characteristics}
The team members estimate that human effort taken to complete the implementation and verify it is around 2 person-years. 
They have ``stress-tested" the core of the system, a first working version, by using different extensions (support for division and mixed-precision computations). Adding division amounted to extending the correctness proofs with a separate case for division but did not require altering the overall structure of the tool. Adding mixed-precision computations was a major rework of the semantics and as such required redoing proofs that deal with finite-precision arithmetic, but, proofs for e.g. real-numbered range inference only needed minor changes because of changes to the semantics, but no major breakages were involved.
In a separate project FloVer's Coq implementation was extended with SMT-based certificates and interval subdivisions \cite{bard2019formally}. This extension essentially required only extending the final soundness proof but did not require significant changes to other parts.

\subsubsection{Lessons}
According to the authors, the language could not have been split between expressions and commands, especially considering that they only have let-bindings and no loops/conditionals.
The fact that the modules simultaneously implement validation of certificate-encoded information and inference of internal information for comparison most likely complicated the proofs. Having two separate functions, one for inference and one for checking would have made FloVer both more widely applicable, and potentially easier to verify. 

The goal to get a fully extensible system seems to be not completely reached: an attempt to support mixed-precision computations required redoing (not all, but) some proofs completely. The addition of an operation like division was not fully modular either because it required addition of a new case to the original proof. 
It therefore seems that a completely modular extension of a verified system is close to impossible.
}

\section{Discussion and lessons learned} \label{discussion_lessons_learned}
The review of systems in the previous section leads, after an analysis of threats to validity (\ref{threats}), to general observations (\ref{observations}), lessons drawn by the project members themselves (\ref{teamlessons}), insights about the penetration of formal techniques (\ref{penetration}), and an overall assessment of the results (\ref{overall}).

\vspace{-0.3cm}
\subsection{Threats to validity} \label{threats}
The analysis of systems leading to this survey is subject to clear limitations:

\begin{itemize}
    \item It is not a full Systematic Literature Review, in particular because  some of the industrial systems surveyed do not appear in the ``literature'' in the sense of scholarly journals or conferences. It follows, however, the recommended review protocol for SLRs \cite{keele2007guidelines}.
    \item We may have overlooked relevant systems for lack of information. We may for example have missed systems from companies that do not think of publishing their experience.
    \item Not all application areas are represented; missing, for example, are desktop applications, malware, software for ML and AI and healthcare systems.
    \item The selected systems were from Europe or the US,  and are described in English. 
    \item A selection bias comes from the papers themselves, which usually report successful verification efforts. Note, however, that the Ironclad authors explain that the method used did not succeed in verifying security but made it possible to find a complementary method that did.
    \item The available information mostly concerned initial system development. It is well known that updates (``maintenance'') are a key part of software engineering. It has been difficult to find information on updates and whether they have been verified.
    \item Contacting people to get information not immediately clear from publications has often been hard (as project members leave the organization or no contact information is available).
\end{itemize}

\vspace{-0.4cm}
 \subsection{Projects and technology: evolution over time} \label{observations}
    
    Fig. \ref{fig:starting_year} presents the timeline of the projects, from 1989 to 2021; note the quasi-constant rate (red line) from 2005 on.
    Proof-based and auto-active approaches had replaced refinement-based ones by 2007. Both kinds continue to strive, but from 2011 on proof-based overtakes auto-active. 
    Model-checking, although not used as a single technique for verifying code, underlies some of the more complete techniques.
   Refinement-based systems are rarely used (Roissy Shuttle is one of example). 
    
        The mostly frequently used programming languages are Coq and C; 8 projects use Coq as their programming language, 6 of which 6 also use it as their specification languages. 
    The most frequently used specification languages are Coq (11 projects) and Z  (4 projects). 7 projects apply annotations in the programs to specify the verified properties. 
    Connections exist between programming and specification languages: a verification using HOL seems to be suitable for systems written in C / SML / OCaml;  SPARK code is specified in Z;  
    Isabelle is used for Scala and C.
    
    The most widely used proof engines are Isabelle, Coq, Z3, and HOL4.
    Correspondingly, the most widely used theoretical framework is Hoare logic.
     
\vspace{-0.3cm}
 \begin{figure} [htb!]
  \centering
    \begin{tikzpicture}
    	\begin{axis}[
    		xtick=data,
    		xticklabels from table={\SystemsByYear}{Project},
    		xtick distance=5,
    		xticklabel style={rotate=90,font=\scriptsize},
    		extra x ticks={4,8,...,36},
    		extra x tick labels={},
    		extra x tick style={grid=major},
    		ylabel={Starting year},
    		ytick={1989,1993,...,2021},
    		yticklabel style={/pgf/number format/1000 sep={}},
    		small,
    		xminorgrids=false,
    		xmajorgrids=false,
    		ymajorgrids=true,
    		yminorgrids=false,
    		ymin={1987},
    		ymax={2021},
    		enlarge x limits={abs={1}},
    		width=0.8\textwidth,
    		height=30ex,
    		scatter src=explicit symbolic,
    		scatter/classes={
    			PB={mark=square*,blue},
    			AA={mark=*,green!60!black},
    			RF={mark=diamond*,purple!75!black}
    		},
    		table/meta={Approach},
    		legend entries={
    			Proof-based,
    			Auto-active,
    			Refinement%
    		},
    		legend style={at={(0.95,0.1)},anchor=south east},
    		legend cell align=left,
    	]
    		\addplot [only marks, black, scatter] table [
    				col sep=tab, x={Project}, x expr= \coordindex,
    				y={Year}
    			] {\SystemsByYear};
    		\addplot [red!90!black!80!white,very thick] table [
    				col sep=tab, x={Project}, x expr= {\coordindex+4},
    				y = {create col/linear regression = {y = {Year}}}
    			] {fs_years.txt};
    	\end{axis}
    \end{tikzpicture}
    \vspace{-0.3cm}
    \caption{Verification approaches by year}
    \label{fig:starting_year}
\end{figure}
\subsection{Lessons stated by project teams} \label{teamlessons}
\noindent Some important general lessons come from comments made by project team members themselves. On the positive side, some  teams (Lockheed-Martin C130J, NATS iFACTS, EuroFighter Typhoon Aircraft FCS, Dutch Tunnel Control) reported measurable reduction of costs after introducing formal verification into the development process.
The NATS iFACTS project actually recommends, as a result, applying formal methods to all phases of the software development life cycle.

Teams also reported limitations. 22 of them had to depend on an unverified code elements, as with the 
 grep and HACL* projects which compile the resulting C code with GCC rather than the verified CompCert compiler, to get more efficient binaries.
Unverified elements can include:
\begin{itemize}
    \item Standard reusable components (in CompCert, addressing parsing, assembling and linking).
    \item Unverified features of the verification tool itself, as with code extraction in Coq, used by many projects.
    In fact, the Ironclad Apps project found an actual bug in the verification tools.
    \item Generated code (such as the output of the Verilog compiler) that requires further processing by trusted tools.
    \item Program design that makes it difficult to specify and verify properties of interest, as in HACL*.
    \item Insufficient computing resources to conduct exhaustive verification, as in SHOLIS.
    \item Trusted but unverified components, as with an API layer in CoCon.
    \item Reliance on the assumption that future users of the system will not attempt certain malicious actions, as in Ironclad Apps.
    \item Bugs in the formal specifications themselves. Manual inspection revealed 3 such bugs  in Ironclad Apps specification. 
    \item Natural language requirements, as were used in the Roissy Shuttle project.
\end{itemize}

\noindent Such recourse to unverified elements is not universal. Some projects --- including CakeML, HACL*, Verve, EuroFighter Typhoon Aircraft FCS and Ynot --- chose instead to sacrifice such functionality.

Members from 14 projects reported performance issues:
\begin{itemize}
    \item Some verified systems perform more slowly than competing non-verified systems. Since verification does not by itself affect performance, one may posit two explanations: some highly efficient algorithms or implementation ideas may raise verification challenges and hence be discarded in favor of less efficient ones; and the effort devoted to verification may mean that less manpower is available for optimizations of the implementation.
 
    \item Some derived systems (such as binaries produced by a verified compiler) may perform more slowly than traditional alternatives.
    
    \item Verification may slow down the development process; in particular, failed verification attempts take a long time to terminate or do not terminate (and have to be interrupted manually).
\end{itemize}

\noindent In the other direction, the HACL* developers (section \ref{system:hacl*}) report that the verification process helped them optimize an existing implementation of the Curve25519 algorithm and prove the optimization correct.
The optimized implementation is about 2.2\% faster than the formerly fastest one.
Optimizations often make the code less readable and thus more likely to contain defects, encouraging developers to be more conservative in optimizing good enough code.
Formal verification tools, as they prove correctness of optimizations, can make the developers more determined and confident about using such optimizations, as observed in the  HACL* project.

Reports from 7 projects (CompCert, CakeML, EiffelBase2, PikeOS, Ynot, Roissy Shuttle, Ironclad) state that their verification technology would be difficult for an average developer to practice.
The Roissy Shuttle reported significant effort invested into validating the formalization of initial natural-language requirements.
Even though the validation was performed rigorously, it was still a manual, human-driven process. 
The researchers were not 100\% sure that the formal specification correctly formalized the actual needs and, as noted above, did find incorrect elements.

\vspace{-0.3cm}
\subsection{Current use of  formal verification in industry, and obstacles to further penetration} \label{penetration}

Many of the verified systems surveyed belong to the category of system software: programs (such as compilers, operating systems and hypervisors, cryptographic and data structure libraries, database management systems) providing a platform for running other software. 
Others are application software in safety-critical areas: aeronautics, rail, nuclear plants. 
Non-safety-critical applications include a web browser and few social applications: a conferencing system, a social media platform.

In spite of the potential of today's verification technology and its successes as documented here, formal verification is still not mainstream. The key obstacle is the high entry level: each project needs one or more verification experts. 
Companies may feel that it will be easier to find many testers than a few verification-savvy developers.
    
Another concern is the rate of change of modern software systems. It is now common to practice ``continuous development'', producing  new versions once or more daily. 
In principle every such version requires re-verification. The good news, reported by several projects 
(seL4, Cocon, Amazon s2n, iFACTS, Dutch tunnel) is that re-verification costs much less than the original verification. seL4 reports that verifying a new and large feature (a new data structure for API calls) cost 
less than 10\% of the initial effort. The re-verification of small changes in iFACTS was completed in seconds, while its initial verification run took three hours. 

An alternative to re-verification is reuse of verified elements. Reviewing the projects revealed a number of cases of successful reuse, for example in OS projects (mCertiKos \ref{mcertikoslessons}, Hyper-V, PikeOS). Particularly interesting is the reuse by VeriCert of part of the verified CompCert compiler: VeriCert's compilation from C to three-address code was adapted from CompCert's compilation from C to Verilog hardware designs, and did not need to be re-verified. Verification reuse is not, however, as widespread as it could be.
It is unclear why Microsoft retired the successful and reusable VCC project tools, initially developed for the verification of Hyper-V and PikeOS.

More generally, projects using continuous development may require continuous verification. Proof-based verification favors this goal: it is possible to develop proofs and code together and keep them in sync. Tools are needed to support such incremental verification processes. 

In practice  it seems that many projects, once verified, do not continue the audit effort, do not transfer their results to other projects, and do not integrate the results into their tools.
A large percentage of the projects initially selected were abandoned immediately or shortly after completion, as these projects, although deployed, were purely for research purposes. Only a few of them, such as Roissy Shuttle and PikeOS, were deployed in public systems for constant use. There seems not to be enough follow-up and scaling-up of even successful results.








We noticed also that some of the verified projects were not accepted by the users even though they have been proven (hence better than their predecessors) mainly due to lack of compatibility of the new verified project with legacy code. An example is ExpressOS.


    
Applications with sophisticated internals but simple input-output streams, without interaction with the (unverified) external world, seem easiest to verify. Systems that do involve many external interactions tend to suffer more from the issue of frequent change. In fact we did not encounter any application with a formally verified graphical user interface, other than research prototypes \cite{declarative-guis-2018}.

Human factors are also at play. Most of the reviewed projects involved researchers for verification support. This approach may not scale well: to increase the use of verification, the industry would need proof-engineering professionals \cite{klein2014proof}, with both general SE competence and a strong formal background. Training engineers does not seem, however, to have been an obstacle in the reviewed projects. 
The iFACTS experience showed that engineers from different programming backgrounds can become familiar with verification languages and tools. 
SHOLIS conducted only two a 5-day Z course and a 10-day SPARK course, for a verification effort of 19 person-years.
\vspace{-0.3cm}
\subsection{Perspectives for formal verification in industry} \label{overall}


\subsubsection{A realistic prospect}

The preceding subsections (\ref {teamlessons} and \ref{penetration}) provide a frank and realistic analysis of the obstacles to generalizing formal verification in industry. It should not, however, lead to a pessimistic assessment, in particular because the projects surveyed include several striking success stories: projects that were closely attuned to market needs and achieved many installations or widespread use; examples include  CompCert, Hyper-V, HACL*, the Roissy shuttle and PikeOS.

These successes demonstrate that the idea of software guaranteed correct (in the precise sense analyzed in sections \ref{limitations} to \ref{expectations}) is no longer a heroic goal, but a distinct practical possibility.

Unlike testing (to any level of depth), formal verification \textit{guarantees} the absence  of bugs of specified kinds\color{black}, dependent only on the quality of the  theory and tools, not on the number of use cases or on whether all were exercised. 
Advances in verification tools have been enormous, enabling the verification of software with thousands of lines of code  \cite{Everest}.

\vspace{-0.2cm}
\subsubsection{Positive effects of formal verification attempts.}
While the prime goal of verification for most projects was to eliminate bugs, many had another incentive: achieving certification for applicable industrial standards.
For example, the verification projects for microkernels and ProvenCore 
aim at the Common Criteria EAL7 standard\footnote{\url{https://www.commoncriteriaportal.org}}, which requires that the system design should be formally verified and tested.
The Lockheed-Martin C130J and PikeOS project relied on the DO-178B \cite{johnson1998178b}
which helps determine whether the software will perform reliably in an airborne environment.
Another potential driver is the needs of product users. In the s2n project, Amazon uses formal analysis to provide  customers with concrete information about how it establishes security.

Some projects (Hyper-V, Amazon s2n, CoCon) report that formal verification revealed  design defects or subtle bugs not detected by the regular quality assurance process. In the Lockheed-Martin C130J project, some errors that formal analysis could establish immediately, such as conditional initialization errors, might only have emerged after very extensive testing. The HACL* project further suggests that formal verification helps identify and verify potential optimizations.

The entry cost for formal methods may seem prohibitive to some companies.  Cost can, however, decrease dramatically for subsequent efforts. As noted in \ref{penetration}, re-verification can be cheaper by an order of magnitude.
Verified software can be reusable and serve as a reliable component in another verified project. While we noted (\ref{penetration}) that more reuse is needed, we also saw notable successes in this area, such as the reuse of the CompCert compiler and its verification in VeriCert.


\section{Conclusion} \label{conclusion}
The survey of 32 software systems, with a detailed analysis of 11 \iftoggle{inBook}{in the shorter version \cite{shortversion} (and}{here (the rest in the extended version \cite{longversion}, with} future entries to be added to the living review  \cite{livingreview}), shows that formally verified software systems use a wide range of methods, tools, specification styles and languages.
While no single solution has achieved dominance for verification (let alone for unified verification and development), some approaches have reached a significant level of successful use, including languages such as Z and B for safety-critical systems, and Hoare-style proof techniques in various areas.

Generalizing verification faces obstacles, of which the main one, per the results of the present study, remains cost, which can be reduced in various ways; for example, training can be inexpensive, and re-verification can cost 90\% less than the original effort; in addition, formal verification considerably reduces testing costs. Even the discovery of a few bugs justifies the effort. Few people who have practiced formal verification on a real system would go back.

\vspace{-0.2cm}
\footnotesize
\subsubsection*{Acknowledgements} We thank Alexander Naumchev and Aliyu Alege for their help towards an earlier version of this work.
\color{black}

\vspace{-0.2 cm}
\bibliographystyle{splncs04}
\bibliography{biblio}

@inproceedings{malecha2010toward,
  title={Toward a verified relational database management system},
  author={Malecha, Gregory and Morrisett, Greg and Shinnar, Avraham and Wisnesky, Ryan},
  booktitle={Proceedings of the 37th annual ACM SIGPLAN-SIGACT symposium on Principles of programming languages},
  pages={237--248},
  year={2010}
}

@article{longversion,
  title={Lessons from Formally Verified Deployed Software Systems (extended version of present paper)},
  author={Huang, Li and Ebersold, Sophie and Kogtenkov, Alexander and Meyer, Bertrand and Liu, Yinglin},
  journal={arXiv},
  url = {https://arxiv.org/abs/2301.02206},
  year={2023, updated Jan. 2026},
  del_publisher={IEEE}
}

@article{shortversion,
  author = {Huang, Li and Ebersold, Sophie and Kogtenkov, Alexander and Meyer, Bertrand and Liu, Yinling},
  title = {Lessons from Formally Verified Deployed Software Systems},
  year = {2026},
  publisher = {ACM},
  hide_publisher = {Association for Computing Machinery},
  hide_address = {New York, NY, USA},
  issn = {0360-0300},
  hide_url = {https://doi.org/10.1145/3785652},
  doi = {10.1145/3785652},
  note = {Just Accepted},
  journal = {ACM Comput. Surv.},
  month = jan
}

@article{ter2024formal,
  title={Formal methods in industry},
  author={Ter Beek, Maurice H. and Chapman, Rod and Cleaveland, Rance and Garavel, Hubert and Gu, Rong and ter Horst, Ivo and Keiren, Jeroen J. A. and Lecomte, Thierry and Leuschel, Michael and Rozier, Kristin Yvonne and Sampaio, Augusto and Seceleanu, Cristina and Thomas, Martyn and Willemse, Tim A. C. and Zhang, Lijuns},
  journal={Formal Aspects of Computing},
  year={2024},
  publisher={ACM New York, NY}
}

@INPROCEEDINGS{protzenko2019signal,
  author={Protzenko, Jonathan and Beurdouche, Benjamin and Merigoux, Denis and Bhargavan, Karthikeyan},
  booktitle={IEEE Symposium on Security and Privacy}, 
  title={Formally Verified Cryptographic Web Applications in {WebAssembly}}, 
  year={2019},
  volume={},
  number={},
  pages={1256-1274}
}

@inproceedings {tao2021dice,
    author = {Zhe Tao and Aseem Rastogi and Naman Gupta and Kapil Vaswani and Aditya V. Thakur},
    title = {{DICE*}: A Formally Verified Implementation of {DICE} Measured Boot},
    booktitle = {USENIX Security Symposium},
    year = {2021},
    pages = {1091--1107},
    publisher = {USENIX Association}
}

@inproceedings{polikarpova2015fully,
  title={A fully verified container library},
  author={Polikarpova, Nadia and Tschannen, Julian and Furia, Carlo A},
  booktitle={Int. Symp. on Formal Methods},
  pages={414--434},
  year={2015},
  organization={Springer}
}

@article{rescorla2018transport,
  title={The transport layer security ({TLS}) protocol version 1.3},
  author={Rescorla, Eric and Dierks, Tim},
  year={2018},
  publisher={RFC 8446}
}

@ARTICLE{mitls,  
author={Bhargavan, Karthikeyan and Fournet, Cédric and Kohlweiss, Markulf},  
journal={IEEE Security Privacy},   
title={{miTLS}: Verifying Protocol Implementations against Real-World Attacks},   
year={2016},  
volume={14},  
number={6},  
pages={18-25}
}

@article{leino2008boogie,
  title={This is {Boogie} 2},
  author={Leino, K Rustan M},
  journal={Manuscript KRML},
  year={2008},
  publisher={Citeseer}
}

@article{Jaroslav2013CompCertTSO,
    author = {\v{S}ev\v{c}\'{\i}k, Jaroslav and Vafeiadis, Viktor and Zappa Nardelli, Francesco and Jagannathan, Suresh and Sewell, Peter},
    title = {{CompCertTSO}: A Verified Compiler for Relaxed-Memory Concurrency},
    year = {2013},
    publisher = {ACM},
    volume = {60},
    number = {3},
    journal = {Journal of the ACM},
    articleno = {22}
}

@misc{wiki2021kernel,
   author = "{Wikipedia}",
   title = "Kernel (operating system)",
   Url = {https://en.wikipedia.org/wiki/Kernel\_(operating\_system)},
    year={2022}
}

@article{Woodcock2009formal,
    author = {Woodcock, Jim and Larsen, Peter Gorm and Bicarregui, Juan and Fitzgerald, John},
    title = {Formal Methods: Practice and Experience},
    year = {2009},
    publisher = {ACM},
    volume = {41},
    number = {4},
journal = {ACM Computing Surveys}
}

@ARTICLE{Abrial07formalmethods,
    author = {Jean-Raymond Abrial},
    title = {Formal Methods: Theory Becoming Practice},
    journal = {Journal of Universal Computer Science},
    volume = {13},
	number = {5},
    year = {2007},
    pages = {619--628}
}

@article{zhangsurvey,
  title={A Survey of Formal Verification Approaches for Practical Systems},
  author={Zhang, Qiao and Zhuo, Danyang and Wilcox, James},
  url={https://courses.cs.washington.edu/courses/cse551/15sp/projects/qz2013-danyangz-jrw12.pdf}
}

@inproceedings{chapman2014we,
  title={Are we there yet? 20 years of industrial theorem proving with {SPARK}},
  author={Chapman, Roderick and Schanda, Florian},
  booktitle={Int. Conf. on Interactive Theorem Proving},
  pages={17--26},
  year={2014},
  organization={Springer}
}

@article{ringer2019qed,
	author = {Talia Ringer and Karl Palmskog and Ilya Sergey and Milos Gligoric and Zachary Tatlock},
	issn = {2325-1107},
	journal = {Foundations and Trends{\textregistered} in Programming Languages},
	number = {2-3},
	pages = {102-281},
	title = {QED at Large: A Survey of Engineering of Formally Verified Software},
	volume = {5},
	year = {2019},
 }

@article{zhao2017high,
  title={High-assurance separation kernels: a survey},
  author={Zhao, Yongwang and San{\'a}n, David and Zhang, Fuyuan and Liu, Yang},
  journal={arXiv:1701.01535},
  year={2017}
}

@article{zhang2019survey,
  title={A survey on formal specification and verification of system-level achievements in industrial circles},
  author={Zhang, Feng and Niu, Wensheng and others},
  journal={Academic Journal of Computing \& Information Science},
  volume={2},
  number={1},
  year={2019},
  publisher={Francis Academic Press}
}

@inproceedings{fonseca2017study,
    author = {Fonseca, Pedro and Zhang, Kaiyuan and Wang, Xi and Krishnamurthy, Arvind},
    title = {An Empirical Study on the Correctness of Formally Verified Distributed Systems},
    year = {2017},
    publisher = {ACM},
    booktitle = {European Conference on Computer Systems},
    pages = {328–343}
}

@misc{cve,
	key= "Common Vulnerabilities and Exposures",
	title = "Common Vulnerabilities and Exposures",
	url = "https://www.cve.org/",
	urldate = "2025-10-31",
	year = {2025}
}

@online{livingreview,
    author={Huang, Li and Ebersold, Sophie and Kogtenkov, Alexander and Meyer, Bertrand and Liu, Yinglin},
	title = "Formally Verified Deployed Software Systems: a living review",
	url = "https://deployed-verified.org/",
	year = {from 2025}
}

@inproceedings{Mai2013Verifying,
    author = {Mai, Haohui and Pek, Edgar and Xue, Hui and King, Samuel Talmadge and Madhusudan, Parthasarathy},
    title = {Verifying Security Invariants in {ExpressOS}},
    year = {2013},
    publisher = {ACM},
    booktitle = {Int. Conf. on Architectural Support for Programming Languages and Operating Systems},
    pages = {293–304}
}

@techreport{seL4-whitepaper,
    Author = {{Gernot Heiser}},
    Institution = {{The se{L}4 Foundation}},
    Title = {The se{L}4 microkernel - an introduction},
    OPTUrl = {https://cdn.hackaday.io/files/1713937332878112/seL4-whitepaper.pdf},
    Year = {2020}
}

@inproceedings{blackham2011timing,
  title={Timing analysis of a protected operating system kernel},
  author={Blackham, Bernard and Shi, Yao and Chattopadhyay, Sudipta and Roychoudhury, Abhik and Heiser, Gernot},
  booktitle={Real-Time Systems Symposium},
  pages={339--348},
  year={2011},
  publisher={IEEE}
}

@article{klein2014comprehensive,
  title={Comprehensive formal verification of an {OS} microkernel},
  author={Klein, Gerwin and Andronick, June and Elphinstone, Kevin and Murray, Toby and Sewell, Thomas and Kolanski, Rafal and Heiser, Gernot},
  journal={Transactions on Computer Systems},
  volume={32},
  number={1},
  pages={1--70},
  year={2014},
  publisher={ACM}
}

@article{sewell2017high,
  title={High-assurance timing analysis for a high-assurance real-time operating system},
  author={Sewell, Thomas and Kam, Felix and Heiser, Gernot},
  journal={Real-Time Systems},
  volume={53},
  number={5},
  pages={812--853},
  year={2017},
  publisher={Springer}
}

@inproceedings{sewell2013translation,
  title={Translation validation for a verified {OS} kernel},
  author={Sewell, Thomas Arthur Leck and Myreen, Magnus O and Klein, Gerwin},
  booktitle={ACM SIGPLAN conference on Programming language design and implementation},
  pages={471--482},
  year={2013}
}

@article{fisher2017hacms,
  title={The {HACMS} program: using formal methods to eliminate exploitable bugs},
  author={Fisher, Kathleen and Launchbury, John and Richards, Raymond},
  journal={Phil. Trans. of the Royal Society A: Mathematical, Physical and Engineering Sciences},
  volume={375},
  number={2104},
  year={2017},
  publisher={The Royal Society Publishing}
}

@online{L4_family,
   author = "{Wikipedia}",
    Url = {https://en.wikipedia.org/wiki/L4_microkernel_family},
    Title = {{L4 microkernel family}},
    key={l4family},
    year={2022}
}

@online{verve_materials,
    Url = {https://www.cs.cmu.edu/~jyang2/talks/pldi2010_verve.pdf},
    Title = {{Safe to the Last Instruction: Automated Verification of a Type-Safe Operating System}},
    year={2021},
    key={-},
    urldate ={2021-08-06},
    date={2021-06-20}
}

@book{nipkow2002isabelle,
  title={Isabelle/HOL: a proof assistant for higher-order logic},
  author={Nipkow, Tobias and Paulson, Lawrence C and Wenzel, Markus},
  volume={2283},
  year={2002},
  publisher={Springer}
}

@online{Common_Criteria,
   author = "{Wikipedia}",
    Key = {commoncriteria},
    Url = {https://en.wikipedia.org/wiki/Common_Criteria},
    Title = {{Common Criteria}},
    year={2022}
}

@online{Rocq:25,
    Key = {Rocq},
    Url = {https://rocq-prover.org/},
    Title = {{Rocq Interactive Theorem Prover}},
    urldate ={2025-10-31},
    date={2025-03-12},
    year= {2025}
}

@book{barnes:03,
  title={High integrity software: the spark approach to safety and security: sample chapters},
  author={Barnes, John Gilbert Presslie},
  year={2003},
  publisher={Pearson Education}
}

@online{POLA,
   author = "{Wikipedia}",
    Key = {POLA},
    Url = {https://en.wikipedia.org/wiki/Principle_of_least_privilege},
    Title = {{Principle of least privilege}},
    year={2022}
}

@inproceedings{mi2019skybridge,
  title={Skybridge: Fast and secure inter-process communication for microkernels},
  author={Mi, Zeyu and Li, Dingji and Yang, Zihan and Wang, Xinran and Chen, Haibo},
  booktitle={EuroSys Conference},
  pages={1--15},
  year={2019}
}

@inproceedings{Sam:19,
  author    = {Sam Lasser and
               Chris Casinghino and
               Kathleen Fisher and
               Cody Roux},
  title     = {A Verified {LL(1)} Parser Generator},
  booktitle = {Int. Conf. on Interactive Theorem Proving},
  volume    = {141},
  pages     = {1--18},
  publisher = {Schloss Dagstuhl - Leibniz-Zentrum f{\"{u}}r Informatik},
  year      = {2019}
}

@inproceedings{yang2010safe,
  title={Safe to the last instruction: automated verification of a type-safe operating system},
  author={Yang, Jean and Hawblitzel, Chris},
  booktitle={SIGPLAN Conference on Programming Language Design and Implementation},
  pages={99--110},
  year={2010},
  publisher={ACM}
}

@inproceedings{Joshua:17,
  author    = {Joshua S. Auerbach and
               Martin Hirzel and
               Louis Mandel and
               Avraham Shinnar and
               J{\'{e}}r{\^{o}}me Sim{\'{e}}on},
  title     = {Handling Environments in a Nested Relat. Algebra with Combinators and an Impl. in a Verified Query Compiler},
  booktitle = {Int. Conf. on Managmt. of Data},
  pages     = {1555--1569},
  publisher = {ACM},
  year      = {2017}
}

@inproceedings{kastner2018compcert,
  title={CompCert: Practical experience on integrating and qualifying a formally verified optimizing compiler},
  author={K{\"a}stner, Daniel and Barrho, J{\"o}rg and W{\"u}nsche, Ulrich and Schlickling, Marc and Schommer, Bernhard and Schmidt, Michael and Ferdinand, Christian and Leroy, Xavier and Blazy, Sandrine},
  booktitle={European Congress Embedded Real-Time Software and Systems},
  pages={1--9},
  year={2018}
}

@inproceedings{leroy2016compcert,
  title={{CompCert} -- a formally verified optimizing compiler},
  author={Leroy, Xavier and Blazy, Sandrine and K{\"a}stner, Daniel and Schommer, Bernhard and Pister, Markus and Ferdinand, Christian},
  booktitle={Embedded Real Time Software and Systems},
  year={2016}
}

@article{leroy2009compcert,
  title={A Formally Verified Compiler Back-end},
  author={Leroy, Xavier},
  journal={Journal of Automated Reasoning},
  volume={43},
  number={4},
  year={2009},
  date={2009-11-04},
  pages={363--446}
 }

@article{song2019compcertm,
author = {Song, Youngju and Cho, Minki and Kim, Dongjoo and Kim, Yonghyun and Kang, Jeehoon and Hur, Chung-Kil},
title = {{CompCertM}: {CompCert} with {C-Assembly} Linking and Lightweight Modular Verification},
year = {2019},
publisher = {ACM},
volume = {4},
journal = {ACM Programming Languages},
articleno = {23}
}

@online{compcertweb,
    Key = {CompCert},
    Url = {https://compcert.org/},
    Title = {{CompCert Webpage}},
    year={2021},
    date ={2021-05-09},
    urldate={2021-08-06},
}

@inproceedings{letouzey2002new,
  title={A new extraction for {Coq}},
  author={Letouzey, Pierre},
  booktitle={Int. Work. on Types for Proofs and Programs},
  pages={200--219},
  year={2002},
  organization={Springer}
}

@inproceedings{yang2011finding,
  title={Finding and understanding bugs in C compilers},
  author={Yang, Xuejun and Chen, Yang and Eide, Eric and Regehr, John},
  booktitle={SIGPLAN Conference on Programming Language Design and Implementation},
  pages={283--294},
  year={2011},
  publisher ={ACM}
}

@article{tan2019verified,
  title={The verified CakeML compiler backend},
  author={Tan, Yong Kiam and Myreen, Magnus O and Kumar, Ramana and Fox, Anthony and Owens, Scott and Norrish, Michael},
  journal={Journal of Functional Programming},
  volume={29},
  year={2019},
  publisher={Cambridge University Press}
}

@online{cakemlrepo,
    Key = {CakeML Github},
    Url = {https://github.com/CakeML/cakeml},
    Title = {{CakeML repository}},
    year= {2021},
    date={2021-06-18},
    urldate={2021-08-06},
}

@article{kumar2014cakeml,
  title={{CakeML}: a verified impl. of {ML}},
  author={Kumar, Ramana and Myreen, Magnus O and Norrish, Michael and Owens, Scott},
  journal={SIGPLAN Not.},
  volume={49},
  number={1},
  pages={179--191},
  year={2014},
  publisher={ACM}
}

@article{abrahamsson2020proof,
  title={Proof-producing synthesis of CakeML from monadic HOL functions},
  author={Abrahamsson, Oskar and Ho, Son and Kanabar, Hrutvik and Kumar, Ramana and Myreen, Magnus O and Norrish, Michael and Tan, Yong Kiam},
  journal={Journal of Automated Reasoning},
  volume={64},
  number={7},
  pages={1287--1306},
  year={2020},
  publisher={Springer}
}

@inproceedings{gueneau2017verified,
  title={Verified characteristic formulae for {CakeML}},
  author={Gu{\'e}neau, Arma{\"e}l and Myreen, Magnus O and Kumar, Ramana and Norrish, Michael},
  booktitle={European Symposium on Programming},
  pages={584--610},
  year={2017},
  organization={Springer}
}

@inproceedings{tan2015verified,
  title={A verified type system for {CakeML}},
  author={Tan, Yong Kiam and Owens, Scott and Kumar, Ramana},
  booktitle={Symposium on the Implementation and Application of Functional Programming Languages},
  pages={1--12},
  year={2015}
}

@article{ericsson2019verified,
  title={A verified generational garbage collector for CakeML},
  author={Ericsson, Adam Sandberg and Myreen, Magnus O and Pohjola, Johannes {\AA}man},
  journal={Journal of Automated Reasoning},
  volume={63},
  number={2},
  pages={463--488},
  year={2019},
  publisher={Springer}
}

@inproceedings{myreen2021minimalistic,
  title={A minimalistic verified bootstrapped compiler (proof pearl)},
  author={Myreen, Magnus O},
  booktitle={Int. Conf. on Certified Programs and Proofs},
  pages={32--45},
  year={2021},
  publisher = {ACM}
}

@inproceedings{bard2019formally,
  title={Formally verified roundoff errors using SMT-based certificates and subdivisions},
  author={Bard, Joachim and Becker, Heiko and Darulova, Eva},
  booktitle={Int. Symp. on Formal Methods},
  pages={38--44},
  year={2019},
  organization={Springer}
}

@inproceedings{zhao2012formalizing,
  title={Formalizing the {LLVM} intermediate representation for verified program transformations},
  author={Zhao, Jianzhou and Nagarakatte, Santosh and Martin, Milo MK and Zdancewic, Steve},
  booktitle={POPL},
  pages={427--440},
  year={2012},
  publisher={ACM}
}

@inproceedings{zhao2012mechanized,
  title={Mechanized verification of computing dominators for formalizing compilers},
  author={Zhao, Jianzhou and Zdancewic, Steve},
  booktitle={Int. Conf. on Certified Programs and Proofs},
  pages={27--42},
  year={2012},
  organization={Springer}
}

@inproceedings{zhao2013formal,
  title={Formal verification of {SSA-based} optimizations for {LLVM}},
  author={Zhao, Jianzhou and Nagarakatte, Santosh and Martin, Milo MK and Zdancewic, Steve},
  booktitle={SIGPLAN conference on Programming Language Design and Implementation},
  pages={175--186},
  year={2013},
  publisher = {ACM}
}

@online{goodin2018failure,
  title={Failure to patch two-month-old bug led to massive Equifax breach},
  Url={https://arstechnica.com/information-technology/2017/09/massive-equifax-breach-caused-by-failure-to-patch-two-month-old-bug/},
  author={Goodin, Dan},
  year={2017},
  date={2017-09-14},
  urldate={2021-08-06}
}

@online{CVE-2017-5638,
  title={CVE-2017-5638},
  Url={https://nvd.nist.gov/vuln/detail/CVE-2017-5638},
  author={National Vulnerability Database},
  key={CVE-2017-5638},
  year={2017},
  date={2021-02-24},
  urldate={2021-08-06}
}

@online{cloudflare,
  title={Cloudflare Reverse Proxies are Dumping Uninitialized Memory},
  Url={https://bugs.chromium.org/p/project-zero/issues/detail?id=1139},
  key={issue of unintialized memory},
  year={2017},
  date={2017-02-20},
  urldate={2021-08-06}
}

@inproceedings{zinzindohoue2017hacl,
  title={{HACL*}: A verified modern cryptographic library},
  author={Zinzindohou{\'e}, Jean-Karim and Bhargavan, Karthikeyan and Protzenko, Jonathan and Beurdouche, Benjamin},
  booktitle={SIGSAC Conference on Computer and Communications Security},
  pages={1789--1806},
  year={2017},
  publisher = {ACM}
}

@online{haclrepo,
  title={HACL*: A High-Assurance Cryptographic Library},
  Url={https://github.com/project-everest/hacl-star},
  key={hacl},
  year={2021},
  date={2021-07-31},
  urldate={2021-08-06}
}

@inproceedings{chudnov2018continuous,
  title={Continuous formal verif. of {Amazon s2n}},
  author={Chudnov, Andrey and Collins, Nathan and Cook, Byron and Dodds, Joey and Huffman, Brian and MacC{\'a}rthaigh, Colm and Magill, Stephen and Mertens, Eric and Mullen, Eric and Tasiran, Serdar and others},
  booktitle={Int. Conf. on Comp.-Aided Verif.},
  pages={430--446},
  year={2018},
  publisher={Springer}
}

@inproceedings{dockins2016constructing,
  title={Constructing semantic models of programs with the software analysis workbench},
  author={Dockins, Robert and Foltzer, Adam and Hendrix, Joe and Huffman, Brian and McNamee, Dylan and Tomb, Aaron},
  booktitle={Working Conference on Verified Software: Theories, Tools, and Experiments},
  pages={56--72},
  year={2016},
  publisher={Springer}
}

@inproceedings{beringer2015verified,
  title={Verified Correctness and Security of OpenSSL $\{$HMAC$\}$},
  author={Beringer, Lennart and Petcher, Adam and Katherine, Q Ye and Appel, Andrew W},
  booktitle={$\{$USENIX$\}$ Security Symposium},
  pages={207--221},
  year={2015}
}

@online{s2nrepo,
  title={s2n Github Repository},
  Url={https://github.com/aws/s2n-tls},
  key={s2n},
  date={2021-08-06},
  year={2021},
  urldate={2021-08-06}
}

@online{saw,
  title={The software analysis workbench},
  Url={https://saw.galois.com/index.html},
  author = {Galois, Inc.},
  key={SAW},
  year={2021},
  date={2021-08-06},
  urldate={2021-08-06}
}

@inproceedings{lewis2003cryptol,
  title={Cryptol: High assurance, retargetable crypto development and validation},
  author={Lewis, Jeffrey R and Martin, Brad},
  booktitle={IEEE Military Communications Conference, 2003. MILCOM 2003.},
  volume={2},
  pages={820--825},
  year={2003},
  organization={IEEE}
}

@online{travisci,
  title={Travis CI Official Website},
  Url={https://www.travis-ci.com/},
  Key={Travis CI},
  year={2021},
  date={2021-08-06},
  urldate={2021-08-06},
  
}

@inproceedings{lescuyer2015provencore,
  title={{ProvenCore}: Towards a Verified Isolation Micro-Kernel.},
  author={Lescuyer, St{\'e}phane},
  booktitle={MILS@ HiPEAC},
  year={2015}
}

@online{provenrun,
  title={ProvenRun Website},
  Url={https://www.provenrun.com/},
  key={ProvenRun},
  year={2021},
  date={2021-07-07},
  urldate={2021-08-06},
  
}

@techreport{xlinx-whitepaper,
    Institution = {{XLINX}},
    Title = {Isolate Security-Critical Applications on Zynq UltraScale+ Devices},
    OPTUrl = {https://www.xilinx.com/support/documentation/white_papers/wp516-security-apps.pdf},
    Year = {2020}
}

@online{trustzone,
  title={TrustZone},
  Url={http://www.arm.com/products/processors/technologies/trustzone/index.php},
  key={TrustZone},
  year={2021},
  date={2021-08-06},
  urldate={2021-08-06}
}

@inproceedings{sabt2015trusted,
  title={Trusted execution environment: what it is, and what it is not},
  author={Sabt, Mohamed and Achemlal, Mohammed and Bouabdallah, Abdelmadjid},
  booktitle={Trustcom/BigDataSE/ISPA},
  volume={1},
  pages={57--64},
  year={2015},
  publisher={IEEE}
}

@article{gu2015deep,
  title={Deep specifications and certified abstraction layers},
  author={Gu, Ronghui and Koenig, J{\'e}r{\'e}mie and Ramananandro, Tahina and Shao, Zhong and Wu, Xiongnan and Weng, Shu-Chun and Zhang, Haozhong and Guo, Yu},
  journal={ACM SIGPLAN Notices},
  volume={50},
  number={1},
  pages={595--608},
  publisher={ACM},  
  year={2015}
}

@inproceedings{gu2016certikos,
  title={Certik{OS}: An extensible architecture for building certified concurrent {OS} kernels},
  author={Gu, Ronghui and Shao, Zhong and Chen, Hao and Wu, Xiongnan Newman and Kim, Jieung and Sj{\"o}berg, Vilhelm and Costanzo, David},
  booktitle={{USENIX} Symp. on OS Design and Implementation},
  pages={653--669},
  year={2016}
}

@article{costanzo2016end,
  title={End-to-end verification of information-flow security for {C} and {Assembly} programs},
  author={Costanzo, David and Shao, Zhong and Gu, Ronghui},
  journal={SIGPLAN Notices},
  volume={51},
  number={6},
  pages={648--664},
  year={2016},
  publisher={ACM}
}

@online{lmbench,
  title={LMbench - Tools for Performance Analysis},
  Url={https://lmbench.sourceforge.net/},
  key={LMbench},
  year={2013},
  date={2013-04-25},
  urldate={2021-08-06}
}

@online{mcertikos,
  title={{mCertiKOS Hypervisor}},
  Url={https://flint.cs.yale.edu/certikos/mcertikos.html\#mcertikos},
  key={mCertikOS},
  year={2021},
  date={2021-08-06},
  urldate={2021-08-06}
}

@inproceedings{owens2009better,
  title={A better x86 memory model: x86-TSO},
  author={Owens, Scott and Sarkar, Susmit and Sewell, Peter},
  booktitle={Int. Conf. on Theorem Proving in Higher Order Logics},
  pages={391--407},
  year={2009},
  organization={Springer}
}

@article{johnson1998178b,
  title={DO-178B: Software considerations in airborne systems and equipment certification},
  author={Johnson, Leslie A and others},
  journal={Crosstalk, October},
  volume={199},
  pages={11--20},
  year={1998}
  }

@article{baumann2012lessons,
  title={Lessons learned from microkernel verification--specification is the new bottleneck},
  author={Baumann, Christoph and Beckert, Bernhard and Blasum, Holger and Bormer, Thorsten},
  journal={arXiv:1211.6186},
  year={2012}
}

@article{deline2005boogiepl,
  title={BoogiePL: A typed procedural language for checking object-oriented programs},
  author={DeLine, Robert and Leino, K Rustan M},
  year={2005},
  publisher={Citeseer}
}

@inproceedings{de2008z3,
  title={Z3: An efficient {SMT} solver},
  author={De Moura, Leonardo and Bj{\o}rner, Nikolaj},
  booktitle={Int. Conf. on Tools and Algorithms for the Construction and Analysis of Systems},
  pages={337--340},
  year={2008},
  organization={Springer}
}

@inproceedings{baumann2011proving,
  title={Proving memory separation in a microkernel by code level verification},
  author={Baumann, Christoph and Bormer, Thorsten and Blasum, Holger and Tverdyshev, Sergey},
  booktitle={Int. Symp. on Object/Component/Service-Oriented Real-Time Distributed Computing Workshops},
  pages={25--32},
  year={2011},
  organization={IEEE}
}

@inproceedings{beckert2011integration,
  title={Integration of bounded model checking and deductive verification},
  author={Beckert, Bernhard and Bormer, Thorsten and Merz, Florian and Sinz, Carsten},
  booktitle={Int. Conf. on Formal Verification of Object-Oriented Software},
  pages={86--104},
  year={2011},
  organization={Springer}
}

@inproceedings{sinz2010precise,
  title={A precise memory model for low-level bounded model checking},
  author={Sinz, Carsten and Falke, Stephan and Merz, Florian},
  booktitle={Int. Conf. on Systems software verification},
  pages={7--7},
  year={2010}
}

@inproceedings{li2021secure,
  title={A Secure and Formally Verified Linux KVM Hypervisor},
  author={Li, Shih-Wei and Li, Xupeng and Gu, Ronghui and Nieh, Jason and Hui, John Zhuang},
  booktitle={IEEE Symposium on Security and Privacy},
  year={2021}
}

@inproceedings{dam2013formal,
  title={Formal verification of information flow security for a simple ARM-based separation kernel},
  author={Dam, Mads and Guanciale, Roberto and Khakpour, Narges and Nemati, Hamed and Schwarz, Oliver},
  booktitle={SIGSAC conference on Computer \& communications security},
  pages={223--234},
  year={2013},
  publisher={ACM}
}

@article{rushby1981design,
  title={Design and verification of secure systems},
  author={Rushby, John M},
  journal={SIGOPS Operating Systems Review},
  volume={15},
  number={5},
  pages={12--21},
  year={1981},
  publisher={ACM}
}

@article{heitmeyer2008applying,
  title={Applying formal methods to a certifiably secure software system},
  author={Heitmeyer, Constance and Archer, Myla and Leonard, Elizabeth and McLean, John},
  journal={Trans. on Software Engineering},
  volume={34},
  number={1},
  pages={82--98},
  year={2008},
  publisher={IEEE}
}

@inproceedings{gehrmann2011there,
  title={Are there good reasons for protecting mobile phones with hypervisors?},
  author={Gehrmann, Christian and Douglas, Heradon and Nilsson, Dennis Kengo},
  booktitle={Consumer Communications and Networking Conference},
  pages={906--911},
  year={2011},
  organization={IEEE}
}

@inproceedings{mcdermott2012separation,
  title={Separation virtual machine monitors},
  author={McDermott, John and Montrose, Bruce and Li, Margery and Kirby, James and Kang, Myong},
  booktitle={Computer Security Applications Conference},
  pages={419--428},
  year={2012}
}

@article{sailer2005shype,
  title={sHype: Secure hypervisor approach to trusted virtualized systems},
  author={Sailer, Reiner and Valdez, Enriquillo and Jaeger, Trent and Perez, Ronald and Van Doorn, Leendert and Griffin, John Linwood and Berger, Stefan and Sailer, Reiner and Valdez, Enriquillo and Jaeger, Trent and others},
  journal={Techn. Rep. RC23511},
  volume={5},
  year={2005}
}

@inproceedings{fox2010trustworthy,
  title={A trustworthy monadic formalization of the ARMv7 instruction set architecture},
  author={Fox, Anthony and Myreen, Magnus O},
  booktitle={Int. Conf. on Interactive Theorem Proving},
  pages={243--258},
  year={2010},
  organization={Springer}
}

@inproceedings{brumley2011bap,
  title={BAP: A binary analysis platform},
  author={Brumley, David and Jager, Ivan and Avgerinos, Thanassis and Schwartz, Edward J},
  booktitle={Int. Conf. on Computer-Aided Verification},
  pages={463--469},
  year={2011},
  organization={Springer}
}

@online{armv7manual,
  title={ARMv7-A architecture reference manual},
  Url={http://infocenter.arm.com/help/index.jsp? topic=/com.arm.doc.ddi0406c},
  key={ARMv7-manual},
  year={2021},
  date={2021-08-06},
  urldate={2021-08-06}
}

@inproceedings{leino2010dafny,
  title={Dafny: An automatic program verifier for functional correctness},
  author={Leino, K Rustan M},
  booktitle={Int. Conf. on Logic for Programming Artificial Intelligence and Reasoning},
  pages={348--370},
  year={2010},
  organization={Springer}
}

@article{mod1991procurement,
  title={The procurement of safety critical software in defence equipment},
  author={MoD, UK},
  journal={Interim Defence Standard 00-55, UK Ministry of Defence, Directorate of Standardization, Kentigern House},
  volume={65},
  year={1991}
}

@article{toyn1995cadi,
  title={CADi: An architecture for Z tools and its implementation},
  author={Toyn, Ian and McDermid, John A},
  journal={Software: Practice and Experience},
  volume={25},
  number={3},
  pages={305--330},
  year={1995},
  publisher={Wiley Online Library}
}

@inproceedings{king1999value,
  title={The value of verification: Positive experience of industrial proof},
  author={King, Steve and Hammond, Jonathan and Chapman, Rod and Pryor, Andy},
  booktitle={Int. Symp. on Formal Methods},
  pages={1527--1545},
  year={1999},
  organization={Springer}
}

@article{white2017formal,
  title={Formal verification: will the seedling ever flower?},
  author={White, Neil and Matthews, Stuart and Chapman, Roderick},
  journal={Phil. Trans. of the Royal Society A: Mathematical, Physical and Engineering Sciences},
  volume={375},
  number={2104},
  year={2017},
  publisher={The Royal Society Publishing}
}

@inproceedings{croxford1995breaking,
  title={Breaking through the {V and V} bottleneck},
  author={Croxford, Martin and Sutton, James},
  booktitle={International Eurospace-Ada-Europe Symposium},
  pages={344--354},
  year={1995},
  organization={Springer}
}

@article{chapman2000industrial,
  title={Industrial experience with SPARK},
  author={Chapman, Roderick},
  journal={SIGAda Ada Letters},
  volume={20},
  number={4},
  pages={64--68},
  year={2000},
  publisher={ACM}
}

@online{natsblog2013,
  url={https://web.archive.org/web/20130801003045/http://nats.aero:80/blog/2013/07/how-technology-is-transforming-air-traffic-management},
  author = {NATS},
  title={How technology is transforming air traffic management},
  key={NATSblog2013},
  date={2013-07-26},
  year={2013},
  urldate={2025-10-31}
}

@inproceedings{chapman2016industrial,
  title={Industrial experience with agile in high-integrity software development},
  author={Chapman, Roderick and White, N},
  booktitle={Safety-Critical Systems Symposium},
  pages={2--4},
  year={2016}
}

@article{authority2001sw01,
  title={{SW01—Regulatory} Objectives for Software Safety Assurance in {ATS} Equipment in Part {B} (Generic Requirements and Guidance) of {CAP670—Air} Traffic Services Safety Requirements},
  author={Authority, Civil Aviation},
  year={2001}
}

@incollection{stich2012clearance,
  title={Clearance of Flight Control Laws for Carefree Handling of Advanced Fighter Aircraft},
  author={Stich, Robert},
  booktitle={Optimization Based Clearance of Flight Control Laws},
  pages={421--442},
  year={2012},
  publisher={Springer}
}

@article{o2013automated,
  title={Automated verification of code autom. generated from {Simulink}},
  author={O’Halloran, Colin},
  journal={Aut. Soft. Eng.},
  volume={20},
  number={2},
  pages={237--264},
  year={2013},
  publisher={Springer}
}

@inproceedings{adams2005clawz,
  title={ClawZ: Cost-effective formal verification for control systems},
  author={Adams, Mark M and Clayton, Philip B},
  booktitle={Int. Conf. on Formal Engineering Methods},
  pages={465--479},
  year={2005},
  organization={Springer}
}

@inproceedings{vernon2010communication,
  title={Communication systems in clawz},
  author={Vernon, Michael and Zeyda, Frank and Cavalcanti, Ana},
  booktitle={Int. Conf. on Abstract State Machines, Alloy, B and Z},
  pages={334--348},
  year={2010},
  organization={Springer}
}

@online{simulink,
  Url={https://www.mathworks.com/products/simulink.html},
  title={Simulink},
  key={Simulink},
  date={2021-08-06},
  year={2021},
  urldate={2021-08-06}
}

@book{abiteboul1995foundations,
  title={Foundations of Databases},
  author={Abiteboul, Serge and Hull, Richard and Vianu, Victor},
  volume={8},
  year={1995},
  publisher={Addison-Wesley Reading}
}

@incollection{floyd1993assigning,
  title={Assigning meanings to programs},
  author={Floyd, Robert W},
  booktitle={Program Verification},
  pages={65--81},
  year={1993},
  publisher={Springer}
}

@inproceedings{jang2012establishing,
  title={Establishing browser security guarantees through formal shim verification},
  author={Jang, Dongseok and Tatlock, Zachary and Lerner, Sorin},
  booktitle={USENIX Security Symposium},
  pages={113--128},
  year={2012}
}

@book{jang2014language,
  title={Language-based Security for Web Browsers},
  author={Jang, Dongseok},
  year={2014},
  publisher={University of California, San Diego}
}

@online{quarkweb,
  Url={https://github.com/Conservatory/quark},
  title={Quark: A Web Browser with a Formally Verified Kernel},
  key={Quark},
  year={2021},
  date={2021-08-06},
  urldate={2021-08-06}
}

@article{popescu2021cocon,
  title={CoCon: A conference management system with formally verified document confidentiality},
  author={Popescu, Andrei and Lammich, Peter and Hou, Ping},
  journal={Journal of Automated Reasoning},
  volume={65},
  number={2},
  pages={321--356},
  year={2021},
  publisher={Springer}
}

@inproceedings{kanav2014conference,
  title={A conference management system with verified document confidentiality},
  author={Kanav, Sudeep and Lammich, Peter and Popescu, Andrei},
  booktitle={Int. Conf. on Computer-Aided Verification},
  pages={167--183},
  year={2014},
  publisher={Springer}
}

@inproceedings{becker2018verified,
  title={A verified certificate checker for finite-precision error bounds in {Coq} and {HOL4}},
  author={Becker, Heiko and Zyuzin, Nikita and Monat, Rapha{\"e}l and Darulova, Eva and Myreen, Magnus O and Fox, Anthony},
  booktitle={Formal Methods in Computer Aided Design},
  pages={1--10},
  year={2018},
  publisher={IEEE}
}

@online{floverrepo,
  Url={https://gitlab.mpi-sws.org/AVA/FloVer},
  title={FloVer: A certificate checker for roundoff error bounds},
  key={Flover},
  year={2021},
  date={2021-07-06},
  urldate={2021-08-06},
  
}

@inproceedings{darulova2018daisy,
  title={Daisy-framework for analysis and optimization of numerical programs (tool paper)},
  author={Darulova, Eva and Izycheva, Anastasiia and Nasir, Fariha and Ritter, Fabian and Becker, Heiko and Bastian, Robert},
  booktitle={Int. Conf. on Tools and Algorithms for the Construction and Analysis of Systems},
  pages={270--287},
  year={2018},
  organization={Springer}
}

@article{cosmed1,
  title={CoSMed: A confidentiality-verified social media platform},
  author={Bauerei{\ss}, Thomas and Gritti, Armando Pesenti and Popescu, Andrei and Raimondi, Franco},
  journal={Journal of Automated Reasoning},
  volume={61},
  number={1},
  pages={113--139},
  year={2018},
  publisher={Springer}
}

@inproceedings{cosmedis,
  title={CoSMeDis: a distributed social media platform with formally verified confidentiality guarantees},
  author={Bauerei{\ss}, Thomas and Gritti, Armando Pesenti and Popescu, Andrei and Raimondi, Franco},
  booktitle={Symposium on Security and Privacy},
  pages={729--748},
  year={2017},
  organization={IEEE}
}

@online{cloudant,
  Url={https://www.ibm.com/cloud/cloudant},
  title={Anatomy of the Cloudant DBaaS},
  key={Cloudant},
  year = {2021},
  date={2021-08-06},
  urldate={2021-08-06}
}

@article{braibant2014implementing,
  title={Implementing and reasoning about hash-consed data structures in Coq},
  author={Braibant, Thomas and Jourdan, Jacques-Henri and Monniaux, David},
  journal={Journal of automated reasoning},
  volume={53},
  number={3},
  pages={271--304},
  year={2014},
  publisher={Springer}
}

@inproceedings{myreen2014reflective,
  title={The reflective Milawa theorem prover is sound},
  author={Myreen, Magnus O and Davis, Jared},
  booktitle={Int. Conf. on Interactive Theorem Proving},
  pages={421--436},
  year={2014},
  organization={Springer}
}

@inproceedings{myreen2011verified,
  title={A verified runtime for a verified theorem prover},
  author={Myreen, Magnus O and Davis, Jared},
  booktitle={Int. Conf. on Interactive Theorem Proving},
  pages={265--280},
  year={2011},
  organization={Springer}
}

@online{mm0repo,
  Url={https://github.com/digama0/mm0},
  title={Metamath Zero},
  key = {mm0},
  year = {2021},
  date={2021-07-04},
  urldate={2021-08-06}
}

@article{carneiro2020metamath,
  author    = {Mario M. Carneiro},
  title     = {Metamath Zero: The Cartesian Theorem Prover},
  journal   = {CoRR},
  volume    = {abs/1910.10703},
  year      = {2020},
  url       = {https://arxiv.org/abs/1910.10703},
  archivePrefix = {arXiv},
  eprint    = {1910.10703},
  primaryClass = {cs.LO}
}

@inproceedings{leinenbach2009verifying,
  title={Verifying the {Microsoft Hyper-V} hypervisor with {VCC}},
  author={Leinenbach, Dirk and Santen, Thomas},
  booktitle={Int. Symp. on Formal Methods},
  pages={806--809},
  year={2009},
  organization={Springer}
}

@inproceedings{cohen2009vcc,
  title={VCC: A practical system for verifying concurrent C},
  author={Cohen, Ernie and Dahlweid, Markus and Hillebrand, Mark and Leinenbach, Dirk and Moskal, Micha{\l} and Santen, Thomas and Schulte, Wolfram and Tobies, Stephan},
  booktitle={Int. Conf. on Theorem Proving in Higher Order Logics},
  pages={23--42},
  year={2009},
  organization={Springer}
}

@inproceedings{dahlweid2009vcc,
  title={VCC: Contract-based modular verification of concurrent C},
  author={Dahlweid, Markus and Moskal, Michal and Santen, Thomas and Tobies, Stephan and Schulte, Wolfram},
  booktitle={Int. Conf. on Software Engineering-Companion Volume},
  pages={429--430},
  year={2009},
  organization={IEEE}
}

@inproceedings{badeau2005using,
  title={Using {B} as a high level programming language in an industrial project: {Roissy VAL}},
  author={Badeau, Fr{\'e}d{\'e}ric and Amelot, Arnaud},
  booktitle={Int. Conf. of B and Z Users},
  pages={334--354},
  year={2005},
  organization={Springer}
}

@book{tanenbaum1997operating,
  title={Operating systems: design and implementation},
  author={Tanenbaum, Andrew S and Woodhull, Albert S},
  volume={68},
  year={1997},
  publisher={Prentice Hall Englewood Cliffs}
}

@inproceedings{10.1007/978-1-4471-2061-2_19,
  title={The rigorous retrospective static analysis of the {Sizewell ‘B’} Primary Protection System software},
  author={Ward, NJ},
  booktitle={Int. Conf. on Computer Safety, Reliability and Security},
  pages={171--181},
  year={1993},
  organization={Springer}
}

@article{chen2008type,
  title={Type-preserving compilation for large-scale optimizing object-oriented compilers},
  author={Chen, Juan and Hawblitzel, Chris and Perry, Frances and Emmi, Mike and Condit, Jeremy and Coetzee, Derrick and Pratikaki, Polyvios},
  journal={SIGPLAN Notices},
  volume={43},
  number={6},
  pages={183--192},
  year={2008},
  publisher={ACM}
}

@inproceedings{fahndrich2006language,
  title={Language support for fast and reliable message-based communication in Singularity OS},
  author={F{\"a}hndrich, Manuel and Aiken, Mark and Hawblitzel, Chris and Hodson, Orion and Hunt, Galen and Larus, James R and Levi, Steven},
  booktitle={SIGOPS/EuroSys European Conference on Computer Systems},
  pages={177--190},
  year={2006},
  publisher={ACM}
}

@InProceedings{hawblitzel2009automated,
author = {Hawblitzel, Chris and Petrank, Erez},
title = {Automated Verification of Practical Garbage Collectors},
booktitle = {POPL},
year = {2009},
publisher = {ACM}
}

@inproceedings{declarative-guis-2018,
author = {Adelsberger, Stephan and Setzer, Anton and Walkingshaw, Eric},
title = {Declarative {GUI}s: Simple, Consistent, and Verified},
year = {2018},
isbn = {9781450364416},
publisher = {ACM}, 
booktitle = {Int. Symp. on Principles and Practice of Declarative Programming}
}

@online{questionnaire,
    author={Huang, Li and Ebersold, Sophie and Kogtenkov, Alexander and Naumchev, Alexandr and Meyer, Bertrand and Liu, Yinglin},
    Url = {https://bit.ly/2LMxbZB},
    Title = {{Questionnaire: Formally Verified Systems}},
    year= {2021},
    date={2021-10-22},
    urldate={2021-10-22}
}

@online{practical-fm,
    Key = {companies that use formal verification methods},
    Url = {https://github.com/ligurio/practical-fm},
    Title = {{List of companies using formal verification methods in soft. eng.}},
    year= {2021},
    date={2021-10-22},
    urldate={2021-10-22}
}

@book{deogun2019secure,
  title={Secure By Design},
  author={Deogun, D. and Johnsson, D.B. and Sawano, D.},
  isbn={9781617294358},
  lccn={2019286713},
  year={2019},
  publisher={Manning}
}

@inproceedings{DutchTunnelControlSystem,
title = "Formal Verification of an Industrial Safety-Critical Traffic Tunnel Control System",
author = "Wytse Oortwijn and Marieke Huisman",
year = "2019",
publisher = "Springer",
pages = "418--436",
booktitle = "Integrated Formal Methods"
}

@InProceedings{10.1007/978-3-030-61467-6_18,
author="Huisman, Marieke
and Monti, Ra{\'u}l E.",
title="On the Industrial Application of Critical Software Verification with VerCors",
booktitle="Leveraging Applications of Formal Methods, Verification and Validation: Applications",
year="2020",
publisher="Springer",
pages="273--292"
}

@inproceedings{klein2014proof,
  title={Proof engineering considered essential},
  author={Klein, Gerwin},
  booktitle={Int. Symp. on Formal Methods},
  pages={16--21},
  year={2014},
  organization={Springer}
}

@article{herklotz2021formal,
  title={Formal verification of high-level synthesis},
  author={Herklotz, Yann and Pollard, James D and Ramanathan, Nadesh and Wickerson, John},
  journal={ACM on Programming Languages},
  volume={5},
  pages={1--30},
  year={2021}
}

@inbook{history2014itp,
author = {Harrison, John and Urban, Josef and Wiedijk, Freek},
year = {2014},
pages = {135--214},
title = {History of Interactive Theorem Proving},
volume = {9},
journal = {Handbook of the History of Logic}
}

@online{hol-itp,
    key = {HOL},
    url = {https://hol-theorem-prover.org/},
    title = {{HOL} Interactive Theorem Prover},
    year= {2021},
    date={2021-12-21},
    urldate={2021-12-21}
}

@online{isabelle,
    key = {Isabelle},
    url = {https://isabelle.in.tum.de/},
    title = {Isabelle},
    year= {2021},
    date={2021-12-12},
    urldate={2021-12-12}
}

@article{Isabelle/Isar,
author = {Wenzel, Markus and Munchen, Tu},
year = {2001},
month = {03},
pages = {},
title = {The Isabelle/Isar Reference Manual}
}

@article{odersky2004overview,
  title={An overview of the Scala programming language},
  author={Odersky, Martin and Altherr, Philippe and Cremet, Vincent and Emir, Burak and Maneth, Sebastian and Micheloud, St{\'e}phane and Mihaylov, Nikolay and Schinz, Michel and Stenman, Erik and Zenger, Matthias},
  year={2004}
}

@inbook{EventB,
author = {Hoang, Thai Son},
year = {2013},
pages = {211-236},
title = {An Introduction to the Event-B Modelling Method}
}

@techreport{keele2007guidelines,
  title={Guidelines for performing systematic literature reviews in soft. eng.},
  author={Keele, Staffs and others},
  year={2007},
  institution={Citeseer}
}

@article{HATCLIFF2012,
author = {Hatcliff, John and Leavens, Gary T. and Leino, K. Rustan M. and M\"{u}ller, Peter and Parkinson, Matthew},
title = {Behavioral Interface Specification Languages},
year = {2012},
issue_date = {June 2012},
publisher = {ACM},
hide_address = {New York, USA},
volume = {44},
number = {3},
issn = {0360-0300},
hide_url = {https://doi.org/10.1145/2187671.2187678},
hide_doi = {10.1145/2187671.2187678},
journal = {ACM Comput. Surv.},
month = {jun},
articleno = {16},
numpages = {58}
}

@inproceedings{Everest,
  TITLE = {{Everest: Towards a Verified, Drop-in Replacement of {HTTPS}}},
  AUTHOR = {Bhargavan, Karthikeyan and Bond, Barry and Delignat-Lavaud, Antoine and Fournet, C{\'e}dric and others},
  BOOKTITLE = {{Summit on Advances in Programming Languages}},
  YEAR = {2017}
}

@misc{
stateflow,
title = {{Matlab Stateflow User Guide}},
howpublished = {\url{https://www.mathworks.com/products/stateflow/}}
}

@article{sanchez_survey_2019,
	title = {A survey of challenges for runtime verification from advanced application domains (beyond software)},
	author = {S{\'a}nchez, César and et al.},
    discard =  { and Schneider, Gerardo and Ahrendt, Wolfgang and Bartocci, Ezio and Bianculli, Domenico and Colombo, Christian and Falcone, Yliès and Francalanza, Adrian and Krstić, Srđan and Lourenço, Joa̋o M. and Nickovic, Dejan and Pace, Gordon J. and Rufino, Jose and Signoles, Julien and Traytel, Dmitriy and Weiss, Alexander},
    journal={Form Methods Syst Des},
    volume = {54},
	number = {3},
    year={2019},
    date = {2019-11-01},
	pages = {279--335},
	}

@article{bartocci_rv,
  title={Introduction to runtime verification},
  author={Bartocci, Ezio and Falcone, Yli{\`e}s and Francalanza, Adrian and Reger, Giles},
  journal={Lectures on Runtime Verification: Introductory and Advanced Topics},
  pages={1--33},
  year={2018},
  publisher={Springer}
}

@article{falcone_rv,
  title={A tutorial on runtime verification},
  author={Falcone, Yli{\`e}s and Havelund, Klaus and Reger, Giles},
  journal={Eng. dependable soft. syst.},
  pages={141--175},
  year={2013},
  publisher={IOS Press}
}

@inproceedings{le2014emi,
author = {Le, Vu and Afshari, Mehrdad and Su, Zhendong},
title = {Compiler validation via equivalence modulo inputs},
year = {2014},
isbn = {9781450327848},
publisher = {ACM},
hide_publisher = {Association for Computing Machinery},
hide_address = {New York, NY, USA},
hide_url = {https://doi.org/10.1145/2594291.2594334},
hide_doi = {10.1145/2594291.2594334},
booktitle = {Proceedings of the 35th ACM SIGPLAN Conference on Programming Language Design and Implementation},
pages = {216–226},
numpages = {11},
location = {Edinburgh, United Kingdom},
series = {PLDI '14}
}

@inproceedings{ynot2008,
author = {Nanevski, Aleksandar and Morrisett, Greg and Shinnar, Avraham and Govereau, Paul and Birkedal, Lars},
title = {Ynot: dependent types for imperative programs},
year = {2008},
isbn = {9781595939197},
publisher = {ACM},
hide_publisher = {Association for Computing Machinery},
hide_address = {New York, NY, USA},
hide_url = {https://doi.org/10.1145/1411204.1411237},
hide_doi = {10.1145/1411204.1411237},
booktitle = {Proceedings of the 13th ACM SIGPLAN International Conference on Functional Programming},
pages = {229–240},
numpages = {12},
location = {Victoria, BC, Canada},
series = {ICFP '08}
}

\end{document}